%% file: main.tex
\documentclass[runningheads]{llncs}
\usepackage[T1]{fontenc}
\usepackage[hyphens]{url}  % DO NOT CHANGE THIS
\usepackage{graphicx} % DO NOT CHANGE THIS
\usepackage{tikz}
\usetikzlibrary{decorations.pathreplacing}
\usetikzlibrary{patterns}
\usepackage[square,sort,comma,numbers]{natbib}  % DO NOT CHANGE THIS AND DO NOT ADD ANY OPTIONS TO IT
\usepackage{caption} % DO NOT CHANGE THIS AND DO NOT ADD ANY OPTIONS TO IT
\usepackage[linesnumbered,ruled,vlined]{algorithm2e} % For algorithms
\usepackage{physics}
\usepackage{amsmath,amsfonts,enumerate}
\usepackage[capitalise, nameinlink]{cleveref}

\usepackage{amssymb}
\usepackage{gensymb}
\usepackage{soul}
\usepackage{xcolor}
\usepackage{subcaption}

\newtheorem{observation}{Observation}
\DeclareMathOperator*{\argmin}{argmin}

\crefname{algorithm}{mechanism}{mechanisms}
\Crefname{algorithm}{Mechanism}{Mechanisms}

\newcommand{\OPT}{\operatorname{OPT}}
\newcommand{\opt}{\operatorname{opt}}
\newcommand{\cost}{\operatorname{cost}}
\newcommand{\COST}{\operatorname{SC}} % RJC changed this as a fast update
\newcommand{\SC}{\operatorname{SC}}

\newcommand{\med}{\operatorname{med}}
\newcommand{\medp}{\operatorname{medp}}
\newcommand{\loc}{\ell}
\newcommand{\inst}{\mathcal{I}}
\newcommand{\error}{\mathcal{E}}
\newcommand{\interval}{\mathcal{L}}
\newcommand{\region}{\mathcal{R}}
\newcommand{\range}{\mathcal{T}}
\newcommand{\rangecost}{\mathcal{C}}
\newcommand{\OC}{\mathcal{O}}
\newcommand{\prob}{\mathbb{P}}
\newcommand{\E}{\mathbb{E}}
\newcommand{\bbar}{\overline{b}}

\newcommand{\ix}{x_i} % \xi is already taken.
\newcommand{\bi}{b_i}

\newcommand{\tilb}{\Tilde{b}}
\newcommand{\tilx}{\Tilde{x}}
\newcommand{\la}{\leftarrow}
\newcommand{\ystar}{\loc^*}

\newcommand{\hide}[1]{}
\title{Facility Location with Public Locations and Private 
  Doubly-Peaked Costs}

\author{Richard Cole\insti{1} \and
Pranav Jangir\insti{1}}
\authorrunning{R. Cole and P. Jangir}
\institute{New York University, NY 10012, USA \\
\email{cole@cs.nyu.edu, pj2251@nyu.edu}}

\begin{document}

\maketitle

\begin{abstract}
In the facility location problem, the task is to place one or more facilities so as to minimize the sum of the  agent costs for accessing their nearest facility. Heretofore, in the strategic version, agent locations have been assumed to be private, while their cost measures have been public and identical.
    For the most part, the cost measure has been the distance to the nearest facility.

    However, in multiple natural settings, such as placing a firehouse or a school, this modeling does not appear to be a good fit. For it seems natural that the agent locations would be known, but their costs  might be private information. In addition, for these types of settings, agents may well want the nearest facility to be at the right distance: near, but not too near. This is captured by the doubly-peaked cost introduced in~\cite{filos2017facility}.

    In this paper, we re-examine the facility location problem from this perspective: known agent locations and private preferred distances to the nearest facility.
    We then give lower and upper bounds on achievable approximations, focusing on the problem in 1D, and in 2D with an $L_1$ distance measure.
\keywords{Facility Location \and Algorithmic Game Theory \and Mechanism Design \and Approximation Algorithms.}

\end{abstract}

\section{Introduction}\label{sec::intro}
The strategic version of the facility location problem was formulated in~\cite{ProcTen2013}; the authors noted that it captured decisions about public goods; indeed, where to locate a facility such as a school, but also, for example, issues such as the level of taxation. In these settings there are $n$ agents; the task is to locate a facility so that the sum of the distances from the facility to the agents is minimized, where distance is appropriately defined.
Each agent's goal is to be as close to the facility as possible.
The agents' positions are private information, creating a need for mechanisms.
The simplest version of the problem has the agents located on the real line.
One simple incentive compatible mechanism is to place the facility at the median of the agents' locations,
as noted in~\cite{Moulin1980} in the context of voting.

In some applications, for example the placement of a firehouse or school, it is often desirable to be close but not too close to the facility. This version of the problem was introduced in~\cite{filos2017facility}. The authors supposed all agents shared the same preferred distance to the facility. For each agent, this was captured by a cost function of the form $c+\Delta$ where $\Delta$ was the distance from the facility to the agent's preferred location for the facility (the reason for the parameter $c$ was to enable finite multiplicative approximation factors). Actually, if the agents are located on the real line, each agent would have two preferred locations and $\Delta$ would be the distance from the facility to the nearer preferred location. All agents had the same cost function. Again, the private information was the agent locations.

In our view, for these sorts of applications, it seems likely that the agent locations will be public information. For the relevant agents for a school or firehouse are the residents, and presumably the locations of their residences are known. On the other hand, it seems plausible that residents will vary regarding how near or far they want the facility to be.
Accordingly, in this paper, we undertake a systematic study of this version of the problem in which the desired distance to the facility is private information.

\subsection{Doubly-Peaked Preferences}

Double-peaked preferences are an important concept in social choice theory, where they describe situations where preferences are not single-peaked, leading to voting cycles and making majority elections difficult (e.g. see \cite{cooter2000strategic}). In political economy, voters or political parties might prefer policies that are either aligned with their ideology or highly effective but ideologically distant, rather than moderate policies perceived as ineffective. For instance, as noted in \cite{egan2014something}, during the Vietnam War, some Americans preferred either complete withdrawal or intensified military action, rather than maintaining the status quo. By incorporating doubly-peaked preferences, economic models can better capture the complex and realistic behavior seen in various social, political, and economic contexts. Apart from locating schools or factories in a neighbourhood, doubly-peaked preferences also capture scenarios where parents might prefer either high public school expenditure (to send their children to good public schools) or low expenditure (to pay less taxes and send children to private schools), rather than moderate spending levels. 
\hide{Many more examples of Doubly-Peaked preferences like forming environmental polices are covered in \pj{unable to cite}.}

These related works highlight that Double-Peaked preferences do have applications in settings more general than literally locating facilities in space. The primary focus of this paper, however, is to study Doubly-Peaked preferences on facility location settings and therefore the modelling assumptions follow the ones normally followed in the facility location literature.

\subsection{Our Results}

Examining the approximation bounds obtained in~\cite{filos2017facility}, we see that they are really additive bounds, where the natural cost function is the distance $\Delta$ from the facility to the nearest preferred location. In order to obtain a multiplicative approximation, Filos-Ratsikas et al.\ used a cost of the form $c+\Delta$, with $c>0$ being a suitable parameter.
We prefer to measure the approximation in terms of the additive error, though our bounds could be converted to multiplicative errors in the same way as in Filos-Ratsikas et al., as shown in the appendix.

To distinguish the original version of the problem introduced by Procaccia et al., we call it the private location setting; in contrast, we call the version we are introducing the private cost setting.
We obtain the following results.

\begin{itemize}
    \hide{\item In 1D, there is a cost-preserving reduction from the private location to the private cost setting, for the $k$-facility problem, for all $k$.} 
    \item In 1D, supposing there is a bound $B$ on the agents' preferred distances, any randomized incentive compatible mechanism cannot achieve an approximate solution with expected cost better than $\OPT + \alpha Bn$, where $n$ is the number of agents, $\alpha>0$ is a constant, and $\OPT$ is the cost of an optimal solution. This bound extends to 2D when using $L_1$ or $L_2$ distances.
    \item For the 1D problem, placing the facility at location $\med$, the median of the agent's locations, achieves an approximation of $\OPT + nB$. A similar bound, $\OPT + 2nB$, holds in 2D when using $L_1$ or $L_2$ distances. We call this the Median algorithm.
    \item In 1D, we give a deterministic mechanism, called Median-Plus, which always achieves a cost as good as the Median algorithm, and can achieve a $\Theta(nB)$ improvement in cost.
    We also extend this to a 2D mechanism, 2D-Median-Plus, when using $L_1$ distances; this extension is a true 2-dimensional computation; it does not reduce to two 1D computations.
    \item When the average distances are less than $B$, we obtain better bounds for both Median-Plus and 2D-Median-Plus: If there are $r$ agents in the range $\med \pm B$, and the average preferred distance for these agents is $\bbar$, then the median algorithm, and hence the Median-Plus mechanisms, achieve a cost of at most $\OPT + 2r\bbar$. 
    
 %   \item In 1D, we show the optimal location for the facility lies in the range $\med \pm B$. \rnote{Not sure about listing this.} \pj{I don't think we should list this.. not a main result.}
%    \item results for two facilities in 1D. \rnote{Do we have anything?} \pj{just one result for two facilities.. we can add as a remark that it works for $k$ facilities, but the two facility result is weaker.}
    \end{itemize}

\subsection{Related Work}

In \cite{ProcTen2013}, the authors used the facility location problem as a vehicle for studying mechanisms that were approximately optimal in settings without money, such as voting or public facility placement.
For as they showed, strategy-proof optimal mechanisms are often not possible in these settings.
They studied both the social cost version of the problem, introduced above, and a max-cost version,
where the goal is to minimize the maximum of the individual agent's costs. For the latter problem
they gave a randomized mechanism achieving a multiplicative approximation factor of $3/2$, and showed a lower bound of 2 for deterministic mechanisms. Earlier, in \cite{Moulin1980}, the author characterized strategy-proof mechanisms in this setting in the context of voting.

Filos-Ratsikas et al.~\cite{filos2017facility} give a randomized mechanism for the private location setting
using the doubly-peaked cost function, achieving cost $\OPT+ nB$; here $B$ is the (same) preferred distance for all the agents. They also show that in their setting, deterministic mechanisms have $\OPT+\Omega(nB)$ cost.

In~\cite{GolTza2017}, the authors argued that measuring the approximation via an additive error rather than a multiplicative factor can be more meaningful in some of these settings. For example, when locating a school, the difference between a 1km and a 2km distance may be much more significant that the difference between 100m and 200m. In \cite{CaiFT16}, the authors gave an approximate algorithm for minimizing the maximum envy with additive error bounds.

\hide{
The problem of locating two facilities, the 2-facility problem, was also considered in~\cite{ProcTen2013}. The authors gave a mechanism achieving an $(n-2)$-approximate factor, and a lower bound of $3/2$. Following a series of works~\cite{LuWZ2009}, \cite{lu2010asymptotically}, a tight $n-2$ lower bound was shown in ~\cite{FotakisTzamos2014}. In~\cite{LuSWZ2010}, the authors gave an ingenious randomized strategy-proof mechanism achieving an approximation factor of 4. The $k$-facility problem was also studied in~\cite{FotakisTzamos2016}; the authors gave a randomized strategy-proof mechanism which achieves both an approximation factor of $n$ for the social cost and of 2 for the maximum cost.
}

Many other variants of the problem have been studied including limiting the possible locations
\cite{FeldmanFiat2016}; verification (meaning forcing an agent to use the closest declared facility)~\cite{FotakisTzamos2010, nissim2012approximately}; public agent locations with private facility preferences \cite{fong2018facility};
obnoxious facilities~\cite{CHENGYZ2013, ChengHYZ2019}; other geometries including the circle~\cite{AlonFPT2010} and $L_p$ norms~\cite{FeigSY2016};
capacitated facilities~\cite{AzizFLLW2021}; reducing variance~\cite{ProcacciaWajcZhang2018}; $\epsilon$-approximate strategy proofness~\cite{SuiBout2015, nissim2012approximately}. 
The problem of locating two facilities, the 2-facility problem, was also considered in~\cite{ProcTen2013};
see also \cite{LuWZ2009,lu2010asymptotically,FotakisTzamos2014,LuSWZ2010}. The $k$-facility problem was studied in~\cite{FotakisTzamos2016}.
An informative recent survey provides a substantially more detailed and thorough overview~\cite{ChanEtal2021}. 

% Doubly peak preferences work of \cite{filos2017facility}.

\subsection{Roadmap}

\hide{We review notation in the next section and explain how to extend a notion of partial group strategy-proofness to our setting. Then, in \Cref{sec::reduction}, we show a cost-preserving reduction from the standard problem to our setting, which implies that the multi-facility hardness results apply in our setting too.
In \Cref{sed::MedAlg} we describe the Median algorithm.
In \Cref{sec::med-plus}, we follow this with a deterministic strategy-proof mechanism, the Median-Plus mechanism, which always performs as well as the Median algorithm and can perform significantly better;
in \Cref{sec::2D-MedPlus}, we create a 2D version of the mechanism for $L_1$ distance measures.
In \Cref{sec::hardness}, we complement these results with a tight lower bound.
We conclude in \Cref{sec::concl}.}
We review notation in the next section and explain how to extend a notion of partial group strategy-proofness to our setting. Then, we show the lower bound for the 1D case, followed by the design and analysis of stratetgyproof mechanisms for the 1D case. The remaining results are deferred to the appendix. 
\section{Preliminaries}

Let $N = \{1, 2, ... n\}$ denote the set of agents and $n = |N|$ be the total number of agents. Let $\Omega$ denote the space of possible agent locations. We consider two cases: when agents are located on a line ($\Omega = \mathbb{R}$) and on a 2D plane ($\Omega = \mathbb{R}^2$).
Agent $i$ is located at a fixed position $\ix \in \Omega$ and has a preferred distance of $\bi \leq B$
between its position and the facility location, where $B \in \mathbb{R}$ is a constant denoting the bound on the preferred distances. We call the tuple $\langle(x_1, b_1), (x_2, b_2), ... , (x_n, b_n)\rangle$ an \emph{instance}. For ease of notation we sometimes define an instance $\inst$ as $\inst = \langle t_1, t_2, ... , t_n \rangle$, where $t_i \in \Omega \cross \mathbb{R}$ for all $i \in [n]$.

A \emph{deterministic mechanism} is a function $f : (\Omega \cross \mathbb{R})^n \rightarrow \Omega$ that maps a given instance to a point in $\Omega$, the location of the facility. 

When agents are located on a line and the facility is placed at $y \in \mathbb{R}$, the cost incurred by agent $i$ is given by
\begin{align*}
    \cost(y, (x_i, b_i)) = \begin{cases}
        |x_i - b_i - y| & \text{if $y \leq x_i$} \\
        |x_i + b_i - y| & \text{if $y > x_i$}
    \end{cases}
\end{align*}

When agents are located on a 2D plane and the facility is placed at $y \in \mathbb{R}^2$, the cost incurred by agent $i$ is given by
\begin{align*}
    \cost(y, (x_i, b_i)) = \begin{cases}
        b_i - \norm{x_i - y} & \text{if $\norm{x_i - y} \leq b_i$} \\
        \norm{x_i - y} - b_i & \text{if $\norm{x_i - y} > b_i$}
    \end{cases}
\end{align*}
where $\norm{.}$ denotes either an $L_1$ or $L_2$ norm, depending on the context.

We call these \emph{doubly-peaked preferences}.

%For two points $x = (x_0, x_1) \in \mathbb{R}^2$ and $y = (y_0, y_1) \in \mathbb{R}^2$, the euclidean distance between them is defined as $\norm{x - y} = \sqrt{(x_0 - y_0)^2 + (x_1 - y_1)^2}$.

When agents are located on a line, agent $i$'s most preferred locations are $x_i + b_i$ and $x_i - b_i$. Likewise, when agents are located on a 2D plane, with the $L_1$ norm,
agent $i$'s most preferred locations are on a diamond centered at $x_i$ with center to corner distances of $b_i$, and with the $L_2$ norm, the preferred locations are on a circle centered at $x_i$ with radius $b_i$.

A \emph{randomized mechanism} is a function $f : (\Omega \cross \mathbb{R})^n \rightarrow \Delta(\Omega)$, where $\Delta(\Omega)$ is the set of probability distributions over $\Omega$. It maps a given instance to probabilistically selected locations for the facility. For an instance $\inst \in (\Omega \cross \mathbb{R})^n$, the expected cost of agent $i$ is $\E_{y \sim f(\inst)}\big[\cost(y, (x_i, b_i))\big]$.

A deterministic mechanism $f$ is strategy-proof if no agent benefits by misreporting their preferred distance. This means that for every instance $\inst = \mathbf{t} = \langle t_1, t_2, ..., t_n\rangle \in (\Omega \cross \mathbb{R})^n$, every $i \in [n]$, and every $b'_i \in \mathbb{R}$, $\cost(f(\mathbf{t}), (x_i, b_i)) \leq \cost(f((x_i, b'_i), \mathbf{t}_{-i}), (x_i, b_i))$, where $\mathbf{t}_{-i} = \langle t_1, ..., t_{i-1}, t_{i+1}, ..., t_{n} \rangle$ (i.e.\ instance $\mathbf{t}$ without $(x_i, b_i)$).

A randomized mechanism $f$ is truthful-in-expectation if it guarantees that every agent always minimizes their expected cost by reporting their preferred distance truthfully. This means that for every instance $\inst = \mathbf{t} = \langle t_1, t_2, ..., t_n\rangle \in (\Omega \cross \mathbb{R})^n$, every $i \in [n]$, and every $b'_i \in \mathbb{R}$, $\E_{y \sim \mathcal{D}_1}\big[\cost(y, (x_i, b_i))\big] \leq \E_{y \sim \mathcal{D}_2}\big[\cost(y, (x_i, b_i))\big]$, where $\mathcal{D}_1 = f(\mathbf{t})$ and $\mathcal{D}_2 = f((x_i, b'_i), \mathbf{t}_{-i})$ are probability distributions and $\mathbf{t}_{-i}$ is as defined earlier. 

Given an instance $\inst = \langle (x_1, b_1), ..., (x_n, b_n)\rangle$ and a location $y \in \Omega$, the social cost of placing the facility at $y$ is defined as:
\begin{equation}
    \COST(y, \inst) = \sum_{i = 1}^n\cost(y, (x_i, b_i)).
\end{equation}

For an instance $\inst$, we will say that $y$ has a social cost of $\COST(y, \inst)$. 
We define $\OPT(\inst)$ to be the minimum achievable social cost on instance $\inst$, and we say  
a location $\loc^*(\inst)$ is optimal if it achieves cost $\OPT(\inst)$:  $\COST(\loc^*(\inst))= \OPT(\inst)$.
In general, there may be more than one optimal location.

If $f$ is a deterministic mechanism, then, for an instance $\inst$, $f$ has a social cost of $\COST(f(\inst), \inst)$. For a randomized mechanism $f$, we are interested in its expected social cost, given by $\E_{y \sim f(\inst)}\big[\COST(y, \inst)\big]$.

We are interested in truthful (or truthful-in-expectation) mechanisms that perform well with respect to the goal of minimizing the social cost $\COST$. We measure the performance of the mechanism by comparing the social cost (or expected social cost) it achieves with the optimal social cost, on any instance $\inst$.

The (additive) approximation error of mechanism $f$ is given by
\begin{equation}
    \error_{f} = \sup_{\inst} \E_{y \sim f(\inst)}\big[\COST(y, \inst) - \OPT(\inst)\big].
\end{equation}
Our goal is to design mechanisms that minimize the error $\error_f$.

The mechanisms and algorithms we consider repeatedly compute medians of sets of $n$ items.
To avoid tie-breaking issues, we define the median to be the $\lfloor (n+1)/2\rfloor$ item.
Furthermore, if several items have the same value, we impose an arbitrary order on the items,
so an item of a given rank is always uniquely and consistently identified.

\subsection{Partial Group Strategy Proofness}

Lu et al.\ introduced the notion of \emph{Partial Group Strategy Proofness} in \cite{lu2010asymptotically}, which is a weaker notion than \emph{Group Strategy Proofness}. This idea was an essential part of their tight\footnote{up to constant multiplicative factors.} lower bounds for the two-facility location problem. In our work, we extend the scope of this definition to include doubly peaked preferences.

\begin{definition}[For facility location mechanisms with doubly-peaked preferences] \label{def:partial-group-strategyproof}
A mechanism is \emph{partial group strategy-proof} if for every set $S$ of agents at the same location $\Tilde{x}$ and having the same distance preference $\Tilde{b}$, no subset of agents in $S$ can benefit by misreporting.

Formally, let $S \subseteq N$ be a group of agents with $\ix = \Tilde{x}$ and $\bi = \Tilde{b} \leq B$ for all $i \in S$. Let $x_S = (\Tilde{x}, \cdots, \Tilde{x}) \in \Omega^{|S|}$ denote the common location of all the agents in this group,  $b_S = (\Tilde{b}, \cdots, \tilb) \in \Omega^{|S|}$ the real preferred distances, and $b'_S \in \Omega^{|S|}$ the potentially misreported (and not necessarily equal) preferred distances.

$f$ is a partial group strategy-proof mechanism if
\begin{align*}
    &\E \big[\cost(f((x_S, b_S), (x_{-S}, b_{-S})), (\tilx, \tilb))\big]  \\ 
    &~~~~\leq \E\big[\cost(f((x_S, b'_S), (x_{-S}, b_{-S})), (\tilx, \tilb))\big]. 
\end{align*}
\end{definition}

Clearly, partial group strategy-proofness implies strategy-proofness. The following theorem states that strategy-proofness implies partial group strategy-proofness as well. The proof can be found in the appendix.

\begin{lemma}\label{thm:grp-strategyproof}
In the doubly-peaked preference facility location game, any strategy-proof mechanism is also partial group strategy-proof.
\end{lemma}

\section{Hardness: Lower bounds on the additive error in ID}
\label{sec::hardness}

%In this section, we prove a strong lower bound. Despite the seemingly straightforward nature of the problem, our analysis reveals the impossibility of attaining an additive approximation better than $\theta(nB)$.
\hide{\begin{theorem}\label{thm:1d-hardness}
    Consider an arbitrary mechanism $f$ for the facility location game with doubly peaked preferences that is truthful-in-expectation. The additive approximation error of $f$ is at least $\alpha nB$, for a constant $\alpha >0$, which depends on the setting (1D, 2D with $L_1$ distances, 2D with $L_2$ distances). That is, for any given $f$, there exists a worst-case instance $\inst$ such that $\E_{y \sim f(\inst)}[\COST(y, \inst)] \geq \OPT(\inst) + \alpha nB$.
\end{theorem}}

We start by analysing the hardness of placing a single facility in the 1D setting when agents have Doubly-Peaked preferences. We show that even when the agents do not have the power to misreport their locations (and can only misreport their preferred distances), the additive approximation error of any strategyproof $f$ is at least $\alpha nB$, for a constant $\alpha >0$, which depends on the setting (1D, 2D with $L_1$ distances, 2D with $L_2$ distances,  universally-truthful mechanism, truthful-in-expectation mechanism, etc.).

In this section, we analyze the hardness for 1D mechanisms. In the 1D case, we prove lower bounds for both deterministic (univerally-truthful) and randomized (truthful-in-expectation) mechanisms. Results for 2D are deferred to the appendix.

\begin{theorem}\label{thm:det-1d-hardness}
    Consider an arbitrary deterministic mechanism $f$ for the facility location game in 1D with doubly peaked preferences that is universally-truthful. The additive approximation error of $f$ is at least $\frac{nB}{24}$. That is, for any given $f$, there exists a worst-case instance $\inst$ such that $\COST(f(\inst), \inst) \geq \OPT(\inst) + \frac{nB}{24}$.
\end{theorem}

\begin{theorem}\label{thm:rand-1d-hardness}
    Consider an arbitrary mechanism $f$ for the facility location game in 1D with doubly peaked preferences that is truthful-in-expectation. The additive approximation error of $f$ is at least $\frac{nB}{300}$. That is, for any given $f$, there exists a worst-case instance $\inst$ such that $\E_{y \sim f(\inst)}[\COST(y, \inst)] \geq \OPT(\inst) + \frac{nB}{300}$.
\end{theorem}

We prove \cref{thm:det-1d-hardness} first. The proof of \cref{thm:rand-1d-hardness} uses similar techniques but is more involved as any randomized mechanism's output is a probability distribution over $\mathbb{R}$ and requires bounding probabilities over different intervals. We detail this new probabilistic analysis in the main paper and defer the remaining details to the appendix.
\hide{We defer the proof of \cref{thm:rand-1d-hardness} to the appendix due to space limitations.}

In the appendix, we also show that with small modifications this construction can be extended to the 2D setting, for both the $L_1$ and $L_2$ distance measures, albeit with smaller constants in the additive term (see Theorems \ref{thm:2d-hardness} and \ref{thm:2d-hardness-L2}).

The proof of \cref{thm:det-1d-hardness} uses two instances each having $n=3m$ agents, partitioned into 3 groups of $m$ agents each. The instances are shown in Figs.~\ref{fig::inst-1} and~\ref{fig::inst-2} below.
The proof applies the partial group strategy proof property to the Group 3 agents, the agents at position $\tfrac 12 B$, in these two instances. More specifically, we will show that if a mechanism $f$ achieves expected total cost of at most $\OPT +nB/24$ on instances of $n$ agents, then in  Instance $\inst_1$, the Group 3 agents benefit by misreporting so as to create Instance $\inst_2$. In particular, in $\inst_1$, each of these agents has cost at least $\tfrac{3}{8}B$ when reporting truthfully, but they have a smaller cost when reporting as if in Instance $\inst_2$. Then \Cref{thm:grp-strategyproof} implies that mechanism $f$ could not be strategy proof, meaning that the assumed approximation factor cannot be achieved.

Let, $\inst_1$ be an instance with $n=3m$ agents
forming three clusters. The first set of $m$ agents, Group 1, are located at $0$ and have a preferred distance of $B$. The next set of $m$ agents, Group 2, are located at $-\frac{B}{4}$ and have a preferred distance of $\frac{B}{2}$, and the last $m$ agents, Group 3, are located at $\frac{B}{2}$ and have a preferred distance of $\frac{3B}{4}$. Formally, let
\begin{align} \label{eq::inst1}
    \begin{split}
        \inst_1 = \langle (x_1, b_1), \ldots, (x_m, b_m), (x_{m+1}, b_{m+1}), \\
        ..., (x_{2m}, b_{2m}), \ldots, (x_{3m}, b_{3m})\rangle
    \end{split}
\end{align}
with
\begin{align*}
     x_i = 0 \text{ and } b_i = B & \hspace*{0.1in}\text{for } 1 \leq i \leq m\\
     x_i = \tfrac{-B}{4} \text{ and } b_i  = \tfrac{B}{2} & \hspace*{0.1in}\text{for } m+1 \leq i \leq 2m \\
     x_i = \tfrac{B}{2} \text{ and } b_i = \tfrac{3B}{4}&\hspace*{0.1in} \text{for } 2m+1 \leq i \leq 3m. 
\end{align*}
\hide{\begin{align*}
     x_i = x_1 = 0 \text{ and } b_i = b_1 = B & \hspace*{0.1in}\text{for } 1 \leq i \leq m\\
     x_i = x_{m+1} = \tfrac{-B}{4} \text{ and } b_i = b_{m+1} = \tfrac{B}{2} & \hspace*{0.1in}\text{for } m+1 \leq i \leq 2m \\
     x_i = x_{2m+1} = \tfrac{B}{2} \text{ and } b_i = b_{2m+1} = \tfrac{3B}{4}&\hspace*{0.1in} \text{for } 2m+1 \leq i \leq 3m. 
\end{align*}}

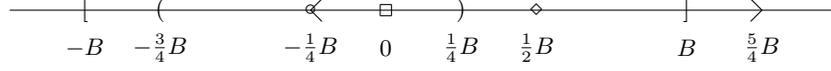
\begin{figure*}[htb]
\begin{center}
\begin{tikzpicture}
\path
(0,0) node (s) {$[$} --
(0,-0.5) node {$-B$} --
(1,0) node (t) {$($} --
(1,-0.5) node {$-\frac 34 B$} --
(3,0) node (u) {$\circ$} --
(3,-0.5) node {$-\frac 14 B$} --
(4,-0.5) node {0} --
(5,0) node (x) {$)$} --
(5,-0.5) node {$\frac 14 B$} --
(6,0) node (y) {$\diamond$} --
(6,-0.5) node {$\frac 12 B$} --
(8,0) node (z) {$]$} --
(8,-0.5) node {$B$} --
(9,-0.5) node {$\frac54 B$};
\begin{scope}[>=latex]
\draw[-] (-1,0) to (10,0);
\draw[-] (3,0) to (3.15,0.15);
\draw[-] (3,0) to (3.15,-0.15);
\draw[-] (9,0) to (8.85,0.15);
\draw[-] (9,0) to (8.85,-0.15);
\draw[-] (3.925,0.075) to (4.075,0.075);
\draw[-] (4.075,0.075) to (4.075,-0.075);
\draw[-] (4.075,-0.075) to (3.925,-0.075);
\draw[-] (3.925,-0.075) to (3.925,0.075);
\end{scope}
\end{tikzpicture}
\end{center}

\caption{\label{fig::inst-1}Instance $\inst_1$. The locations are indicated by the circle, diamond and square respectively, and the corresponding preferred locations are indicated by the round brackets, the angle brackets, and the square brackets.}
\end{figure*}

Let $\inst_2$ be an instance which is identical to $\inst_1$, except that the last set of $m$ agents, located at $\frac{B}{2}$, have a preferred distance of $\frac{B}{2}$. Formally, let\\
\begin{align} \label{eq::inst2}
    \begin{split}
        \inst_2 = \langle (x'_1, b'_1), ..., (x'_m, b'_m), (x'_{m+1}, b'_{m+1}), \\
        ..., (x'_{2m}, b'_{2m}), ..., (x'_{3m}, b'_{3m})\rangle
    \end{split}
\end{align}
with 
\hide{\begin{align*}
     x'_i = x'_1 = 0 \text{ and } b'_i = b'_1 = B &\hspace*{0.1in} \text{for  } 1 \leq i \leq m \\
      x'_i = x'_{m+1} = \tfrac{-B}{4} \text{ and } b'_i = b'_{m+1} = \tfrac{B}{2} & \hspace*{0.1in}\text{for } m+1 \leq i \leq 2m\\
     x'_i = x'_{2m+1} = \tfrac{B}{2} \text{ and } b'_i = b'_{2m+1} = \tfrac{B}{2} & \hspace*{0.1in}\text{for } 2m+1 \leq i \leq 3m.
\end{align*}}

\begin{align*}
     x'_i = 0 \text{ and } b'_i = B &\hspace*{0.1in} \text{for  } 1 \leq i \leq m \\
      x'_i = \tfrac{-B}{4} \text{ and } b'_i = \tfrac{B}{2} & \hspace*{0.1in}\text{for } m+1 \leq i \leq 2m\\
     x'_i = \tfrac{B}{2} \text{ and } b'_i = \tfrac{B}{2} & \hspace*{0.1in}\text{for } 2m+1 \leq i \leq 3m.
\end{align*}

Next, via a series of observations, we proceed to prove \cref{thm:det-1d-hardness}.

Let $\interval_1 = (-\tfrac 78 B, -\tfrac 58 B)$ and $\interval_2 = (\tfrac 78 B, \tfrac{25}{24} B)$ be two intervals.

\begin{observation} \label{obs:det-opt1}
$\loc^*(\inst_1) = -\frac{3B}{4}$ and $\OPT(\inst_1) = \frac{3mB}{4}$.
\end{observation}

\begin{proof}
To see this note that if the facility is moved away in either direction from location $-\frac{3B}{4}$, at least two agent groups are increasing in cost and at most one is decreasing, until location $0$ is reached. Then, as the facility continues moving to the right, two agent groups have reducing cost and one has increasing cost, until location $\tfrac B4$ is reached, and then the change in cost switches to at least two increasing and at most one reducing till $\tfrac B2$ is reached. Finally, as one moves the facility towards $B$, the cost reduces for two agent groups and increases for one agent group. Placing the facility beyond $B$ results in increasing the cost for at least two agent groups.
Calculating the costs at $-\frac{3B}{4}$, $\tfrac B4$ and $B$ concludes the argument.
\end{proof}

\begin{observation}\label{obs:determinstic-interval1}
For $y \in \interval_1$, $\COST(y, \inst_1) < \frac{3mB}{4} + \frac{mB}{8} = \frac{7mB}{8}$ and for $y \notin \interval_1$, $\COST(y, \inst_1) \geq \frac{7mB}{8}$.
\end{observation}

\begin{proof}
Any point $y \in \interval_1$ can be written as $y = \frac{-3B}{4} - \delta$ where $-\frac B8 < \delta < \frac{B}{8}$. Thus,
\begin{align*}
    \cost(y, (x_1, b_1)) &= \tfrac{1}{4}B - \delta\\
    \cost(y, (x_{m+1}, b_{m+1})) &= |\delta|\\
    \cost(y, (x_{2m+1}, b_{2m+1})) &= \tfrac{1}{2}B + \delta\\
    \implies \COST(y, \inst_1) &= m\cdot\big[\tfrac{3}{4}B + |\delta|\big] ~~ \text{for all } y \in \interval_1.
\end{align*}
Therefore, $\frac{3mB}{4}\leq \COST(y, \inst_1) < \frac{3mB}{4} + \frac{mB}{8} = \frac{7mB}{8}$.
\medskip
If $y \notin \interval_1$, then $y \in \interval_{11} \cup \interval_{12} \cup \interval_{13} \cup \interval_{14}$ where $\interval_{11} = (-\infty, -\frac{7B}{8}] \cup [-\frac{5B}{8}, -\frac B4]$, $\interval_{12} = (-\frac B4, 0] $, $\interval_{13} = (0, \frac B2]$, and $\interval_{14} = (\frac B2, \infty)$.

For a facility located at $y \in \interval_{11}$, the Group 1 agents ($1, \ldots, m$) preferred location is $-B$, the Group 2 agents ($m+1, \ldots, 2m$) preferred location is $-\tfrac{3}{4}B$, and the Group 3 agents ($2m+1, \ldots, 3m$) preferred location is $-\tfrac{1}{4}B$. Therefore, $\COST(y, \inst_1)$ takes its infimum at $-\tfrac{7}{8}B$ and $-\tfrac{5}{8}B$ with $\COST(-\tfrac{7}{8}B, \inst_1 = \COST(-\tfrac{5}{8}B, \inst_1) = \tfrac{7mB}{8}$ and so the cost in $\interval_{11}$ is greater than or equal to $\tfrac{7mB}{8}$.

For a facility located at $y \in \interval_{12}$, the Group 1 agents ($1, \ldots, m$) preferred location is $-B$, the Group 2 agents ($m+1, \ldots,2m$) preferred location is $\tfrac{1}{4}B$, and the Group 3 agents ($2m+1, \ldots, 3m$) preferred location is $-\tfrac{1}{4}B$. Therefore, $\COST(y, \inst_1)$ is minimized at the median of these three preferred locations, namely at $y = -\tfrac 14 B$, and $\COST(-\tfrac 14 B, \inst_1) = \tfrac{5mB}{4} > \tfrac{7mB}{8}$.

Similarly, for $y \in \interval_{13}$, the preferred locations for the three groups are at $B$, $\tfrac{1}{4}B$  and  $-\tfrac{1}{4}B$, respectively. $\COST(y, \inst_1)$ is minimized at $y = \tfrac 14 B$ with $\COST(y, \inst_1) = \tfrac{5mB}{4} > \tfrac{7mB}{8}$.

Finally, for $y \in \interval_{14}$, the preferred locations for the three groups are at $B, \tfrac{1}{4}B \text{ and } \tfrac{5}{4}B$, respectively. $\COST(y, \inst_1)$ is minimized at $y = B$ with $\COST(y, \inst_1) = mB > \tfrac{7mB}{8}$.

This proves that $\COST(y, \inst_1) < \frac{7mB}{8}$ for $y \in \interval_1$, and $\COST(y, \inst_1) \geq \frac{7mB}{8}$ otherwise.
% \begin{align*}
%     \cost(y, (x_1, b_1)) &= \tfrac{1}{4}B + \delta\\
%     \cost(y, (x_{n+1}, b_{n+1})) &= \delta\\
%     \cost(y, (x_{2n+1}, b_{2n+1})) &= \tfrac{1}{2}B - \delta\\
%     \implies \COST(y, \inst_1) &= n\cdot\big[\tfrac{3}{4}B + \delta\big] ~~ \text{for all } y \in \interval_{11}. 
% \end{align*}
\end{proof}

We state the following observations without proof as their proofs are similar to \cref{obs:det-opt1} and \cref{obs:determinstic-interval1}.

\begin{observation}
$\loc^*(\inst_2) = B$ and $\OPT(\inst_2) = \frac{3mB}{4}$.
\end{observation}

\begin{observation}\label{obs:determinstic-interval2}
For $y \in \interval_2$, $\COST(y, \inst_2) < \frac{3mB}{4} + \frac{mB}{8} = \frac{7mB}{8}$ and for $y \notin \interval_2$, $\COST(y, \inst_2) \geq \frac{7mB}{8}$.
\end{observation}

We are now ready to prove our main theorem.

\begin{proof}[Proof of \cref{thm:det-1d-hardness}]
    Assume for the sake of contradiction that $f$ is a deterministic mechanism such that $\SC(f(\inst), \inst) < \OPT(\inst) + \tfrac{nB}{24}$ for all instances $\inst$. Since $f$ is strategyproof, by \cref{thm:grp-strategyproof}, $f$ must also be group strategyproof.
    
    Consider the instance $\inst_1$ (\cref{eq::inst1}). Since $\OPT(\inst_1) = \tfrac 34 nB$ and since $f$ has an error of less than $\tfrac{nB}{24}$, by \cref{obs:determinstic-interval1}, we must have that $f(\inst_1) \in \interval_1$.

    Similarly, since $\OPT(\inst_2) = \tfrac 34 nB$ and since $f$ has an error of less than $\tfrac{nB}{24}$, by \cref{obs:determinstic-interval2}, we must have that $f(\inst_2) \in \interval_2$.

    Let $y_1$ be any point in $\interval_1$. Clearly, $\cost(y_1, (x_{2m+1}, b_{2m+1})) > -\tfrac 14 B - (-\tfrac 58 B) = \tfrac 38 B$. Let $y_2$ be any point in $\interval_2$. Clearly, $\cost(y_2, (x_{2m+1}, b_{2m+1})) < \tfrac 54 B - \tfrac 78 B = \tfrac 38 B$.

    Suppose that the Group 3 agents in $\inst_1$ misreport their preferred distances so that the resulting instance looks like $\inst_2$. This moves the mechanism's facility location from $\interval_1$ to $\interval_2$, reducing the Group 3 agent costs. This contradicts the fact that $f$ is group strategyproof and therefore a deterministic strategyproof mechanism $f$ cannot have an error less than $\tfrac{nB}{24}$.
\end{proof}

\begin{figure*}[htb]
\begin{center}
\begin{tikzpicture}
\path
(0,0) node (s) {$[$} --
(0,-0.5) node {$-B$} --
(1,0) node (t) {$($} --
(1,-0.5) node {$-\frac 34 B$} --
(3,0) node (u) {$\circ$} --
(3,-0.5) node {$-\frac 14 B$} --
(4,-0.5) node {0} --
(5,0) node (x) {$)$} --
(5,-0.5) node {$\frac 14 B$} --
(6,0) node (y) {$\diamond$} --
(6,-0.5) node {$\frac 12 B$} --
(8,0) node (z) {$]$} --
(8,-0.5) node {$B$};
\begin{scope}[>=latex]
\draw[-] (-1,0) to (9,0);
\draw[-] (4,0) to (4.15,0.15);
\draw[-] (4,0) to (4.15,-0.15);
\draw[-] (8,0) to (7.85,0.15);
\draw[-] (8,0) to (7.85,-0.15);
\draw[-] (3.925,0.075) to (4.075,0.075);
\draw[-] (4.075,0.075) to (4.075,-0.075);
\draw[-] (4.075,-0.075) to (3.925,-0.075);
\draw[-] (3.925,-0.075) to (3.925,0.075);
\end{scope}
\end{tikzpicture}
\end{center}
\caption{\label{fig::inst-2}Instance $\inst_2$}
\end{figure*}

\subsubsection{Proof of \cref{thm:rand-1d-hardness}}
Consider the instances $\inst_1$ (\cref{eq::inst1}) and $\inst_2$ (\cref{eq::inst2}) with $n = 3m$ agents from \cref{thm:det-1d-hardness}. The only difference between the two instances is in the preferred distance of Group 3 agents (the last $m$ agents). We will prove \cref{thm:rand-1d-hardness} by contradiction: If there exists a truthful-in-expectation randomized mechanism $f : (\mathbb{R} \times \mathbb{R})^n \rightarrow \mathbb{R}$ with an additive approximation error smaller than $\tfrac{nB}{300}$, then Group 3 agents in $\inst_1$ can reduce their expected cost by misreporting their preferred distances so that the instance looks like $\inst_2$.

Let $\interval_1 = \big[\frac{-19B}{20}, \frac{-11B}{20}\big]$ and $\interval_2 = \big[\frac{4B}{5}, \frac{16B}{15}\big]$.

We state the following observations without proof as their proofs are similar to the observations for \cref{thm:det-1d-hardness}.

\begin{observation}\label{obs:opt1}
$\loc^*(\inst_1) = -\frac{3B}{4}$ and $\OPT(\inst_1) =\OC_1 = \frac{3mB}{4}$.
\end{observation}

\begin{observation} \label{obs:interval1}
For $y \in \interval_1$, $\COST(y, \inst_1) \leq \OC_1-\frac{4mB}{20}$ and for $y \notin \interval_1$, $\COST(y, \inst_1) > \OC_1+\frac{4mB}{20}$.
\end{observation}

\begin{observation}\label{obs:opt2}
$\loc^*(\inst_2) = B$ and $\OPT(\inst_2) =\OC_2= \frac{3mB}{4}$.
\end{observation}

\begin{observation}\label{obs:interval2}
For $y \in \interval_2$, $\COST(y, \inst_2) \leq \OC_2-\frac{4mB}{20}$, and for $y \notin \interval_2$, $\COST(y, \inst_2) \geq \OC_2+\frac{4mB}{20}$.
\end{observation}

\hide{The proof of \cref{obs:interval2} is similar to the proof of \cref{obs:interval1}, and can be found in the appendix. \\}

\begin{observation}\label{obs:gp3}
For all $y\in \interval_1$, 
\begin{align*}
    \cost(y, (x_{2m+1}, b_{2m+1})) \geq \tfrac{3B}{10}.
\end{align*}
\end{observation}
\begin{proof}
Since $\interval_1 = \big[\frac{-19B}{20}, \frac{-11B}{20}\big]$, $x_{2m+1} = \frac{B}{2}$, and $x_{2m+1} - b_{2m+1} = -\frac{B}{4}$, we obtain, for all $y \in \interval_1$, 
\begin{align*}
    \cost(y, (x_{2m+1}, b_{2m+1})) &\geq -\tfrac{B}{4} - (-\tfrac{11B}{20}) 
    = \tfrac{3B}{10}.
\end{align*}
\end{proof}

\begin{observation}\label{obs:gps1-2}
For all $y \in \interval_2$, 
\begin{align*}
    \cost(y, (x'_{1}, b'_{1})) + \cost(y, (x'_{m + 1}, b'_{m + 1})) \geq \frac{3B}{4}.
\end{align*}
Also, for $y \notin \interval_2$,
\begin{align*}
    \cost(y, (x'_{1}, b'_{1})) + \cost(y, (x'_{m + 1}, b'_{m + 1})) \geq \frac{B}{4}.
\end{align*}
\end{observation}

The proof of \Cref{obs:gps1-2} can be found in the appendix.

\hide{ %TODO(pranav): Unhide this.
\begin{proof}
The first claim holds because the nearest preferred locations for $x'_1$ and $x'_{m+1}$ are at
$\tfrac B4$ and $B$, respectively.
For the second claim, note that $\cost(y, (x'_{1}, b'_{1})) + \cost(y, (x'_{m + 1}, b'_{m + 1}))$ achieves the minimum value of $\frac{B}{4}$ when $y \in [-B, \frac{-3}{4}B]$. Therefore, when $y \notin \interval_2$, $\cost(y, (x'_{1}, b'_{1})) + \cost(y, (x'_{m + 1}, b'_{m + 1})) \geq \frac{B}{4}$.
\end{proof}
}

In the following lemmas we prove some important results that hold for any truthful-in-expectation mechanism $f$ given instances $\inst_1 \text{ and } \inst_2$. For ease of notation, let $\OC_i = \COST(\loc^*(\inst_i), \inst_i)$ for $i=1,2$.
\begin{lemma}\label{lemma:prob1}
    Suppose that $\E_{y \sim f(\inst_1)}\big[\COST(y, \inst_1)\big] < \OC_1 + \frac{mB}{100}$. Then, $\prob\big[y \in \interval_1 ~|~ y \sim f(\inst_1)\big] \geq \frac{19}{20}$.
\end{lemma}
\begin{proof}
    Let $p = \prob\big[y \in \interval_1 ~|~ y \sim f(\inst_1)\big]$. Assume, for a contradiction, that $p < \frac{19}{20}$. \\

    Observe that $\E_{y \sim f(\inst_1)}\big[\COST(y, \inst_1)\big] = $
    \begin{align}
    \begin{split}
        % &\E_{y \sim f(\inst_1)}\big[\COST(y, \inst_1)\big] \\
        & \prob\big[y \in \interval_1 ~|~ y \sim f(\inst_1) \big]\cdot\E_{y \sim f(\inst_1)}\big[\COST(y, \inst_1) ~|~ y \in \interval_1\big] \\
        &+ \prob\big[y \notin \interval_1 ~|~ y \sim f(\inst_1) \big]\cdot\E_{y \sim f(\inst_1)}\big[\COST(y, \inst_1) ~|~ y \notin \interval_1\big] \\
        &= p\cdot\E_{y \sim f(\inst_1)}\big[\COST(y, \inst_1) ~|~ y \in \interval_1\big] \\
        &+ (1-p)\cdot\E_{y \sim f(\inst_1)}\big[\COST(y, \inst_1) ~|~ y \notin \interval_1\big].
        \label{eq:prob1}
    \end{split}
    \end{align}
%where the last equality follows from the fact that
%\begin{align*}
%    \prob\big[y \notin \interval_1 ~|~ y \sim f(\inst_1) \big] = 1 - \prob\big[y \in \interval_1 ~|~ y \sim f(\inst_1) \big] = 1 - p.
%\end{align*}

By \cref{obs:opt1}, $\OC_1 = \frac{3mB}{4}$. Therefore, $\E_{y \sim f(\inst_1)}\big[\COST(y, \inst_1) ~|~ y \in \interval_1\big] \geq \frac{3mB}{4}$. By \cref{obs:interval1},  $\E_{y \sim f(\inst_1)}\big[\COST(y, \inst_1) ~|~ y \notin \interval_1\big] \geq \frac{19mB}{20}$. Substituting these inequalities in equation \eqref{eq:prob1} gives
\begin{align*}
    \E_{y \sim f(\inst_1)}\big[\COST(y, \inst_1)\big] &\geq p\cdot\frac{3mB}{4} + (1-p)\cdot\frac{19mB}{20} \\
    &\geq \frac{19mB}{20} - p\cdot \frac{mB}{5} \\
    &> \frac{19mB}{25}.
\end{align*}

However, this is a contradiction because of our assumption that 
\begin{align*}\E_{y \sim f(\inst_1)}\big[\COST(y, \inst_1)\big] < \OC_1 + \frac{mB}{100} = \frac{76mB}{100} = \frac{19mB}{25}.
\end{align*}

Therefore, $p=\prob[y \in \interval_1 ~|~ y \sim f(\inst_1)] \geq \frac{19}{20}$.
\end{proof}

\begin{lemma}[Analog of \cref{lemma:prob1} for $\inst_2$] \label{lemma:prob2}
    Suppose that $\E_{y \sim f(\inst_2)}\big[\COST(y, \inst_2)\textbf{}] < \OC_2 + \frac{mB}{100}$. Then, $\prob\big[y \in \interval_2 ~|~ y \sim f(\inst_2)\big] \geq \frac{19}{20}$.
\end{lemma}

The proof of \Cref{lemma:prob2} can be found in the appendix.

\begin{lemma}\label{lemma:dist1}
    Let $f$ be a truthful-in-expectation mechanism such that $\E_{y \sim f(\inst_1)}\big[\COST(y, \inst_1)\big] < \OC_1 + \frac{mB}{100}$. Then, $\E_{y \sim f(\inst_1)}\big[\cost(y, (x_{2m+1}, b_{2m+1}))\big] \geq \frac{57B}{200}$.
\end{lemma}

\begin{proof}
Throughout this proof, $y \sim \inst_1$.  
In the following argument, the expectation is w.r.t.\ $y \sim \inst_1$, which we omit for brevity. 
Let $p = \prob\big[y \in \interval_1 ~|~ y \sim f(\inst_1)\big]$. Now,
\begin{align}\label{eqn::lower-bdd-cost-in-Ione}
    \begin{split}
        &\E\big[ \cost(y, (x_{2m+1}, b_{2m+1})) \big] \\
        % &= p\cdot\E\big[ \cost(y, (x_{2n+1}, b_{2n+1})) ~|~ y \in \interval_1 \big] \\
        % &~~~~+ (1-p)\cdot\E\big[ \cost(y, (x_{2n+1}, b_{2n+1})) ~|~ y \notin \interval_1 \big] \\
        &\hspace*{0.3in} \geq p\cdot\E\big[ \cost(y, (x_{2m+1}, b_{2m+1})) ~|~ y \in \interval_1 \big].
    \end{split}
\end{align}
%where the last inequality follows from the fact that $(1-p)\cdot\E\big[ \cost(y, (x_{2n+1}, b_{2n+1})) ~|~ y \notin \interval_1 \big] \geq 0$.

% Since $\interval_1 = \big[\frac{-19B}{20}, \frac{-11B}{20}\big]$, $x_{2m+1} = \frac{B}{2}$, and $x_{2m+1} - b_{2m+1} = -\frac{B}{4}$, we obtain, for all $y \in \interval_1$, 
% \begin{align*}
%    \cost(y, (x_{2m+1}, b_{2m+1})) &\geq -\tfrac{B}{4} - (-\tfrac{11B}{20}) 
%    = \tfrac{3B}{10}.
% \end{align*}

By \Cref{obs:gp3}, $\E\big[ \cost(y, (x_{2m+1}, b_{2m+1})) ~|~ y \in \interval_1 \big] \geq \frac{3B}{10}$. By \cref{lemma:prob1}, $p \geq \frac{19}{20}$.
Substituting these lower bounds in \Cref{eqn::lower-bdd-cost-in-Ione} yields:
%Combining the lower bounds on $\E\big[ \cost(y, (x_{2m+1}, b_{2m+1})) ~|~ y \in \interval_1 \big]$ and $p$ yields:
\begin{align*}
    \E\big[ \cost(y, (x_{2m+1}, b_{2m+1})) \big] &\geq \tfrac{19}{20}\cdot\tfrac{3B}{10}
    =\tfrac{57B}{200}.
\end{align*}
\end{proof}

\begin{lemma}[Triangle Inequality for $\cost$] \label{thm:triangle-inequality}
For all $x, y \in \Omega$ and $b, b' \in \mathbb{R}$,
\begin{equation*}
    \cost(y, (x, b)) \leq \cost(y, (x, b')) + |b - b'|.
\end{equation*}
    
\end{lemma}
% The proof of \cref{thm:triangle-inequality} is straigtforward and can be found in the appendix.

\begin{lemma}\label{lemma:dist2}
    Let $f$ be a truthful-in-expectation mechanism such that $\E_{y \sim f(\inst_2)}\big[\COST(y, \inst_2)\big] < \OC_2 + \frac{mB}{100}$. Then, $\E_{y \sim f(\inst_2)}\big[\cost(y, (x_{2m+1}, b_{2m+1}))\big] < \frac{57B}{200}$.
\end{lemma}
\begin{proof}
    Throughout this proof, $y \sim \inst_2$. While taking expectations over $\cost(y, (x_{2m+1}, b_{2m+1}))$ and $\COST(y, \inst_2)$, we omit $y \sim \inst_2$. Let $p = \prob\big[y \in \interval_2 ~|~ y \sim f(\inst_2)\big]$. Note the difference between $x_i$ and $x'_i$ (and between $b_i$ and $b'_i$). The former refers to the locations (or preferred distances) of agents in $\inst_1$ and the latter refers to those of agents in $\inst_2$.

    \cref{obs:opt2} states that $\OC_2 = \frac{3mB}{4}$. Therefore, 
    \begin{equation}
        \E \big[\COST(y, \inst_2)\big] < \OC_2 + \tfrac{mB}{100} = \tfrac{3mB}{4} + \tfrac{mB}{100}.
    \end{equation}
    Since $\inst_2$ has $3m$ agents forming three groups, each group with the same values of $x'_i$ and $b'_i$, by linearity of expectation, 
    \begin{align*}
        &\E \big[\COST(y, \inst_2)\big]\\
        &~~~~= m\cdot\E\big[\cost(y, (x'_{1}, b'_{1}))\big] + m\cdot\E\big[\cost(y, (x'_{m+1}, b'_{m+1}))\big] \\
        &~~+ m\cdot\E\big[\cost(y, (x'_{2m+1}, b'_{2m+1}))\big] \\
        &~~~~< \tfrac{3mB}{4} + \tfrac{mB}{100},
    \end{align*}
    which implies that 
    \begin{align} \label{eq:Csum}
        \begin{split}
        &\E\big[\cost(y, (x'_{1}, b'_{1}))\big] + \E\big[\cost(y, (x'_{m + 1}, b'_{m + 1}))\big] \\
        &\hspace*{0.3in}+ \E\big[\cost(y, (x'_{2m + 1}, b'_{2m + 1}))\big] \\ 
        &\hspace*{0.5in}< \tfrac{3B}{4} + \tfrac{B}{100}.
        \end{split}
    \end{align}
    By linearity of expectation and the fact that the events $y\in\interval_2$ and $y \notin \interval_2$ are mutually exclusive,
    \begin{align}
    \begin{split} \label{eq:c1plusc2}
        &\E\big[\cost(y, (x'_{1}, b'_{1}))\big] + \E\big[\cost(y, (x'_{m + 1}, b'_{m + 1}))\big] \\
        &\hspace*{0.1in}= p\cdot\E\big[\cost(y, (x'_{1}, b'_{1})) + \cost(y, (x'_{m + 1}, b'_{m + 1})) ~|~ y \in \interval_2\big] \\
        &\hspace*{0.4in}+ (1-p)\cdot\E\big[\cost(y, (x'_{1}, b'_{1})) \\
        &\hspace*{0.5in}+ \cost(y, (x'_{m + 1}, b'_{m + 1})) ~|~ y \notin \interval_2\big].
        \end{split}
    \end{align}

\hide{
    Observe that $\cost(y, (x'_{1}, b'_{1})) + \cost(y, (x'_{m + 1}, b'_{m + 1})) \geq \frac{3B}{4}$ for all $y \in \interval_2$. Therefore, $\E\big[\cost(y, (x'_{1}, b'_{1})) + \cost(y, (x'_{m + 1}, b'_{m + 1})) ~|~ y \in \interval_2\big] \geq \frac{3B}{4}$.

    Also observe that $\cost(y, (x'_{1}, b'_{1})) + \cost(y, (x'_{m + 1}, b'_{m + 1}))$ achieves the minimum value of $\frac{B}{4}$ when $y \in [-B, \frac{-3}{4}B]$. Therefore, $\E\big[\cost(y, (x'_{1}, b'_{1})) + \cost(y, (x'_{m + 1}, b'_{m + 1})) ~|~ y \notin \interval_2\big] \geq \frac{B}{4}$.
}

    Substituting the lower bounds from \Cref{obs:gps1-2} in equation \eqref{eq:c1plusc2} gives
    \begin{align}
    \begin{split} \label{eq:c1plusc2_2}
        \E\big[\cost(y, (x'_{1}, b'_{1}))\big] &+ \E\big[\cost(y, (x'_{m + 1}, b'_{m + 1}))\big] \\
        &\geq p\cdot \tfrac{3B}{4} + (1-p)\cdot \tfrac{B}{4} \\
        &= p\cdot\tfrac{B}{2} + \tfrac{B}{4}\\
        &\geq \tfrac{29B}{40},
    \end{split}
    \end{align}
    where the last inequality follows from the fact that $p \geq \frac{19}{20}$ (\cref{lemma:prob2}).

    Using inequality \eqref{eq:c1plusc2_2} in \eqref{eq:Csum} gives 
\begin{equation} \label{eqn::hardness-ub-i2-dist}
    \E\big[\cost(y, (x'_{2m + 1}, b'_{2m + 1}))\big] < \tfrac{3B}{4} + \tfrac{B}{100} - \tfrac{29B}{40} 
    = \tfrac{7B}{200}.
\end{equation}

Finally, by invoking the triangle inequality for $\cost$ (\cref{thm:triangle-inequality}), we get
\begin{align}\label{eq::hardness-ub-no-expec}
    \begin{split}
        \cost(y, (x_{2m + 1}, b_{2m + 1})) &\leq \cost(y, (x'_{2m + 1}, b'_{2m + 1}))\\
        &\hspace*{0.3in}+ |b_{2m+1} - b'_{2m+1}|.
    \end{split}
\end{align}

Taking expectation, using \eqref{eqn::hardness-ub-i2-dist} and the values of $b_{2m+1}, b'_{2m+1}$ in \cref{eq::hardness-ub-no-expec}, we get
\begin{align*}
    \begin{split}
        \E\big[\cost(y, (x_{2m + 1}, b_{2m + 1}))\big] &\leq \E\big[\cost(y, (x'_{2m + 1}, b'_{2m + 1}))\big] \\
        &\hspace*{0.3in}+ \frac{B}{4} \\
        &<\frac{7B}{200} + \frac{B}{4} = \frac{57B}{200}.
    \end{split}
\end{align*}
\end{proof}

We are now ready to prove our main result.
\begin{proof}[Proof of \cref{thm:rand-1d-hardness}]
    Assume for the sake of contradiction that $f$ is a truthful-in-expectation strategy-proof mechanism such that, for all instances $\inst$,
    \begin{align*}
        \E_{y \sim f(\inst)}\big[\COST(y, \inst)\big] < \OPT(\inst) + \frac{mB}{100}.
    \end{align*}
    By \cref{thm:grp-strategyproof}, $f$ must also be partial group strategy-proof. Since the only difference between $\inst_1$ and $\inst_2$ is the preferred distance of the last $m$ agents, this property means that
    \begin{align}
        \begin{split}
            \E_{y \sim f(\inst_1)}&\big[\cost(y, (x_{2m+1}, b_{2m+1}))\big] \\
            &\leq \E_{y \sim f(\inst_2)}\big[\cost(y, (x_{2m+1}, b_{2m+1}))\big]. \label{eq:strategyproof-cond}
        \end{split}
    \end{align}
    By assumption,
    \begin{align*}
        \E_{y \sim f(\inst_i)}\big[\COST(y, \inst_i)\big] < \OC_i + \frac{mB}{100} \hspace*{0.2in}\text{for  } i=1,2,
    \end{align*}
    which, by \cref{lemma:dist1}, implies that in $\inst_1$, 
    \begin{align*}\E_{y \sim f(\inst_1)}\big[\cost(y, (x_{2m+1}, b_{2m+1}))\big] \geq \frac{57B}{200},\end{align*} 
    and, by \cref{lemma:dist2}, implies that in $\inst_2$, 
    \begin{align*}\E_{y \sim f(\inst_2)}\big[\cost(y, (x_{2m+1}, b_{2m+1}))\big] < \frac{57B}{200}.\end{align*}
    Therefore,
    \begin{align*}
        \E_{y \sim f(\inst_1)}&\big[\cost(y, (x_{2m+1}, b_{2m+1}))\big] \\
        &> \E_{y \sim f(\inst_2)}\big[\cost(y, (x_{2m+1}, b_{2m+1}))\big],
    \end{align*}
    which contradicts the fact that $f$ is partial group strategy-proof (equation \eqref{eq:strategyproof-cond}).
    Therefore,
    \begin{align*}
        \E_{y \sim f(\inst)}\big[\COST(y, \inst)\big] &\geq \OPT(\inst) + \frac{mB}{100}
         \geq \OPT(\inst) + \frac{nB}{300}.
    \end{align*}
    
\end{proof}

\section{The Median Algorithm}
\label{sed::MedAlg}

In this section we analyze a simple algorithm that uses only public information, and consequently is trivially strategy proof, the Median algorithm. It simply places the facility at the median of the agent positions
(recall that this is the rank $\lfloor (n+1)/2)\rfloor$ position).
In 2D we consider both the $L_1$ and $L_2$ metrics. With the $L_1$ metric, this means the facility location
has $x$-coordinate equal to the median of the agent's $x$-coordinates, and $y$-coordinate equal to the median of the agents' $y$ coordinates. With the $L_2$ metric the median employed is the geometric median,
the location minimizing the sum of the Euclidean distances from the agents.

\begin{theorem}\label{thm::med-mech-perf}
    In all three settings, on any instance $\inst$, the Median algorithm has a social cost bounded by $\OPT(\inst)+2nB$. 
\end{theorem}

Rather than prove this bound directly, we start with a tighter bound in the case that the average preferred distances are smaller.
Suppose $r$ agent locations are within distance $B$ of the median location, and for these agents let
$\bbar$ denote the average of their $b_i$ values.

\begin{theorem}\label{thm::av-dist-bound}
In all three settings, suppose there are $r$ agents within distance $B$ of the median location,
and suppose the average $b_i$ value for these agents is $\bbar$.
Then the Median algorithm achieves a cost of at most $\OPT(\inst) +2r\bbar$.
\end{theorem}
\begin{proof}
It's helpful to introduce some more notation.
$\SC_{\med}(b)$ denotes the cost of the solution produced by the Median algorithm with preferred distances $b$ (an $n$-vector), and $\SC_{\med}(0)$ denotes the cost of the solution produced by the Median algorithm if all the preferred distances are 0.
Similarly $\SC_{\opt}(b)$ denotes the cost of the optimal solution with preferred distances $b$,
and $\SC_{\opt}(0)$ denotes the cost of this solution with the preferred distances reset to 0.

Recall that $\med$ is the facility location chosen by the Median algorithm.
Then,
\begin{align*}
    \SC_{\med}(b) \le \SC_{\med}(0) +\sum_{|x_i-\med|< b_i} b_i - \sum_{|x_i-\med|\ge b_i} b_i,
\end{align*}
since all agents with $|x_i-b_i| \ge b_i$ reduce their distance to $\med$ by $b_i$, when the preferred distance is $b_i$ rather than 0; and the other agents increase their distance by at most $b_i$.
Similarly,
\begin{align*}
    \SC_{\opt}(0) \le \SC_{\opt}(b) + \sum_i b_i.
\end{align*}
We know that $\SC_{\med}(0)= \SC_{\opt}(0)$, as $\med$ is the optimal solution when the preferred distances are all zero.
Therefore,
\begin{align*}
\SC_{\med}(b) &\le \SC_{\med}(0) +\sum_{|x_i-\med|< b_i} b_i - \sum_{|x_i-\med|\ge b_i} b_i\\
              &\le \SC_{\opt}(0) +\sum_{|x_i-\med|< b_i} b_i - \sum_{|x_i-\med|\ge b_i} b_i\\
              & \le \SC_{\opt}(b) +2\sum_{|x_i-\med| < b_i|} b_i\\
              & \le \SC_{\opt}(b) +2\sum_{|x_i-\med| < B|} b_i = \OPT + 2r\bbar.\\
\end{align*}
\end{proof}

\hide{
\begin{proof}
 Let $y$ be the location returned by the median mechanism and let $y^*$ be a location with social cost equal to $\OPT$. Let $\SC(y,0)$ be the social cost at location $y$ if all the preferred distances were 0. Clearly,
 $\SC(y,b)$, the social cost with preferred distances $b$ is at most $\SC(y,b)\le \SC(y,0)+nB$, as the distance from $y$ to each agent's preferred location increases by at most $B$ is each of these settings.
 Similarly, $\SC(y^*,0) \le \SC(y^*,b) +nB$. Finally, $\SC(y,0) \le \SC(y^*,0)$, as the median is the optimal solution when $b=0$.
 Putting everything together yields
 \begin{align*}
      \SC(y,b) \le \SC(y,0) +nB \le \SC(y^*,0)+nB\le SC(y^*,b)+2nB.
 \end{align*}
\end{proof}
}

\Cref{thm::med-mech-perf} follows immediately as $\OPT(\inst) + 2r\bbar \le \OPT(\inst) +2n B$.
\hide{\pj{following theorem is not required}
\begin{theorem}\label{thm:1d-hardness}
    Consider an arbitrary mechanism $f$ for the facility location game with doubly peaked preferences that is truthful-in-expectation. The additive approximation error of $f$ is at least $\alpha nB$, for a constant $\alpha >0$, which depends on the setting (1D, 2D with $L_1$ distances, 2D with $L_2$ distances). That is, for any given $f$, there exists a worst-case instance $\inst$ such that $\E_{y \sim f(\inst)}[\COST(y, \inst)] \geq \OPT(\inst) + \alpha nB$.
\end{theorem} }

A simple generalization of the Median algorithm, the $k$-Median algorithm, achieves the same approximation bound for the $k$-facility problem in the 1D setting. The algorithm simply places the facilities at the $k$ agent locations that minimize the sum of the distances from the agents to their nearest facilities (this placement is readily computed in $O(n^2 k)$ time using dynamic programming). The argument in \Cref{thm::av-dist-bound} is unchanged.

\begin{theorem}\label{thm::av-dist-bound-kmed}
Suppose $k$ facilities are placed at the agent locations that minimize the social cost if all agents had a preferred distance of 0 (these are the locations provided by the $k$-Median algorithm).
Suppose there are $r$ agents within distance $B$ of the facilities,
and suppose the average $b_i$ value for these agents is $\bbar$.
Then the $k$-Median algorithm achieves a cost of at most $\OPT(\inst) +2r\bbar$.
\end{theorem}
\begin{proof}
  It suffices to show that the locations chosen by the $k$-Median algorithm with the preferred 0 distance setting are actually optimal for these distances. To see this, consider the locations in an optimal solution with the preferred 0 distances;
  they provide a partitioning of the agents, with each agent assigned to one of its closest facilities. Each partition now forms a 1-facility problem, for which an optimal solution is to place the facility at the median agent's location. Thus there is an optimal solution in which all the facilities are at agent locations, and therefore the minimization performed by the $k$-Median algorithm selects these locations or another set of $k$ locations having this minimum cost.
  Applying the argument from \Cref{thm::av-dist-bound} completes the proof.
\end{proof}
Next, we describe the Median-Plus mechanism which always achieves a cost as good as the Median algorithm, and can out-perform it by a $\Theta(nB)$ factor (as shown in the appendix's section \emph{Improvement with Median-Plus}).

\section{The Median-Plus Mechanism}
\label{sec::med-plus}

%In this section, we design mechanisms for placing facilities on the real line. We begin with a mechanism for placing one facility and then show how we can use that to solve the problem of placing $k$ facilities. 
Throughout this section we use $\mathbb{R}$ instead of $\Omega$ for clarity.

%\subsubsection{Placing one facility}

The optimal solution for Instance $\inst=\langle(x_1, b_1), $\ldots$, (x_n, b_n) \rangle$ has the following
characteristic. We set $q_i=\argmin_{y_i=x_i\pm b_i} \big|\loc^*(\inst) -y_i\big|$, namely Agent $i$'s preferred location that is nearer to the optimal facility location $\loc^*(\inst)$,
breaking ties arbitrarily.
When $n$ is odd, $\loc^*(\inst)$ is the median of $\{q_1,q_2,\ldots,q_n\}$, and when $n$ is even, there is a tie-breaking
for which $\loc^*(\inst)$ is a median of the resulting locations.

As our mechanism does not know $\loc^*(\inst)$, for each Agent $i$ it selects $i$'s preferred location $p_i$ using a rule based on the position of the agent compared to the median agent position. Then the mechanism takes the  median of the preferred locations as the facility location. 
We will show this mechanism achieves at least as good an approximation to the optimal solution as the median mechanism described in the previous section.

\begin{algorithm}
\caption{Median-Plus Mechanism for one facility in $\mathbb{R}$}
\label{mech::OneFacR1}

\KwIn{$\inst = \langle(x_1, b_1), $\ldots$, (x_n, b_n) \rangle$.}
\KwOut{The facility location} %: the median (rank $m=\lfloor (n+1)/2\rfloor$) of the locations $\{p_1,p_2,\ldots,p_n\}$.

Let $\med$ be the median of $\{x_1,x_2,\ldots,x_n\}$, the rank $m=\lfloor (n+1)/2\rfloor$ agent position.\\
\For{each agent $i$}{
    (* set $p_i$, $i$'s preferred facility location *) \footnote{If $x_i=\med$, we use the ordering imposed on the locations $x_i$ to resolve the comparison.}\\
    \uIf{$x_i\le \med$}{
        $p_i\la x_i +b_i$
    }
    \Else{
        $p_i\la x_i -b_i$
    }
}
Output: the median (rank $m=\lfloor (n+1)/2\rfloor$) of the locations $\{p_1,p_2,\ldots,p_n\}$.
\end{algorithm}

\medskip

For the following lemmas and theorems, we assume that agents in $\inst = \langle(x_1, b_1), $\ldots$, (x_n, b_n) \rangle$ are sorted in increasing order by their locations $x_i$. This doesn't affect the optimum facility location and clearly, doesn't affect our mechanism.

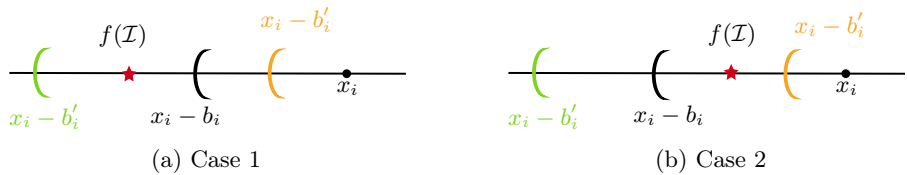
\begin{figure} 
    \centering
    \begin{subfigure}[b]{0.45\textwidth} 
        \centering
        \input{1d-strategyproof-case1.tex}
        \caption{Case 1}
        \label{fig::1d-strategyproof-case1}
    \end{subfigure}
    \hfill % Optional space between the figures
    \begin{subfigure}[b]{0.45\textwidth}
        \input{1d-strategyproof-case2.tex}
        \caption{Case 2}
        \label{fig::1d-strategyproof-case2}
    \end{subfigure}
    \caption{Two cases when $x_i \geq \med$. The red star represents $f(\inst)$---the output of Median-Plus when agents report truthfully. The black bracket shows agent $i$'s true preferred facility location. The orange bracket is agent $i$'s apparent preferred facility location when she misreports $b'_i < b_i$. The green bracket shows the location when she misreports $b'_i > b_i$.}
    \label{fig::1d-strategyproof}
\end{figure}

\begin{lemma}\label{lem::basic-mech-str-prf}
    Mechanism Median-Plus is strategy proof.
\end{lemma}

\begin{proof} 
    Since the positions of the agents are fixed, $\med$ is fixed and therefore for every agent, $p_i$ is either $x_i - b_i$ or $x_i + b_i$ depending on its position.

    Let $f(\inst)$ denote the output of Median-Plus on $\inst$. Let $\inst' = \langle (x_1, b_1) \ldots (x_i, b'_i) \ldots (x_n, b_n) \rangle$ be the instance when agent $i$ misreports their preferred distance as $b'_i \leq B$.
    We will show that 
    \begin{equation*}
        \cost(f(\inst), (x_i, b_i)) \leq \cost(f(\inst'), (x_i, b_i)).
    \end{equation*}

    Without loss of generality, assume $x_i > \med$, which implies $p_i = x_i - b_i$. There are two cases (as shown in \cref{fig::1d-strategyproof}).\\
\noindent
    {\sc Case 1} (\cref{fig::1d-strategyproof-case1}): $f(\inst) \leq p_i$. If $b'_i < b_i$ (orange bracket),  $p'_i = x_i - b'_i > p_i$ and $f(\inst') = f(\inst)$. 
    If $b'_i > b_i$ (green bracket),  $p'_i = x_i - b'_i < p_i$; thus $f(\inst') \leq f(\inst)$ and therefore $\cost(f(\inst), (x_i, b_i)) \leq \cost(f(\inst'), (x_i, b_i))$.\\
\noindent \newline
    {\sc Case 2} (\cref{fig::1d-strategyproof-case2}): $f(\inst) > p_i$. If $b'_i > b_i$ (green bracket), then $p'_i = x_i - b'_i < p_i$ and $f(\inst') = f(\inst)$. If $b'_i < b_i$ (orange bracket), then $p'_i = x_i - b'_i > p_i$. Observe that since $b'_i \geq 0$, if $x_i \leq f(\inst)$, then $p'_i \leq x_i$ implies that $f(\inst') = f(\inst)$. Otherwise, if $x_i > f(\inst)$, then $f(\inst') \in [f(\inst), x_i]$ and therefore, $\cost(f(\inst), (x_i, b_i)) \leq \cost(f(\inst'), (x_i, b_i))$.
\end{proof}

Next, we show that Median-Plus always achieves at least as good a cost as Median.
Let $SC_{\medp}(\inst)$ denote the cost achieved by Median-Plus on $\inst$, and $\SC_{\med}(\inst)$ the cost achieved by the Median algorithm.
\begin{lemma}\label{lem::med-plus-better-than-med}
 On any instance $\inst$, $\SC_{\medp}(\inst)\le \SC_{\med}(\inst)$.
\end{lemma}
\begin{proof}
     Consider the $n$ preferred locations used by Mechanism Median-Plus.
    Let $y_{\medp}$ be the facility location it chooses.
    We pair the locations with one location on either side of $y_{\medp}$
    (one of these locations is $y_{\medp}$, and in the event that $n$ is even, one of the other locations
    is paired with $y_{\medp}$. Let $p$ and $q$ be paired locations.
    Then the contribution of $p$ and $q$ to $\SC_{\medp}(\inst)$ is $|p-q|$, while their contribution to 
    $\SC_{\med}(\inst)$ is at least $|p-q|$. If $n$ is odd, the unpaired location $y_{\medp}$ contributes 0 to $\SC_{\medp}(\inst)$ and at least 0 to $\SC_{\med}(\inst)$. Summing over all paired locations yields the claim.
\end{proof}

The following bound is immediate from \Cref{thm::av-dist-bound}.

\begin{corollary}
    Suppose that in instance $\inst$ there are $r$ agent locations in the range $\med\pm B$, and suppose their average $b_i$ value
    is $\bbar$. Then the Median-Plus mechanism achieves a cost of at most $\OPT(\inst)+2r\bbar$.
\end{corollary}

This also implies that Median-Plus always achieves a cost of at most $\OPT(\inst)+2nB$.
In the appendix, we show how to improve this bound to $\OPT(\inst)+nB$ (see \Cref{lem::basic-mech-perf}). \newline
\noindent
In the appendix, we generalize the Median-Plus mechanism to handle settings with agents located in 2D,
when using $L_1$ distances as the cost metric.

\section{Improvement with Median-Plus}
\label{app::MedP-is-better}

In Lemmas \ref{lem::med-plus-better-than-med} and \ref{lem::2d-med-plus-better-than-med} we prove that Median-Plus and 2D-Median-Plus perform at least as well as the median algorithm. In this section we demonstrate how the potential improvements to the social cost can be $\Theta(nB)$ on \emph{strongly skewed instances}.

\begin{figure}
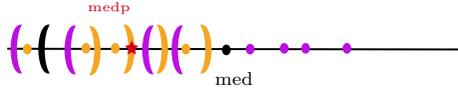

    \centering
    \include{skewed-instance}
    \caption{Skewed Instance. Median and one of its preferred location is marked with black color. Agents to the left of median are colored orange along with their preferred locations and agents to the right of median are colored purple. The output of Median-Plus $\medp$ is marked in a red star.}
    \label{fig::skewed-instance}
\end{figure}

In a skewed instance all agents on one side of the median (say $x_i < \med$) have preferred distances such that both the preferred locations of the agents lie on the same side of the median as the agent location (if $x_i < \med$, then $x_i + b_i < \med$; for the instance to be strongly skewed we require $x_i + b_i < \med -\Theta(B)$). On the other hand, agents on the other side of the median (say $x_i > \med$), have their two preferred distances on different side of the median (if $x_i > \med$, then $x_i - b_i < \med$; for the instance to be strongly skewed we require $x_i + b_i < \med -\Theta(B)$).
Consider a skewed instance in 1D, as shown in \cref{fig::skewed-instance}. 

If the instance is strongly skewed, it can be readily observed that moving the median from $\medp$ to $\med$ increases the social cost by $\Theta(nB)$, when all the $b_i = \Theta(B)$.

Note that an instance need not be \emph{perfectly strongly skewed} for $\medp$ to be a strictly better solution. 

\section{Conclusion}
\label{sec::concl}

We have introduced and investigated a new version of the facility location problem with public locations and private costs.
We showed a $\Theta(nB)$ lower bound on the additive approximation factor, gave a simple algorithm which uses just public information to match this bound up to constant factors, and developed a non-trivial mechanism that matches and can substantially outperform this mechanism, both for the 1D and 2D with $L_1$ costs versions.

As we indicated, an analog of the median algorithm solves the $k$-facility problem, and again achieves a $\Theta(nB)$ additive approximation faction. This leaves open the question of whether there are any strategy-proof mechanisms which leverage the bids to improve on this performance.

\bibliographystyle{splncs04}
\bibliography{main.bib}

\newpage

\appendix

\hide{
\section{Triangle Inequality}
\begin{theorem}[Triangle Inequality for $\cost$] \label{thm:triangle-inequality}
For all $x, y \in \Omega$ and $b, b' \in \mathbb{R}$,
\begin{equation*}
    \cost(y, (x, b)) \leq \cost(y, (x, b')) + |b - b'|.
\end{equation*}
    
\end{theorem}
\begin{proof}
    Observe that $\cost(y, (x, b))$ can be written as
    \begin{equation} \label{eq::triangle-ineq-dist}
        \cost(y, (x, b)) = \lvert d(x, y) - b \rvert
    \end{equation}
    where $d(x, y)$ denotes the distance between $x$ and $y$. If $x, y \in \mathbb{R}$, then $d(x, y) = |x - y|$ and if $x, y \in \mathbb{R}^2$, then $d(x, y)$ can be the $L_1$ distance or the $L_2$ distance between the points. \\
    \noindent
    {\sc Case 1}: $d(x, y) \leq \min(b, b')$. In this case, by \eqref{eq::triangle-ineq-dist}, $\cost(y, (x, b)) - \cost(y, (x, b')) = b - b' \leq |b - b'|$. \\
    \noindent \newline
    {\sc Case 2}: $d(x, y) \geq \max(b, b')$. Similar to Case 1, here $\cost(y, (x, b)) - \cost(y, (x, b')) = b' - b \leq |b - b'|$. \\
    \noindent \newline
    {\sc Case 3}: $\min(b, b') < d(x, y) < \max(b, b')$. If $b < b'$, then $\cost(y, (x, b)) - \cost(y, (x, b')) = 2d(x, y) - b - b' < b' - b = |b' - b|$. Similarly, if $b > b'$, $\cost(y, (x, b)) - \cost(y, (x, b')) = b + b' - 2d(x, y) < b - b' = |b' - b|$.
\end{proof}
}

\section{Missing Proofs}

\begin{proof}[Proof Of \Cref{thm:grp-strategyproof}]
We use the same notation as in \cref{def:partial-group-strategyproof}. Let $S = (s_1, s_2, \cdots, s_k)$, where $k = |S|$. For all $i\in [k]$, let $\bi$ denote the real distance preferences, $b'_i$ the reported distance preferences, and let $f$ denote a truthful-in-expectation strategy-proof mechanism.

Consider the following sequence of instances:
\begin{align*}
    \inst_i: ~~~~& s_j \text{ reports } b'_j \text{ for } 1 \leq j \leq i \text{ (misreports)}\\
    & s_j \text{ reports } b_j \text{ for } i < j \leq k \text{ (truthful reports)}\\
    & \text{other agents in $N \setminus S$ report $b_{-S}$}.
\end{align*}
In all the instances, the location $x_j$ of all agents is fixed and publicly known.

By definition, $\inst_0$ is the instance where every agent in $S$ reports truthfully, and $\inst_k$ is the instance where every agent in $S$ misreports.
We will prove that
\begin{align*}
    \E_{y \sim f(\inst_0)} \big[\cost(y, (\tilx, \tilb))\big] \leq \E_{y \sim f(\inst_k)} \big[\cost(y, (\tilx, \tilb))\big]. 
\end{align*}

Observe that for all $1 \leq j \leq k-1$, the one difference between $\inst_j$ and $\inst_{j+1}$ is in the preferred distance reported by agent $j+1$.  Since $f$ is strategy-proof, 
\begin{align*}
    \E_{y \sim f(\inst_j)} \big[\cost(y, (\tilx, \tilb))\big] \leq \E_{y \sim f(\inst_{j+1})} \big[\cost(y, (\tilx, \tilb))\big].
\end{align*}
Combining all the inequalities together yields
\begin{align*}
    \E_{y \sim f(\inst_0)} \big[\cost(y, (\tilx, \tilb))\big] &\leq\cdots 
    \leq  \E_{y \sim f(\inst_{k-1})} \big[\cost(y, (\tilx, \tilb))\big] \\
    &\leq \E_{y \sim f(\inst_{k})} \big[\cost(y, (\tilx, \tilb))\big].
\end{align*}
Thus, 
\begin{align*}
    \E_{y \sim f(\inst_0)} \big[\cost(y, (\tilx, \tilb))\big] \leq \E_{y \sim f(\inst_k)} \big[\cost(y, (\tilx, \tilb))\big], 
\end{align*}
which means that $f$ is partial group strategy-proof.
\end{proof}

\subsection{Multiplicative Bounds from Additive Error Bounds} \label{app::multiplicative-bounds}

In our analysis, we used natural cost functions such as the $L1$ and $L2$ distances from the facility. However, this means that the optimal social cost of placing a facility in some instances can be zero.
In order to obtain a multiplicative approximation, Filos-Ratsikas et al.\ used a cost of the form $c + \Delta$, with $c>0$ being a suitable parameter and $\Delta$ being the distance from the facility. This ensured that for any instance $\inst$ with $n$ agents, $\OPT(\inst) \geq nc$.
In this section, we demonstrate how we can turn our bounds into multiplicative bounds by using the same cost function as \cite{filos2017facility}. We show this for the upper bound on the approximation of the Median-Plus mechanism (\cref{mech::OneFacR1}) but the conversion holds for all the results we have analysed in this paper.

Consider the modified cost function,

\begin{align*}
    \cost(y, (x_i, b_i)) = \begin{cases}
        |x_i - b_i - y| + c & \text{if $y \leq x_i$} \\
        |x_i + b_i - y| + c & \text{if $y > x_i$}
    \end{cases}
\end{align*}

The multiplicative approximation error of mechanism $f$ is given by
\begin{equation*}
  %  \error^{M}_{f} = \sup_{\inst} \frac{\E_{y \sim f(\inst)}\big[\COST(y, \inst)\big] }{ \OPT(\inst)}.
    \error^{M}_{f} = \sup_{\inst}\frac{\min_y\COST(y, \inst) }{ \OPT(\inst)}.
\end{equation*}

\begin{theorem}
    Mechanism Median-Plus has a constant multiplicative approximation error of at most $1 + \tfrac{B}{c}$. 
\end{theorem}

\begin{proof}
    By \cref{lem::basic-mech-perf}, % we get $\COST(y, \inst) - \OPT(\inst) \leq nB$, therefore, 
    $\COST(y, \inst) \leq \OPT(\inst) + nB$.

    Now, because $\cost(y, (x_i, b_i)) \geq c$, we have that $\OPT(\inst) \geq nc$. Therefore, 
    \begin{align*}
        \error^{M}_{f} &= \sup_I\frac{\COST(y, \inst)} { \OPT(\inst)} \\
        &\leq \frac{\OPT(\inst) + nB}{\OPT(\inst)} \\
        &\leq 1 + \frac{nB}{nc} \\
        &\leq 1 + \frac{B}{c}.
    \end{align*}
\end{proof}

\subsection{Missing Proofs for the Median-Plus Mechanism}

\begin{lemma}\label{lem::basic-mech-perf}
    On any instance $\inst$, Mechanism Median-Plus has cost at most $\OPT(\inst) + nB$. 
\end{lemma}
To prove this, we first need to show that $|\loc^*-\med| \le B$.

\begin{lemma}\label{lem::pos-opt-soln}
    $\loc^*(\inst)\in \med\pm B$, where $\med$ is the median of $\{x_1,x_2,\cdots,x_n\}$. If there are multiple optimal solutions, we mean that at least one is in this range.
\end{lemma}
\begin{proof}
    Let $x_m = \med$ where $m = \lfloor \frac{n+1}{2} \rfloor$. Assume for the sake of contradiction that the optimal solution $y$ closest to $x_m$ lies outside $x_m \pm B$ (if there are two such solutions choose either one). 

    Suppose WLOG that $x_m > y+B$. Then there are at least $\lfloor (n+1)/2\rfloor$ agents $i$ with $x_i\ge x_m$. For all these agents their preferred location $p_i \ge x_i -B > x_m -B >y$. Then moving $y$ to the next larger preferred location would reduce the cost of solution, or maintain its value while moving it closer to $x_m$, in both cases contradicting the fact that $y$ was the optimal solution closest to $x_m$.

   \hide{    
    If $y < x_m - B$,
    then there are $\lfloor n/2\rfloor +1$  agents with $x_i - b_i \geq x_m - B > y$. Therefore, moving $y$ closer to $x_m - B$ decreases the cost for these agents while potentially increasing the cost for the remaining $\lfloor n/2\rfloor$ agents, with the magnitude of the changes being the same for every agent. 
    This yields a lower cost solution, contradicting the optimality of $y$.

    If $y > x_m + B$ and $n$ is odd, then we can use a symmetric argument to obtain a contradiction.
    If $n$ is even, then the median is at $x_{n/2}$ and there are exactly $\tfrac{n}{2}$ agents with $x_i + b_i \leq x_m + B < y$. Therefore, moving $y$ closer to $x_m + B$ would strictly decrease the cost for $\tfrac n2$ agents while potentially increasing the cost for the remaining $n/2$ agents, again by equal amounts. 
    This yields an optimal solution closer to $x_m$, a contradiction.
    }
\end{proof}

\begin{proof}[Proof of \cref{lem::basic-mech-perf}]
    % Follows from \cref{lem::med-plus-better-than-med} and \cref{lem::basic-mech-perf}.
    Let $\ystar$ be the optimal location that minimizes the social cost. By \cref{lem::pos-opt-soln}, $\ystar \in \med \pm B$. Let $x_m$ denote the median location. Clearly, $\lvert \ystar - x_m \rvert \leq B$. Observe that for all $i$,
    \begin{align*}
        \cost(x_m, (x_i, b_i)) &\leq \cost(\ystar, (x_i, b_i)) + \lvert x_m - \ystar \rvert \\ 
        &\leq \cost(\ystar, (x_i, b_i)) + B.
    \end{align*}
    Summing over $i$ gives
    \begin{align*}
        \SC_{\med}(\inst) \leq \SC_{\ystar}(\inst) + nB = \OPT(\inst) + nB 
     \end{align*}
     and by \cref{lem::med-plus-better-than-med},
     \begin{align*}
        \SC_{\medp}(\inst) \le \SC_{\med(\inst)} \leq \OPT(\inst) + nB.
    \end{align*}
\end{proof}

% \subsection{Missing Proofs, \Cref{sec::2D-MedPlus}} \label{app::split-line-intersection-proof}

\subsection{Missing Proofs for the Error Lower Bound}

% The analysis uses a triangle inequality which we present first.

% \begin{lemma}[Triangle Inequality for $\cost$] \label{thm:triangle-inequality}
% For all $x, y \in \Omega$ and $b, b' \in \mathbb{R}$,
% \begin{equation*}
%     \cost(y, (x, b)) \leq \cost(y, (x, b')) + |b - b'|.
% \end{equation*}
    
% \end{lemma}
\begin{proof}[Proof of the triangle inequality (\Cref{thm:triangle-inequality})]
    Observe that $\cost(y, (x, b))$ can be written as
    \begin{equation} \label{eq::triangle-ineq-dist}
        \cost(y, (x, b)) = \lvert d(x, y) - b \rvert
    \end{equation}
    where $d(x, y)$ denotes the distance between $x$ and $y$. If $x, y \in \mathbb{R}$, then $d(x, y) = |x - y|$ and if $x, y \in \mathbb{R}^2$, then $d(x, y)$ can be the $L_1$ distance or the $L_2$ distance between the points. \\
    \noindent
    {\sc Case 1}: $d(x, y) \leq \min(b, b')$. In this case, by \eqref{eq::triangle-ineq-dist}, $\cost(y, (x, b)) - \cost(y, (x, b')) = b - b' \leq |b - b'|$. \\
    \noindent \newline
    {\sc Case 2}: $d(x, y) \geq \max(b, b')$. Similar to Case 1, here $\cost(y, (x, b)) - \cost(y, (x, b')) = b' - b \leq |b - b'|$. \\
    \noindent \newline
    {\sc Case 3}: $\min(b, b') < d(x, y) < \max(b, b')$. If $b < b'$, then $\cost(y, (x, b)) - \cost(y, (x, b')) = 2d(x, y) - b - b' < b' - b = |b' - b|$. Similarly, if $b > b'$, $\cost(y, (x, b)) - \cost(y, (x, b')) = b + b' - 2d(x, y) < b - b' = |b' - b|$.
\end{proof}

\begin{figure*}[htb]
\begin{center}
\begin{tikzpicture}
\path
(0,0) node (s) {$[$} --
(0,-0.5) node {$-B$} --
(1,0) node (t) {$($} --
(1,-0.5) node {$-\frac 34 B$} --
(3,0) node (u) {$\circ$} --
(3,-0.5) node {$-\frac 14 B$} --
(4,-0.5) node {0} --
(5,0) node (x) {$)$} --
(5,-0.5) node {$\frac 14 B$} --
(6,0) node (y) {$\diamond$} --
(6,-0.5) node {$\frac 12 B$} --
(8,0) node (z) {$]$} --
(8,-0.5) node {$B$};
\begin{scope}[>=latex]
\draw[-] (-1,0) to (9,0);
\draw[-] (4,0) to (4.15,0.15);
\draw[-] (4,0) to (4.15,-0.15);
\draw[-] (8,0) to (7.85,0.15);
\draw[-] (8,0) to (7.85,-0.15);
\draw[-] (3.925,0.075) to (4.075,0.075);
\draw[-] (4.075,0.075) to (4.075,-0.075);
\draw[-] (4.075,-0.075) to (3.925,-0.075);
\draw[-] (3.925,-0.075) to (3.925,0.075);
\end{scope}
\end{tikzpicture}
\end{center}
\caption{Instance $\inst_2$}
\end{figure*}

\begin{proof}[Proof of \Cref{obs:gps1-2}]
The first claim holds because the nearest preferred locations for $x'_1$ and $x'_{m+1}$ are at
$\tfrac B4$ and $B$, respectively.
For the second claim, note that $\cost(y, (x'_{1}, b'_{1})) + \cost(y, (x'_{m + 1}, b'_{m + 1}))$ achieves the minimum value of $\frac{B}{4}$ when $y \in [-B, \frac{-3}{4}B]$. Therefore, when $y \notin \interval_2$, $\cost(y, (x'_{1}, b'_{1})) + \cost(y, (x'_{m + 1}, b'_{m + 1})) \geq \frac{B}{4}$.
\end{proof}

\begin{proof}[Proof of \Cref{lemma:prob2}]
    Let $p = \prob\big[y \in \interval_2 ~|~ y \sim f(\inst_2)\big]$. Assume, for a contradiction, that $p < \frac{19}{20}$. \\

    Observe that $\E_{y \sim f(\inst_2)}[\COST(y, \inst_2)] = $
    \begin{align}
    \begin{split}
         &\prob[y \in \interval_2 ~|~ y \sim f(\inst_2) ]\cdot\E_{y \sim f(\inst_2)}[\COST(y, \inst_2) ~|~ y \in \interval_2] \\
        &~~~~+ \prob[y \notin \interval_2 ~|~ y \sim f(\inst_2) ]\cdot\E_{y \sim f(\inst_2)}[\COST(y, \inst_2) ~|~ y \notin \interval_2] \\
        &= p\cdot\E_{y \sim f(\inst_2)}[\COST(y, \inst_2) ~|~ y \in \interval_2] \\
        &~~~~+ (1-p)\cdot\E_{y \sim f(\inst_2)}[\COST(y, \inst_2) ~|~ y \notin \interval_2]. \label{eq:prob2}
    \end{split}
    \end{align}

%where the last equality follows from the fact that
%\begin{align*}
%    \prob[y \notin \interval_2 ~|~ y \sim f(\inst_2) ] = 1 - \prob[y \in \interval_2 ~|~ y \sim f(\inst_2) ] = 1 - p.
%\end{align*}

By \cref{obs:opt2}, $\OC_2 = \frac{3mB}{4}$. Therefore, $\E_{y \sim f(\inst_2)}\big[\COST(y, \inst_2) ~|~ y \in \interval_2\big] \geq \frac{3mB}{4}$. By \cref{obs:interval2}, $\E_{y \sim f(\inst_2)}\big[\COST(y, \inst_2) ~|~ y \notin \interval_2\big] \geq \frac{19mB}{20}$. Substituting these inequalities in equation \eqref{eq:prob2} gives 
\begin{align*}
    \E_{y \sim f(\inst_2)}\big[\COST(y, \inst_2)\big] &\geq p\cdot\frac{3mB}{4} + (1-p)\cdot\frac{19mB}{20} \\
    &\geq \frac{19mB}{20} - p\cdot \frac{mB}{5}
    > \frac{19mB}{25}.
\end{align*}

However, this is a contradiction because, by assumption, $\E_{y \sim f(\inst_2)}\big[\COST(y, \inst_2)\big] < \OC_2 + \frac{mB}{100} = \frac{76mB}{100} = \frac{19mB}{25}$.

Therefore, $p=\prob\big[y \in \interval_1 ~|~ y \sim f(\inst_1)\big] \geq \frac{19}{20}$.
\end{proof}

\subsection{Lower Bounds for 2D}

\begin{theorem}\label{thm:2d-hardness}
%{\bf \Cref{thm:2d-hardness}}~
  Consider an arbitrary mechanism $f$ for the facility location game in 2D with the $L_1$ distance measure and with doubly peaked preferences that is truthful-in-expectation. The additive approximation error of $f$ is at least $\frac{nB}{500}$. 
  %That is, for any given $f$, there exists a worst-case instance $\inst$ such that $\E_{y \sim f(\inst)}[\COST(y, \inst)] \geq \OPT(\inst) + \frac{nB}{300}$.
\end{theorem} 

\begin{figure} 
    \centering
    \begin{subfigure}[b]{0.45\textwidth} 
        \centering
        \input{2d-l1-lower-bound-instance1.tex}
        \caption{Modified $\inst_1$. $\region_1$ is shaded in gray for $\beta = \tfrac 25$. Point X is $(-\tfrac{19B}{20}, 0)$, Y is $(-\tfrac{B}{2}, \tfrac{B}{4})$ and Z is $(-\tfrac{11}{20}B, 0)$.}
        \label{fig::2d-hardness-inst1}
    \end{subfigure}
    \hfill % Optional space between the figures
    \begin{subfigure}[b]{0.45\textwidth}
        \input{2d-l1-lower-bound-instance2.tex}
        \caption{Modified $\inst_2$. $\region_2$ is shaded in gray for $\beta = \tfrac 45$. Point X is $(\tfrac{4B}{5}, 0)$, Y is $(\tfrac{7B}{8}, \tfrac{B}{8})$ and Z is $(\tfrac{16}{15}B, 0)$.}
        \label{fig::2d-hardness-inst2}
    \end{subfigure}
    \caption{An extended construction of the 1D setting. Agent group $1, 2$ and $3$ marked by black points. New groups of $\beta m$ agents marked by red points at $(-\tfrac 34 B, 0)$ and $(B, 0)$. Note that in the two instances, the value of $\beta$ is different so that the regions are defined as stated in the proof of \cref{thm:2d-hardness}. In the final construction we use $\beta = 1$.}
    \label{fig::2d-hardness}
\end{figure}
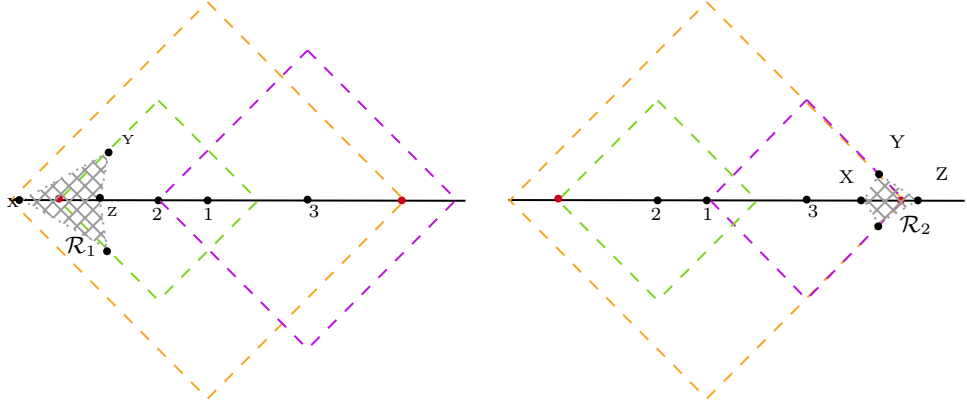

\begin{proof}
    We will be using the same construction as in the 1D setting with the same groups of agents arranged in the same locations on a line. We will introduce a further two groups, each of $\beta m$ agents, at locations $-\tfrac 34B$ and $B$, resp., with each new agent having its $b_i=0$. This increases both optimal social costs $\OC_1$ in $\inst_1$
    and $\OC_2$ in $\inst_2$ to $\tfrac 34 Bm + \tfrac 74 \beta Bm$.

    The goal is to identify regions $\region_1$ and $\region_2$ which will play the same roles as $\interval_1$ and $\interval_2$ did in the proof of \Cref{thm:rand-1d-hardness}. 
    We then reprove Observations \ref{obs:interval1}--\ref{obs:gps1-2} (with the values of $\OC_1$ and $\OC_2$ appropriately updated). Lemmas \ref{lemma:prob1}--\ref{lemma:dist1} now follow with unchanged proofs, while the proof of \Cref{lemma:dist2} needs a minor change to account for the new agents; the concluding proof of the theorem is also unchanged. This yields an error lower bound of $\frac{Bm}{100}$, as in \Cref{thm:rand-1d-hardness}.
    All that changes is the value of $n$, which is now $3m+2\beta m$.
    
Specifically, we need the following to hold: 
    \begin{itemize}
    \item As in \Cref{obs:interval1}, for
    $y\in \region_1$, $\SC(y,\inst_1)\le \OC_1 + \frac{4mB}{20}$ and for $y\notin \region_1$, $\SC(y,\inst_2)\ge \OC_1 +\frac{4mB}{20}$. 
    \item As in \Cref{obs:interval2}, for
    $y\in \region_2$, $\SC(y,\inst_2)\le \OC_2 +\frac{4mB}{20}$, and for $y\notin \region_2$, $\SC(y,\inst_2)\ge \OC_2 + \frac{4mB}{20}$.
    \item As in \Cref{obs:gp3}, for Group 3 agents, their cost in $\region_1$ is at least $\frac {3B}{10}$.
    \item As in \Cref{obs:gps1-2}, for all $y \in \region_2$, 
    $\cost(y, (x'_{1}, b'_{1})) + \cost(y, (x'_{m + 1}, b'_{m + 1}))  \geq \tfrac{3B}{4}$ and 
    for $y \notin \region_2$,
    $\cost(y, (x'_{1}, b'_{1})) + \cost(y, (x'_{m + 1}, b'_{m + 1}))  \geq \tfrac{B}{4}$. 
    \end{itemize}

    \hide{
    Then \rjc{Lemmas~\ref{lemma:prob1} and} \ref{lemma:dist1} hold, with $\region_1$ replacing
    $\interval_1$ in the proof, and similarly for \rjc{Lemmas~\ref{lemma:prob2} and} \ref{lemma:dist2}. With these  results in hand, as in the proof of \Cref{thm:rand-1d-hardness}, we deduce that the error bound is at least $\tfrac {mB}{100}$.
}

    Next, we specify the regions $\region_1$ and $\region_2$. We only describe these regions for $y \geq 0$ as the regions for $y < 0$ are symmetric (see \cref{fig::2d-hardness}).
    $\region_1$ will comprise a triangle 
    with base $\interval_1$ and apex at height $\overline h$ on the left boundary of the Group 2 agent diamonds (see \cref{fig::2d-hardness-inst1}), where $\overline h$ is a value we will specify shortly.
  %  \rjc{More precisely, $\region_1$ with comprise exactly those points $y$ in the triangle for which $\SC(y,\inst_1)\le \OC_1 + \frac{4mB}{20}$.}
    
    Now we show the analog of \Cref{obs:interval1}, as specified above.
    We want that the cost of the boundary $\region_1$ be exactly $\OC_1+ \tfrac {mB}5$.
    The cost at the triangle apex is $\OC_1+ 2mB\beta\cdot \overline h$, as long as $\overline h \le \tfrac B4$
    (the height of the point at which the Group 2 and Group 3 diamonds intersect).
    Thus it suffices to have $\beta\overline h= \tfrac B{10}$.
    As we will see when we define $\region_2$, we will want $\beta=\tfrac 25$, and therefore
    $\overline h = \tfrac B4$.
    By inspection, the cost strictly inside the boundary of $\region_1$ is always less that $\OC_1+ \tfrac {mB}5$.
    This can be seen by considering the change in costs as one moves horizontally from the boundary.
    Similarly, the cost outside of $\region_1$ is always strictly greater than $\OC_1+ \tfrac {mB}5$.
    Again, this can be verified by considering the change in cost due to horizontal moves away from this boundary
    (this is somewhat more painstaking).

    \hide{
    Note that the new agents cause the cost to increase by $2\beta m\cdot h$ at height $h$ above the line on which all the agents are located.
    At height $h$, placing the facility at location $(\tfrac {-3B}4+h,h)$ incurs the following costs: Groups 1 and 2 contribute $\tfrac {mB}4$; Group 3 contributes $\tfrac {2mB}4$; and the new agents contribute
    $\beta m\cdot \rjc{(\tfrac{7B}{4} +2h)}$. 
    As moving \rjc{to the left} from this location by horizontal distance $d$ increases the Group 3 cost by $m\cdot d$, to keep the social cost at most
    $\OC_1+\tfrac {4mB}{20}=\OC_1 + \tfrac {mB}5$, we need $m d \le  \tfrac{mB}{5} - 2\beta m \cdot h$;
    \rjc{similarly, moving to the right increases the Group 1 and 2 cost by $2m\cdot d$ and reduces the Group 3 cost by $m\cdot d$, for the same net effect.}
    Note that if the RHS is negative, then there is no value of $d$ for which the social cost is within this bound.
    To ensure that $h > \tfrac B4$ is not feasible, 
    we require $2\beta m \cdot \tfrac B4  \rjc{\geq} \tfrac {mB}5$; $\beta \rjc{\geq} \tfrac 25$ suffices. 
    We conclude that to have social cost at most $\OC_1 + \tfrac {mB}5$, the facility needs to be in the interval $\tfrac {-3B}4+h \pm(\tfrac{mB}{5} - 2\beta m \cdot h)$ and $\tfrac{mB}{5} \ge 2\beta m \cdot h$. 
    Finally, we note that with the facility at height greater than $\tfrac B4$, the three groups cost at least $\tfrac 34 mB$,
    and thus the total social cost in these locations is more than $\OC_1 + \tfrac {mB}5$.
    }

    To verify the analog of \Cref{obs:gp3}, we note that in region $\region_1$, at height $h$, Group 3 agents will have cost at least 
    $(-\tfrac B4 +h) -(\tfrac {-3B}4 +h) -(\tfrac{B}{5} - 2\beta  \cdot h) \ge \tfrac {3}{10}B$, as desired.

    We turn to $\inst_2$ and $\region_2$. $\region_2$ will comprise a triangle with base $\interval_2$ and apex at height $\tilde h$ on the right boundary of the Group 1 agent diamonds (see \cref{fig::2d-hardness-inst2}). 
    $\tilde h$ will be chosen so that the cost on the boundary is exactly $\OC_2+\tfrac {mB}5$.
    At the triangle apex, the cost is $\OC_2+2mB\beta\cdot \tilde h$, so long as $\tilde h\le \tfrac B8$
    (this is the height of the intersection of the diamonds for Groups 2 and 3).
    We choose $\tilde h= \tfrac B8$, and therefore $\beta = \tfrac 25$.
    We note that the point $(\tfrac B8,\tfrac B8)$ (the previously mentioned intersection) also has cost
    $\OC_2+ 2mB\beta \tilde h = \OC_2 +\tfrac {mb}5$, so we add it to $\region_2$.
    Again, by inspection, all points strictly inside $\region_2$ have cost less than $\OC_2 +\tfrac {mb}5$,
    and all points outside this region have strictly larger cost.
    
    Finally, the analog of \Cref{obs:gps1-2} is really verified by inspection: in $\region_2$, in combination, a Group 1 and a Group 2 agent have cost at least $\tfrac 34 B$,
    and outside this region, they have cost at least $\tfrac 14 B$.

    \hide{
    This has been designed so that no point on the right boundary of the Group 2 agents has cost less that $\OC_2+ \tfrac 15 mB$ \pj{this is unclear}. To see this, note that at height $h$, such a location \pj{what coordinates?} has social cost $\OC_2 + \tfrac 14 mB -2h \cdot mB +2\beta m\cdot h$. So we need $\tfrac 14\beta mB > \tfrac 4{20} mB$. $\beta=1$ suffices. \pj{for clarity: mention where $\OC_2$ is achieved.}
    The three groups of agents combined always have social cost at least $\tfrac 34 mB$, which ensures that above this height the overall social cost is more than $\OC_2+ \frac 4{20}mB$.
    }
However, the proof of \Cref{lemma:dist2} needs a few changes (basically, to add in and then cancel the contribution of the new agents to the social cost). We now have
    \begin{align*}
        \E \big[\COST(y, \inst_2)\big] < \OC_2 + \tfrac{mB}{100} = \tfrac{3mB}{4} + \tfrac74 mB\beta + \tfrac{mB}{100},
    \end{align*}
    yielding
        \begin{align}\label{eqn:new-lem10-cost-bdd}
        &\E \big[\COST(y, \inst_2)\big]\\ \nonumber
        &~~= m\cdot\E\big[\cost(y, (x'_{1}, b'_{1}))\big] + m\cdot\E\big[\cost(y, (x'_{m+1}, b'_{m+1}))\big] \\ \nonumber
        &~~~~~~+ m\cdot\E\big[\cost(y, (x'_{2m+1}, b'_{2m+1}))\big] + \E\big[\text{new agent costs}\big] \\ \nonumber
        &~~< \tfrac{3mB}{4} + \tfrac74 mB\beta + \tfrac{mB}{100},
    \end{align}
    
    As in the proof of \Cref{lemma:dist2}, using \Cref{obs:gps1-2}, we obtain
    \begin{align*}
        m\cdot\E\big[\cost(y, (x'_{1}, b'_{1}))\big] + m\cdot\E\big[\cost(y, (x'_{m+1}, b'_{m+1}))\big]\\
        ~~~~\ge \tfrac {29mB}{40}.
    \end{align*}
    
    We also observe that $\E\big[\text{new agent costs}\big] \ge \tfrac74 mB\beta$.

    Substituting these two bounds into \eqref{eqn:new-lem10-cost-bdd} gives
    \begin{align*}
       &m\cdot \E\big[\cost(y, (x'_{2m+1}, b'_{2m+1}))\big]\\
        &~~~~< \tfrac{3mB}{4} + \tfrac74 mB\beta + \tfrac{mB}{100} - \tfrac{29mB}{40} - \tfrac74 mB\beta\\
        &~~~~= m\cdot\tfrac{57B}{200}.
    \end{align*}
    On dividing by $m$ this proves \Cref{lemma:dist2}.

    To conclude the proof, we note that the approximation factor is at least $\tfrac 1{100}mB$. And $n= 3m+2\beta m=3.8m$. %, taking into account the additional $2\beta m$ agents (recall that $\beta=1$ suffices). 
    This yields an approximation factor of $\frac 1{380}nB$. (Strictly speaking, we need $m\beta$ to be an integer, as this is the number of new agents at each of the new agent locations; it suffices to have $m$ be an integer multiple of 5.)
    \end{proof}

\begin{theorem}\label{thm:2d-hardness-L2}
%{\bf \Cref{thm:2d-hardness}}~
  Consider an arbitrary mechanism $f$ for the facility location game in 2D with the $L_2$ distance measure and with doubly peaked preferences that is truthful-in-expectation. The additive approximation error of $f$ is at least $\frac{nB}{630,500}$. 
  %That is, for any given $f$, there exists a worst-case instance $\inst$ such that $\E_{y \sim f(\inst)}[\COST(y, \inst)] \geq \OPT(\inst) + \frac{nB}{300}$.
\end{theorem} 
\begin{proof} %(Sketch.)~  
    As in the proof of \Cref{thm:2d-hardness} we introduce an additional $\beta m$ agents at each of the locations $(\tfrac{-3B}4,0)$ and $(B,0)$. The shape of regions $\region_1$ and $\region_2$ will change a bit, but we will achieve the same bounds on the costs that are needed to deduce that the approximation factor must be at least $\frac{mB}{100}$. 
    To this end, we define $\region_1$ to comprise those points in $\inst_1$ with social cost at most $\OC_1+\tfrac{mB}5$, and $\region_2$ to comprise those points in $\inst_2$ with social cost at most $\OC_2+\tfrac{mB}5$.
    In addition, we need that
    %Namely, we need that in $\inst_1$, the social cost outside $\region_1$ be at least $\OC_1+\frac{mB}5$, and 
    for each Group 3 agent, the cost inside $\region_1$ be at least $\frac{3B}{10}$. Similarly, in $\inst_2$, the social cost outside $\region_2$ is at least $\OC_2+\frac{mB}5$, 
    and
    we will need to modify \Cref{obs:gps1-2} to account for the new agents, again causing minor changes to the proof of \Cref{lemma:dist2}. The modified \Cref{obs:gps1-2} states that excluding Group 3 agents, the cost
    inside $\region_2$ is at least $\tfrac{3mB}{4}+ \tfrac{7mB\beta}4$ and outside $\region_2$ is at least $\tfrac {mB}4+\tfrac{7mB\beta}4$.

    This leads to the following modification of the proof of \Cref{lemma:dist2}. Letting $p$ denote the probability that the mechanism chooses a facility location in $\region_2$, we bound
    \begin{align*}
        &m\cdot\E\big[\cost(y, (x'_{1}, b'_{1}))\big] + m\cdot\E\big[\cost(y, (x'_{m+1}, b'_{m+1}))\big] \\ 
        &~~~~~~+ \E\big[\text{new agent costs}\big]\\
        &~~\le p\big(\tfrac{3mB}{4}+ \tfrac{7mB\beta}4\big) +(1-p)\cdot\big(\tfrac {mB}4+\tfrac{7mB\beta}4\big)\\
        &~~\le \tfrac{7mB\beta}4 + \tfrac {29B}{40},
    \end{align*}
    using the fact that $p\ge \tfrac{19}{20}$, as in the proof of \Cref{lemma:dist2}.

Now, as in the proof of \Cref{thm:2d-hardness}, we substitute this bound into \eqref{eqn:new-lem10-cost-bdd}, which yields the bound stated in \Cref{lemma:dist2}.
  
  The additional cost due to the new agents at a location $(Bx,By)$ with $-1\le x \le 1$ and $y\ge 0$
  is 
  \begin{align*}
  \beta Bm\big[\big((1-x)^2 +y^2\big)^{1/2} + \big(\tfrac 34 +x)^2 + y^2\big)^{1/2}\big].
  \end{align*}
  We first ensure that when $By> \tfrac B{60}$, the contribution of this term is at least $\OC_1+\tfrac {Bm}5$, so these values of $By$ lie outside both $\region_1$ and $\region_2$. This function is minimized when $1-x = \tfrac 34 +x$, i.e.\ $ x = \tfrac 18$, so it suffices to have $\beta(\tfrac {49}{64} + \tfrac{1}{3600}) \ge \tfrac 34 + \tfrac {7\beta}4 + \tfrac 15$; $\beta \ge 3151$ suffices.

  We turn to the analysis of $\inst_1$. First, we observe that if $Bx\le \tfrac{-11}{20}B$, then
  each Group 3 agent has cost at least $\tfrac 3{10}B$, so for any location in this range in $\region_1$,
  the Group 3 agents have the desired cost. We conclude the analysis of $\inst_1$ by showing that all locations with $Bx> \tfrac{-11}{20}B$ have cost greater than $\OC_1 +\tfrac{4}{20}mB$ and hence lie outside
  $\region_1$.
  
  We begin by noting that if the social cost due to the agents in Groups 1--3 at location $(Bx,0)$ is at least $mB$, then the social cost at location $(Bx,By)$ for $0\le y \le 1/60$ is at least $\tfrac{19}{20}mB$, as the cost for each agent in Groups 1--3 decreases by at most $B/60$, and the cost for the new agents only increases. Therefore only values of $x$ we still need to consider are $\tfrac{-11}{20}<x<\tfrac{-1}2$.
  The social cost in this range is given by the following expression (with $mB$ factored out to reduce clutter).
  \begin{align*}
    \SC(x,y) &=  1 -\big[x^2 +y^2\big]^{1/2} + \big[(\tfrac 12-x\big)^2+y^2\big]^{1/2}-\tfrac 34 \\
    &+\tfrac 12 - \big[(\tfrac{-1}{4}-x)^2 +y^2\big]^{1/2} + \beta\big[(1-x)^2+y^2\big]^{1/2}\\\nonumber
     &~~~~+\beta\big[(\tfrac {3}4+x)^2+y^2\big]^{1/2}.
  \end{align*}
  We perform a power series expansion to allow us to lower bound this expression.
  \begin{align*}
      \SC(x,y) & \ge \frac 34 - (-x)\Big(1 +\frac{y^2}{2x^2}\Big) \\
      &~~+ (\tfrac 12 -x) \cdot \Big(1 + \frac {y^2}{2(\tfrac12 -x)^2} - \frac{y^4}{8(\tfrac12-x)^4}\Big) \\
      &~~~-(\tfrac{-1}{4} -x)\cdot\Big(1 + \frac{y^2}{2(\tfrac14 +x)^2} \Big)\\
      &~~~~+\beta(1-x)\Big(1 + \frac{y^2}{(1-x)^2} - \frac{y^4}{8(1-x)^4}\Big) \\
            &~~~~~+\beta(\tfrac34 +x)\Big(1 + \frac{y^2}{2(\tfrac34 +x)^2} - \frac{y^4}{8(\tfrac34 + x)^4}\Big)\\
  \end{align*}
 We use the bounds $y\le 1/60$, $-11/20\le x \le -1/2$, to absorb the $y^4$ terms into the $y^2$ terms, as follows. 
 % \pj{minor fix. x = -11/20}. rnote{Done}
\begin{align*}
    \SC(x,y) &\ge \frac32 + x -\frac{y^2}{2(-x)} + \frac {y^2}{2(\tfrac 12 -x)} -\frac {y^2}{8\cdot 60^2} \\
                  &~~~~-\frac{y^2}{2(\tfrac {-1}{4} -x))}+\tfrac 74 \beta +\frac{\beta y^2}{1-x} - \frac{\beta y^2}{ 60^2 \cdot 27}\\
              &~~~~
              + \frac {\beta y^2}{2(\tfrac34 +x)} -\frac {5^3\beta y^2}{ 60^2 \cdot 8}\\
              & \ge \frac 32 + \frac 74 \beta +x - {y^2} + \frac {10y^2}{21} -\frac {y^2}{8\cdot 60^2}
                    -  {2y^2} \\
                    &~~~~+ \frac {20 \beta y^2}{31} - \frac{\beta y^2}{27 \cdot 60^2} + 2\beta y^2 - \frac{5^3\beta y^2}{8\cdot 60^2}. \\
                & > \frac{19}{20} + \frac 74 \beta =\OC_1 + \frac {4}{20} + \frac 74 \beta,
                %> \frac 34 + \frac 74 \beta.
\end{align*}
if $\beta \ge 2$.

This demonstrates that we achieve the required bounds in $\inst_1$.

Next, we analyze $\inst_2$. Now the only values of $x$ of concern are $x>\tfrac 34 B$ (because the points $(x,0)$ with $x>\tfrac 34 B$ are the only ones with Groups 1--3 contributing social cost less than $Bm$). Note that we are only interested in those points in $\region_2$, whichever they are. So it is enough to lower bound the Group 1 plus Group 2 cost plus new agent cost by $(\tfrac 34  +\frac 74\beta) mB$ in the range $x>\tfrac 34 B$. 
We carry this out via two cases: $x^2+y^2 \ge 1$ and $x^2+y^2<1$, again using a power series in $x^2/y^2$.

{\sc Case 1}: $x^2+y^2 < 1$.\\
The Group 1, Group 2 and new agents contribute the following cost to the social cost.
  \begin{align*}
      \SC(x,y) & \ge  1-x \cdot\Big(1 + \frac{y^2}{2x^2}\Big) \\
      &~~~~+(x+\tfrac{1}{4})\cdot\Big(1 + \frac{y^2}{2(x+\tfrac14)^2} - \frac{y^4}{8(x+\tfrac14)^4} \Big) \\
      &~~~~- \frac 12 +\beta(1-x) \\
            &~~~~+\beta(\tfrac34 +x)\Big(1 + \frac{y^2}{2(\tfrac34 +x)^2} - \frac{y^4}{8(\tfrac34 + x)^4}\Big)\\
  \end{align*}
We use the bounds $y\le 1/60$, $\tfrac 34\le x \le 1$, yielding:
\begin{align*}
  \SC(x,y) & \ge \frac 34 + \frac 74 \beta - \frac{2y^2}{3} + \frac{2y^2}{5} - \frac{y^2}{8.60^2}\\
    &~~~~+\frac{2\beta y^2}{7} - \frac{\beta y^2}{27\cdot 60^2} \ge \frac 34 + \frac 74 \beta ,
\end{align*}
if $\beta \ge 1$.

\smallskip\noindent
{\sc Case 2}: $x^2+y^2 \ge 1$.\\
We break this case into two different cases: \\
\noindent
{\sc Case 2a: $\tfrac 34 \leq x \leq 1$}. \\
The Group 1, Group 2 and new agents contribute the following cost to the social cost.
  \begin{align} \label{eq:2d-L2-R2-cost}
  \begin{split}
      \SC(x,y) & \ge  (x^2+y^2)^{1/2}-1 \\
      &~~~~+(x+\tfrac{1}{4})\cdot\Big(1 + \frac{y^2}{2(x+\tfrac14)^2} - \frac{y^4}{8(x+\tfrac14)^4} \Big) \\
      &~~~~- \frac 12 +\beta(1-x) \\
            &~~~~+\beta(\tfrac34 +x)\Big(1 + \frac{y^2}{2(\tfrac34 +x)^2} - \frac{y^4}{8(\tfrac34 + x)^4}\Big)\\
    \end{split}
  \end{align}
We use the bounds $x^2+y^2\ge 1$, $y\le 1/60$, $\tfrac 34\le x \le 1$. Let $x = 1 - \delta$ for $\delta < \tfrac 14$. Since $x^2 + y^2 \geq 1$, we have $y^2 \geq \delta(2 - \delta)$. Substituting these values yields:
\begin{align*}
  \SC(x,y) & \ge x - \frac 14 + \frac 74 \beta  + \frac{2y^2}{5} - \frac{y^2}{8.60^2}
    +\frac{2\beta y^2}{7} \\
    &~~~~- \frac{\beta y^2}{27\cdot 60^2} \\
    &\ge \frac 34 + \frac 74 \beta + \frac{y^2}{5} + \frac{2\beta \delta (2 - \delta)}{7} -\frac{\beta \delta (2 - \delta)}{27 \cdot 60^2} -\delta,
\end{align*}
Note that $\frac{2\beta \delta (2 - \delta)}{7} -\frac{\beta \delta (2 - \delta)}{27 \cdot 60^2} -\delta$
\begin{align*}
    &= \delta \big(\beta (2-\delta)\cdot(\tfrac 27 - \tfrac{1}{27\cdot60^2}) - 1\big) \\
    &\geq \delta \big(\beta (2-\delta)\cdot(\tfrac 17) - 1\big) \\
    &\geq 0
\end{align*}
if $\beta > 4$ (since $\delta < \tfrac 14$).

So $\SC(x,y) \ge \frac 34 + \frac 74 \beta$, as desired.

\medskip
\noindent
{\sc Case 2b: $x > 1$}. \\
Now 
\begin{align*}
    \SC(x,y) &= (x^2+y^2)^{1/2}-1 + ((x+\tfrac 14)^2 +y^2)^{1/2}\\ 
    &~~~~- \frac 12 + ((x-\tfrac 12)^2+ y^2)^{1/2} -\frac 12\\
    &~~~~ + \beta((x+\tfrac 34)^2+y^2)^{1/2} 
     + \beta((x-1)^2+y^2)^{1/2}\\
    &> \frac 34 + \frac 74 \beta,   
\end{align*}
as desired.

\hide{
{\sc Case 2b: $1 \leq x \leq 2$} \\
Using \cref{eq:2d-L2-R2-cost} and the bounds on $x$ and $y$, we get 

\begin{align*}
  \SC(x,y) & \ge \frac 34 + \frac 74 \beta + \frac{y^2}{2(2+\tfrac 14)} \\
  &~~~~ -\frac{y^2}{8\cdot 60^2\cdot(1 + \tfrac 14)^3} \\
  &~~~~ +\frac{\beta y^2}{2(2 + \tfrac 34)} - \frac{\beta y^2}{8\cdot60^2(1 + \tfrac 34)^3} \\
  &> \frac{3}{4} + \frac{7}{4} \beta
\end{align*}
for $\beta \geq 0$.

\noindent
{\sc Case 2c: $x > 2$} \\
Using \cref{eq:2d-L2-R2-cost} and the bounds on $x$ and $y$, we get
\begin{align*}
      \SC(x,y) & \ge  (x^2+y^2)^{1/2}-1 \\
      &~~~~+(x+\tfrac{1}{4}) - \frac 12 +\beta(1-x) \\
            &~~~~+\beta(\tfrac34 +x)\\
            & > \frac{11}{4} + \frac 74 \beta
  \end{align*}
for $\beta \geq 0$.
}

Finally, by inspection, the Group 1 plus Group 2 costs are always at least $\tfrac {mB}4$ and the new agents always contribute at least $\tfrac 74 mB\beta$, verifying the remaining part of the analog of \Cref{obs:gps1-2}.

Since $\beta = 3151$, we conclude that $n =6305m$, and as the additional cost in the approximation is $100Bm$,
this yields the claimed lower bound.
\end{proof}

We note that the constant in the last theorem could be improved substantially, but at the cost
of having many more cases to analyze.

\subsection{2D Median Plus}
\label{sec::2D-MedPlus}

In this section, we generalize the Median-Plus mechanism to handle settings with agents located in 2D,
when using $L_1$ distances as the cost metric. Throughout this section $\Omega = \mathbb{R}^2$, and the facility location instance is $\inst = \langle ((x_1, y_1), b_1), \ldots, ((x_n, b_n), b_n) \rangle$.

Placing the facility at the median of agent's $x$ and $y$ coordinates gives an approximation within $2nB$ of the optimal cost $\OPT$. In this section, we design a mechanism that performs at least as well as the Median algorithm for every instance $\inst$, and potentially improves the cost by $\Theta(nB)$ by utilizing agent's preferred distances. We begin by defining some notation we will be using.

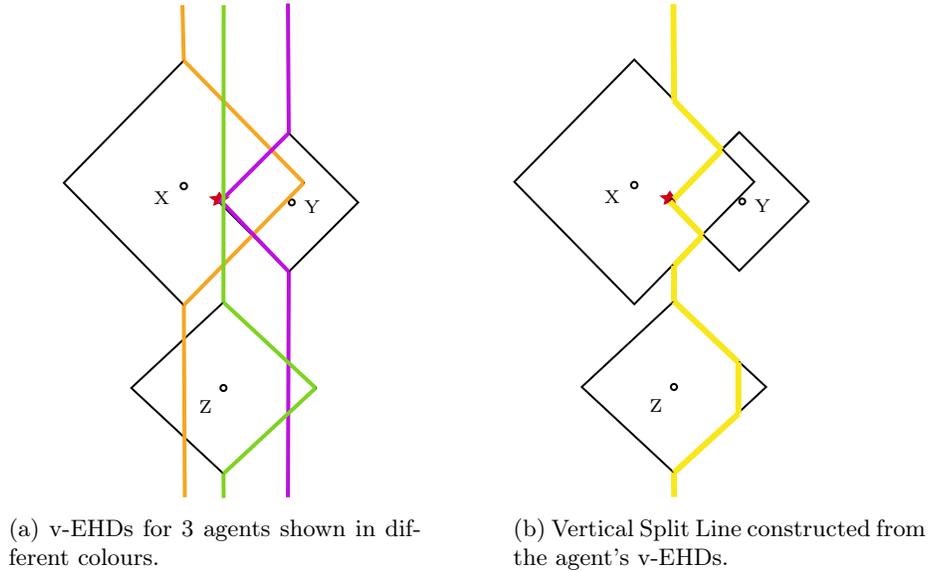
\begin{figure} 
    \centering
    \begin{subfigure}[b]{0.45\textwidth} \label{fig::vertical-split-line}
        \centering
        \input{vertical-extended-half-diamond.tex}
        \caption{v-EHDs for 3 agents shown in different colours.}
        \label{fig::vertical-ehd}
    \end{subfigure}
    \hfill % Optional space between the figures
    \begin{subfigure}[b]{0.45\textwidth}
        \input{vertical-split-line.tex}
        \caption{Vertical Split Line constructed from the agent's v-EHDs.}
        \label{fig::vertical-splitline}
    \end{subfigure}
    \caption{Agent X, Y and Z with their preferred locations. Locations where an agent has zero cost form a diamond centered at their location. The red star marks the coordinate-wise median of the agent locations.}
    \label{fig::vertical-lines}
\end{figure}

Let $\med = (\med_x, \med_y)$ denote the coordinate-wise median of the agent's locations. $\med_x$ is the median of the points $(x_1, x_2, \ldots, x_n)$, and $\med_y$ is the median of the points $(y_1, y_2, \ldots, y_n)$. The optimal facility location for an agent $i$ lies on a diamond at a distance $b_i$ from $(x_i, y_i)$, as shown in \cref{fig::vertical-lines}. 

\begin{definition}[Vertical Extended Half Diamond (v-EHD)] \label{def::vehd}
    Let Agent $i$ be located at $(x_i, y_i)$ and have a preferred distance $b_i$. We define two zig-zag lines, named $L$ and $R$. Each consists of half the boundary of a diamond at distance $b_i$ from $(x_i,y_i)$, plus vertical extensions to $y=\pm \infty$. More precisely:
    \begin{align*}
        L = \{ (x, y) ~|~ |x - x_i| + |y - y_i| = b_i \text{ and } x \leq x_i\} \\ \cup \{(x_i, y) ~|~ |y - y_i| > b_i\} \\
        R = \{ (x, y) ~|~ |x - x_i| + |y - y_i| = b_i \text{ and } x \geq x_i\} \\ \cup \{(x_i, y) ~|~ |y - y_i| > b_i\} \\
    \end{align*}
    If $x_i \leq \med_x$, then v-EHD for agent $i$ is defined as the set of points in $R$ and otherwise it is defined as the points in $L$. For an instance $\inst$, v-EHD for an agent $i$ can also be interpreted as a function $v_i$ of the $y$ coordinate such that the points of the form $(v_i(y), y)$ define the v-EHD of agent $i$ (\cref{fig::vertical-ehd}). 
\end{definition}

\begin{definition}[Vertical Split Line]
    Let $v_i$ denote the v-EHD of agent $i$. For  $y \in \mathbb{R}$, let $\Vec{v}(y) = (v_1(y), \ldots, v_n(y))$. The Vertical split line is defined by the function $\mathcal{V}(y) = \operatorname{median}(\Vec{v}_y)$ (see \cref{fig::vertical-splitline}).
\end{definition}

\begin{figure}
    \centering
    \begin{subfigure}[b]{0.45\textwidth}
        \centering
        \input{horizontal-extended-half-diamond.tex}
        \caption{h-EHDs for 3 agents shown in different colours.}
        \label{fig::horizontal-ehd}
    \end{subfigure}
    \hfill % Optional space between the figures
    \begin{subfigure}[b]{0.45\textwidth}
        \input{horizontal-split-line.tex}
        \caption{Horizontal Split Line constructed from the agent's h-EHDs.}
        \label{fig::horizontal-splitline}
    \end{subfigure}
    \caption{h-EHDs and Horizontal Split Lines for agents X, Y and Z (same instance as \cref{fig::vertical-lines}).}
    \label{fig::horizontal-lines}
\end{figure}
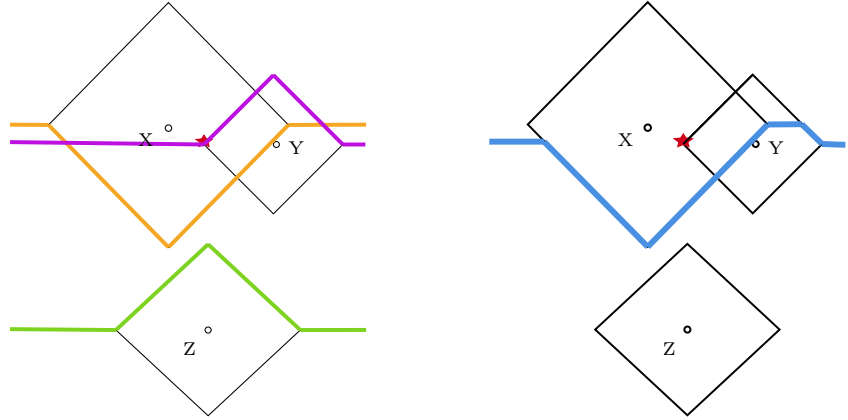

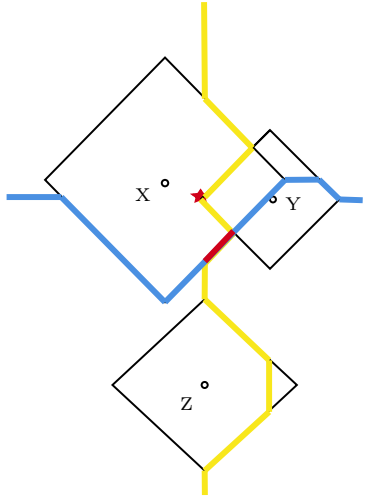
\begin{figure}
    \input{split-line-intersection.tex}
    \caption{Intersection of vertical and horizonal split lines. In this instance, the intersection is a line segment on agent X's diamond and is highlighted in red.}
    \label{fig::splitline-intersection}
\end{figure}

The Horizontal Extended Half Diamonds (h-EHDs) and Horizontal Split Line are defined analogously, and illustrated in Figures \ref{fig::horizontal-splitline} and \ref{fig::splitline-intersection}.

\hide{
\begin{definition}[Horizontal Extended Half Diamond (h-EHD)]\label{def::hehd}
    For an agent $i$ located at $(x_i, y_i)$ with a preferred distance $b_i$, define $U$ and $D$ to be the set of points as follows:
    \begin{align*}
        D &= \{ (x, y) ~|~ |x - x_i| + |y - y_i| = b_i \text{ and } y \leq y_i\} \cup \{(x, y_i) ~|~ |x - x_i| > b_i\} \\
        U &= \{ (x, y) ~|~ |x - x_i| + |y - y_i| = b_i \text{ and } y \geq y_i\} \cup \{(x, y_i) ~|~ |x - x_i| > b_i\} \\
    \end{align*}
    If $y_i < \med_y$, then h-EHD for agent $i$ is defined as the set of points in $D$ and otherwise it is defined by the points in $U$. For an instance $\inst$, h-EHD for an agent $i$ can also be interpreted as a function $h_i$ of the $x$ coordinate such that the points of the form $(x, h_i(x))$ define the h-EHD of agent $i$ (\cref{fig::horizontal-ehd}).
\end{definition}

\begin{definition}[Horizontal Split Line]
    Let $h_i$ denote the h-EHD of agent $i$. For  $x \in \mathbb{R}$, let $\Vec{h}(x) = (h_1(x), \ldots, h_n(x))$. The Horizontal split line is defined by the function $\mathcal{H}(x) = \operatorname{median}(\Vec{h}_x)$ (\cref{fig::horizontal-splitline}).
\end{definition}
}

The following observation follows from the construction.
\begin{observation} \label{obs::split-line-intersection}
    For any instance $\inst$, the intersection of the Vertical Split Line and the Horizontal Split Line is either a single point or a single line segment with a slope of $45\degree /135\degree$ (which will be along the edge of some agent's diamond, as shown in \cref{fig::splitline-intersection}).
\end{observation}

\begin{figure}
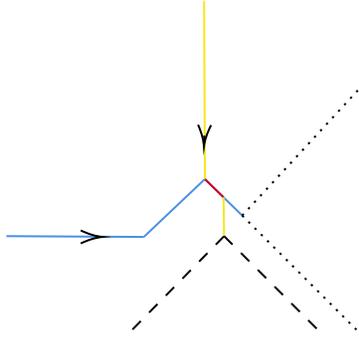

    \centering
    \include{split-line-intersection-proof}
    \caption{Vertical (yellow) and Horizontal (blue) split lines intersecting on a line segment (red) and separating thereafter. The regions that bound the two split lines are shaded. Note that the regions that bound the split lines are disjoint after the two lines get separated.}
    \label{fig::split-line-intersection-proof}
\end{figure}

\begin{proof} (Sketch.)~
    In order to see that there can only be one intersection (at a single point or a line segment), we traverse through the split lines simultaneously. Starting from the uppermost point of intersection, we traverse the vertical split line from top to bottom and traverse the horizontal split line as follows: \footnote{We do this in order to make sure that the traversal on the vertical and horizontal split lines can be made simultaneously in the same direction.}
    \begin{itemize}
        \item \emph{left to right} if the intersection is at a single point or if it's on an edge that is going to the right and down. (see \cref{fig::split-line-intersection-proof})
        \item \emph{right to left} if the intersection edge is going to the left and down.
    \end{itemize}
    
    % starting from the upper-leftmost point of intersection between the two split lines.
    Now, without loss of generality, assume that the intersection is either a single point or an edge that is going to the right and down.
    Note that from any given point on the horizontal split line, its continuation must lie in a region
    bounded by two lines, one heading up and right at 45\degree and one heading down and right at 45\degree  (\cref{fig::split-line-intersection-proof}).
    Similarly the continuation of the vertical split line is bounded by two lines, one heading down and right at 45\degree and one heading down at left at 45\degree (\cref{fig::split-line-intersection-proof}). This implies that once the two split lines intersect they can only continue their intersection along a single line heading down and to the right.
    And once they separate, the regions for their continuation will become disjoint (\cref{fig::split-line-intersection-proof}). Consequently, they intersect either at a single point or along a single edge.

    We also note that this single edge is formed by a single diamond. For example, if we consider the vertical split line, if two diamonds share an edge, when the corresponding v-EHDs separate, the left-to-right order is maintained (modulo tie-breaking along the shared edge). 

    \hide{
    While traversing the horizontal split line the $y$ coordinate can decrease only when moving along the edge of a diamond, at a $135 \degree$ angle. Similarly, while traversing along the vertical split line the $x$ coordinate can only increase while moving along the edge of a diamond at a $135 \degree$ angle.

    Since the split lines intersect at either a point or on the edge of a diamond, the intersection region is finite. At some point through the traversal, the split lines will separate. From that point onwards, traversing the horizontal split line always increases the $x$ coordinate and traversing the vertical split line always decreases the $y$ coordinate. While traversing the horizontal split line the $y$ coordinate can decrease only when moving along the edge of a diamond, at a $135 \degree$ angle. Similarly, while traversing along the vertical split line the $x$ coordinate can only increase while moving along the edge of a diamond at a $135 \degree$ angle. Therefore, the two split lines can at-most be parallel to each other after separating at the first intersection region and hence will never intersect again.
    }
\end{proof}

With the help of these definitions, we design the following mechanism that uses agents' preferred distances to improve over the Median algorithm.

\begin{algorithm}
\caption{2D-Median-Plus mechanism for one facility in $\mathbb{R}^2$ with $L_1$ distance metric}
\label{mech::2dl1}

\KwIn{$\inst = \langle((x_1, y_1), b_1), $\ldots$, ((x_n, y_n), b_n) \rangle$.}

%Plot the horizontal split line. \\
%Plot the vertical split line.
Compute the horizontal and vertical split lines.

\KwOut{$\med^+$, the intersection point of the vertical and horizontal split line. If there are multiple such points, return the lowest cost point, and if there are several, output the one nearest to $\med$.}
\end{algorithm}

% Note that in the 2D-Median-Plus mechanism, if the vertical and horizontal split lines intersect in a line segment, then the facility location is not yet fully determined.
% can be placed at any deterministically chosen location on the line segment as long as the choice is independent of the agent's preferred distances.

\hide{
\begin{figure}
    \input{2d-l1-strategyproof.tex}
    \caption{}
    \label{fig::2d-l1-strategproof}
\end{figure}
}

\begin{theorem}
    2D-Median-Plus is strategyproof.
\end{theorem}
\begin{proof}
\hide{
    Let $t = (t_x, t_y)$ be the output of \cref{mech::2dl1} on instance $\inst$. We use $v_i$ and $h_i$ to denote the v-EHD and h-EHD of agent $i \in [n]$ as defined in \cref{def::vehd} and \cref{def::hehd}. We will prove that for any agent $i$ located at $(x_i, y_i)$ with a preferred distance of $b_i$, truthfully reporting its preferred distance is the best strategy.
    }
     The key observation is that the only agents who can affect the location $\med^+$ computed by the mechanism are those for which at least one of their reported v-EHD or h-EHD include the location $\med^+$.
     Otherwise a small change to an agent's declarations does not affect the median of the v-EHDs and h-EHDs at $\med^+$, and so this remains the location reported by the mechanism.

     A little more care is needed to justify this claim in the event that the intersection of the two split lines is an edge rather than a point. But even then, a small change to an agent's declarations does not affect the relative costs on this edge because changing its $b_i$ declaration has the same effect on all the points on the edge, as the edge on the intersection is either completely outside the agent's declared diamond or completely inside, and so the location $\med^+$ is still unchanged.

     Now if the change is sufficient to touch the location $\med^+$, then further changes of the declared preferred distance could move the location $\med^+$. First, let's suppose the correct declaration has $\med^+$ outside Agent $i$'s diamond. We will describe Agent $i$'s bid as if she is continuously increasing her bid to bring her diamond to the actual $\med^+$ and then beyond it; at first, this increase could drag the computed $\med^+$ with the expanding diamond; eventually, the computed $\med^+$ moves no further and as the diamond boundary is beyond the computed $\med^+$, further expansion has no effect.
     But note that the expansion only moves the computed $\med^+$ further from Agent $i$'s preferred locations, namely the diamond boundary with the truthful declaration.

     A similar argument applies if $\med^+$ lies inside Agent's $i$'s ``correct'' diamond. Here a shrinking diamond may drag the computed $\med^+$ away from the boundary of its correct diamond. So in both cases, Agent $i$ has no incentive to make an incorrect declaration.
%    
   % WLOG, assume that agent $i$ is located at $(0, 0)$. Consider the case when $|t_x| + |t_y| \geq b_i$ (mechanism output is outside agent $i$'s diamond). Let $ $
\end{proof}

\begin{lemma} \label{lem::2d-med-plus-better-than-med}
    The social cost of 2D-Median-Plus is at least as good as the social cost of the Median algorithm for every instance $\inst$.
\end{lemma}
\begin{proof}
 Let $\med=(\med_x,\med_y)$ denote the median location, and let $\med^+=(\med^+_x,\med^+_y)$ denote the location returned by the 2D-Median-Plus Mechanism. 
 To facilitate the remaining discussion, WLOG suppose that $\med$ is above and to the left of $\med^+$.
 We imagine continuously moving the facility from location $\med$ to location $\med^+$ in such a way that the social cost is always non-increasing. To this end, the overall move is made in a series of alternating horizontal and vertical moves,
 always moving away from $\med$ and toward $\med^+$. The moves are further constrained as follows. Each horizontal move stops no later than the vertical split line (i.e.\ it does not cross this split line) and each vertical move stops no later than the horizontal split line.
 Note that a horizontal move could cross the horizontal split line, and similarly for a vertical move.

There is possibly one additional final move along a diagonal edge. In the event that the two split lines intersect in an edge, this edge could have a uniform cost, or could have increasing cost in one direction or the other. 
As $\med^+$ is the low-cost endpoint, the move along this edge will either reduce the cost or leave it unchanged.
\hide{
In the event the endpoint farther from $\med$ is the strictly low cost endpoint, and the vertical and horizontal moves reach this edge before the far endpoint, we complete the traversal by moving along the edge.
But in this case the cost of this move is decreasing, and so our claim that the whole traversal has non-increasing cost throughout is maintained.
\rjc{Otherwise, the low cost endpoint, which is the location $\med^+$, is where the edge is first reached,
as the series of moves never goes to the right of $\med^+$.}
}

 Now we justify the claim regarding the social cost being non-increasing. Consider Agent $i$'s cost.
Let $d_i$ be the distance from the facility to Agent $i$ and let $b_i$ be Agent $i$'s preferred distance.
Then, if the facility is outside Agent $i$'s diamond, its cost is $d_i-b_i$, and if inside its cost is $b_i-d_i$. Suppose the facility moves distance $\delta$ horizontally. If the facility is outside $i$'s diamond and the move is away from the diamond, then $d_i$ increases by $\delta$. If the facility is inside, and the move is away from the nearest boundary, then $d_i$ decreases by $\delta$. In both cases, the cost increases by $\delta$. With moves in the opposite directions, the cost decreases by $\delta$.

The next observation is that a horizontal move is toward at least $\lfloor (n+1)/2\rfloor$ v-EHDs and away from at most $\lceil (n-1)/2\rceil$ (these two numbers could be equal if $n$ is even). Furthermore, as we will show shortly, the toward moves are all cost decreasing. Note that if this claim is true then the social cost must be non-increasing. 

Clearly, if the facility is on the outside of the diamond defining a v-EHD, and the move is toward the v-EHD, then the cost is decreasing. If inside the diamond, the issue is whether the facility is moving toward or away from the agent: moving away reduces the cost, while moving toward would increase it. 
Consider an R-type v-EHD. This can occur only when the agent is aligned with or to the left of $\med$, and as the facility is aligned with or to the right of location $\med$ and moving to the right in a horizontal move,
it must be moving away from the agent, i.e.\ cost reducing.
Now consider an L-type v-EHD. In this case the facility is moving away from the v-EHD; so this v-EHD is on the minority side of $\med$.  While the cost may increase for this v-EHD, as the movement is toward the side with the majority of v-EHDs, overall the movement does not increase the cost.

An analogous argument applies to the vertical moves.

\end{proof}

\end{document}

%% file: 1d-strategyproof-case1.tex
% \documentclass[tikz,border=10pt]{standalone}
% \usepackage{tikz}
% % Other packages or TikZ libraries
% \usetikzlibrary{decorations.pathreplacing}

% \begin{document}

\tikzset{every picture/.style={line width=0.75pt}} %set default line width to 0.75pt        

\begin{tikzpicture}[x=0.75pt,y=0.75pt,yscale=-1,xscale=1]
%uncomment if require: \path (0,310); %set diagram left start at 0, and has height of 310

%Straight Lines [id:da6080280675160796] 
\draw    (149.96,179.93) -- (349.96,180.33) ;
%Shape: Circle [id:dp3875348189266262] 
\draw  [fill={rgb, 255:red, 0; green, 0; blue, 0 }  ,fill opacity=1 ] (318.4,180.3) .. controls (318.4,179.36) and (319.16,178.6) .. (320.1,178.6) .. controls (321.04,178.6) and (321.8,179.36) .. (321.8,180.3) .. controls (321.8,181.24) and (321.04,182) .. (320.1,182) .. controls (319.16,182) and (318.4,181.24) .. (318.4,180.3) -- cycle ;
%Shape: Moon [id:dp4980308427335908] 
\draw  [fill={rgb, 255:red, 0; green, 0; blue, 0 }  ,fill opacity=1 ] (249,192.6) .. controls (245.8,192.6) and (243.2,187.23) .. (243.2,180.6) .. controls (243.2,173.97) and (245.8,168.6) .. (249,168.6) .. controls (246.22,169.51) and (244.1,174.54) .. (244.1,180.6) .. controls (244.1,186.66) and (246.22,191.69) .. (249,192.6) -- cycle ;
%Shape: Star [id:dp10060896827609234] 
\draw  [color={rgb, 255:red, 208; green, 2; blue, 27 }  ,draw opacity=1 ][fill={rgb, 255:red, 208; green, 2; blue, 27 }  ,fill opacity=1 ] (210.32,177.4) -- (211.16,179.34) -- (213.02,179.65) -- (211.67,181.17) -- (211.99,183.3) -- (210.32,182.29) -- (208.65,183.3) -- (208.97,181.17) -- (207.62,179.65) -- (209.48,179.34) -- cycle ;
%Shape: Moon [id:dp1576120615794635] 
\draw  [color={rgb, 255:red, 245; green, 166; blue, 35 }  ,draw opacity=1 ][fill={rgb, 255:red, 245; green, 166; blue, 35 }  ,fill opacity=1 ] (286.6,192.2) .. controls (283.4,192.2) and (280.8,186.83) .. (280.8,180.2) .. controls (280.8,173.57) and (283.4,168.2) .. (286.6,168.2) .. controls (283.82,169.11) and (281.7,174.14) .. (281.7,180.2) .. controls (281.7,186.26) and (283.82,191.29) .. (286.6,192.2) -- cycle ;
%Shape: Moon [id:dp5671448831386761] 
\draw  [color={rgb, 255:red, 126; green, 211; blue, 33 }  ,draw opacity=1 ][fill={rgb, 255:red, 126; green, 211; blue, 33 }  ,fill opacity=1 ] (168.2,191.8) .. controls (165,191.8) and (162.4,186.43) .. (162.4,179.8) .. controls (162.4,173.17) and (165,167.8) .. (168.2,167.8) .. controls (165.42,168.71) and (163.3,173.74) .. (163.3,179.8) .. controls (163.3,185.86) and (165.42,190.89) .. (168.2,191.8) -- cycle ;

% Text Node
\draw (313.6,183) node [anchor=north west][inner sep=0.75pt]   [align=left] {$\displaystyle x_{i}$};
% Text Node
\draw (219.6,195.2) node [anchor=north west][inner sep=0.75pt]   [align=left] {$\displaystyle x_{i} -b_{i}$};
% Text Node
\draw (193.48,153.6) node [anchor=north west][inner sep=0.75pt]   [align=left] {$\displaystyle f(\mathcal{I})$};
% Text Node
\draw (148.4,194.4) node [anchor=north west][inner sep=0.75pt]  [color={rgb, 255:red, 126; green, 211; blue, 33 }  ,opacity=1 ] [align=left] {$\displaystyle x_{i} -b'_{i}$};
% Text Node
\draw (275.2,145.2) node [anchor=north west][inner sep=0.75pt]  [color={rgb, 255:red, 245; green, 166; blue, 35 }  ,opacity=1 ] [align=left] {$\displaystyle x_{i} -b'_{i}$};

\end{tikzpicture}

% \end{document}

%% file: 1d-strategyproof-case2.tex
% \documentclass[tikz,border=10pt]{standalone}
% \usepackage{tikz}
% % Other packages or TikZ libraries
% \usetikzlibrary{decorations.pathreplacing}

% \begin{document}

\tikzset{every picture/.style={line width=0.75pt}} %set default line width to 0.75pt        

\begin{tikzpicture}[x=0.75pt,y=0.75pt,yscale=-1,xscale=1]
%uncomment if require: \path (0,310); %set diagram left start at 0, and has height of 310

%Straight Lines [id:da6080280675160796] 
\draw    (149.96,179.93) -- (349.96,180.33) ;
%Shape: Circle [id:dp3875348189266262] 
\draw  [fill={rgb, 255:red, 0; green, 0; blue, 0 }  ,fill opacity=1 ] (318.4,180.3) .. controls (318.4,179.36) and (319.16,178.6) .. (320.1,178.6) .. controls (321.04,178.6) and (321.8,179.36) .. (321.8,180.3) .. controls (321.8,181.24) and (321.04,182) .. (320.1,182) .. controls (319.16,182) and (318.4,181.24) .. (318.4,180.3) -- cycle ;
%Shape: Moon [id:dp4980308427335908] 
\draw  [fill={rgb, 255:red, 0; green, 0; blue, 0 }  ,fill opacity=1 ] (228.6,193) .. controls (225.4,193) and (222.8,187.63) .. (222.8,181) .. controls (222.8,174.37) and (225.4,169) .. (228.6,169) .. controls (225.82,169.91) and (223.7,174.94) .. (223.7,181) .. controls (223.7,187.06) and (225.82,192.09) .. (228.6,193) -- cycle ;
%Shape: Star [id:dp10060896827609234] 
\draw  [color={rgb, 255:red, 208; green, 2; blue, 27 }  ,draw opacity=1 ][fill={rgb, 255:red, 208; green, 2; blue, 27 }  ,fill opacity=1 ] (262.32,176.6) -- (263.16,178.54) -- (265.02,178.85) -- (263.67,180.37) -- (263.99,182.5) -- (262.32,181.49) -- (260.65,182.5) -- (260.97,180.37) -- (259.62,178.85) -- (261.48,178.54) -- cycle ;
%Shape: Moon [id:dp1576120615794635] 
\draw  [color={rgb, 255:red, 245; green, 166; blue, 35 }  ,draw opacity=1 ][fill={rgb, 255:red, 245; green, 166; blue, 35 }  ,fill opacity=1 ] (295,192.6) .. controls (291.8,192.6) and (289.2,187.23) .. (289.2,180.6) .. controls (289.2,173.97) and (291.8,168.6) .. (295,168.6) .. controls (292.22,169.51) and (290.1,174.54) .. (290.1,180.6) .. controls (290.1,186.66) and (292.22,191.69) .. (295,192.6) -- cycle ;
%Shape: Moon [id:dp5671448831386761] 
\draw  [color={rgb, 255:red, 126; green, 211; blue, 33 }  ,draw opacity=1 ][fill={rgb, 255:red, 126; green, 211; blue, 33 }  ,fill opacity=1 ] (168.2,191.8) .. controls (165,191.8) and (162.4,186.43) .. (162.4,179.8) .. controls (162.4,173.17) and (165,167.8) .. (168.2,167.8) .. controls (165.42,168.71) and (163.3,173.74) .. (163.3,179.8) .. controls (163.3,185.86) and (165.42,190.89) .. (168.2,191.8) -- cycle ;

% Text Node
\draw (313.6,183) node [anchor=north west][inner sep=0.75pt]   [align=left] {$\displaystyle x_{i}$};
% Text Node
\draw (211.6,194.8) node [anchor=north west][inner sep=0.75pt]   [align=left] {$\displaystyle x_{i} -b_{i}$};
% Text Node
\draw (249.08,153.2) node [anchor=north west][inner sep=0.75pt]   [align=left] {$\displaystyle f(\mathcal{I})$};
% Text Node
\draw (148.4,194.4) node [anchor=north west][inner sep=0.75pt]  [color={rgb, 255:red, 126; green, 211; blue, 33 }  ,opacity=1 ] [align=left] {$\displaystyle x_{i} -b'_{i}$};
% Text Node
\draw (292.8,148) node [anchor=north west][inner sep=0.75pt]  [color={rgb, 255:red, 245; green, 166; blue, 35 }  ,opacity=1 ] [align=left] {$\displaystyle x_{i} -b'_{i}$};

\end{tikzpicture}

% \end{document}

%% file: skewed-instance.tex
% \documentclass[tikz,border=10pt]{standalone}
% \usepackage{tikz}
% \usepackage{amsmath}
% % Other packages or TikZ libraries
% \usetikzlibrary{decorations.pathreplacing}

% \begin{document}

\tikzset{every picture/.style={line width=0.75pt}} %set default line width to 0.75pt        

\begin{tikzpicture}[x=0.75pt,y=0.75pt,yscale=-1,xscale=1]
%uncomment if require: \path (0,310); %set diagram left start at 0, and has height of 310

%Straight Lines [id:da3973759827258103] 
\draw    (169.88,190.23) -- (400.05,190.6) ;
%Shape: Moon [id:dp3426491839373218] 
\draw  [color={rgb, 255:red, 189; green, 16; blue, 224 }  ,draw opacity=1 ][fill={rgb, 255:red, 189; green, 16; blue, 224 }  ,fill opacity=1 ] (257.65,202.6) .. controls (255.22,202.6) and (253.25,197.32) .. (253.25,190.8) .. controls (253.25,184.29) and (255.22,179) .. (257.65,179) .. controls (256.33,181.36) and (255.45,185.76) .. (255.45,190.8) .. controls (255.45,195.85) and (256.33,200.25) .. (257.65,202.6) -- cycle ;
%Shape: Moon [id:dp3122830131689215] 
\draw  [color={rgb, 255:red, 245; green, 166; blue, 35 }  ,draw opacity=1 ][fill={rgb, 255:red, 245; green, 166; blue, 35 }  ,fill opacity=1 ] (212.05,203) .. controls (214.48,203) and (216.45,197.72) .. (216.45,191.2) .. controls (216.45,184.69) and (214.48,179.4) .. (212.05,179.4) .. controls (213.36,181.76) and (214.25,186.16) .. (214.25,191.2) .. controls (214.25,196.25) and (213.36,200.65) .. (212.05,203) -- cycle ;
%Flowchart: Connector [id:dp4206023517482159] 
\draw  [color={rgb, 255:red, 0; green, 0; blue, 0 }  ,draw opacity=1 ][fill={rgb, 255:red, 0; green, 0; blue, 0 }  ,fill opacity=1 ] (278.68,190.82) .. controls (278.68,189.61) and (279.43,188.63) .. (280.36,188.63) .. controls (281.29,188.63) and (282.05,189.61) .. (282.05,190.82) .. controls (282.05,192.02) and (281.29,193) .. (280.36,193) .. controls (279.43,193) and (278.68,192.02) .. (278.68,190.82) -- cycle ;
%Flowchart: Connector [id:dp9460775466835731] 
\draw  [color={rgb, 255:red, 245; green, 166; blue, 35 }  ,draw opacity=1 ][fill={rgb, 255:red, 245; green, 166; blue, 35 }  ,fill opacity=1 ] (178.28,190.42) .. controls (178.28,189.21) and (179.03,188.23) .. (179.96,188.23) .. controls (180.89,188.23) and (181.65,189.21) .. (181.65,190.42) .. controls (181.65,191.62) and (180.89,192.6) .. (179.96,192.6) .. controls (179.03,192.6) and (178.28,191.62) .. (178.28,190.42) -- cycle ;
%Flowchart: Connector [id:dp8777461191083638] 
\draw  [color={rgb, 255:red, 245; green, 166; blue, 35 }  ,draw opacity=1 ][fill={rgb, 255:red, 245; green, 166; blue, 35 }  ,fill opacity=1 ] (207.88,190.02) .. controls (207.88,188.81) and (208.63,187.83) .. (209.56,187.83) .. controls (210.49,187.83) and (211.25,188.81) .. (211.25,190.02) .. controls (211.25,191.22) and (210.49,192.2) .. (209.56,192.2) .. controls (208.63,192.2) and (207.88,191.22) .. (207.88,190.02) -- cycle ;
%Flowchart: Connector [id:dp8555892927020221] 
\draw  [color={rgb, 255:red, 245; green, 166; blue, 35 }  ,draw opacity=1 ][fill={rgb, 255:red, 245; green, 166; blue, 35 }  ,fill opacity=1 ] (222.68,190.02) .. controls (222.68,188.81) and (223.43,187.83) .. (224.36,187.83) .. controls (225.29,187.83) and (226.05,188.81) .. (226.05,190.02) .. controls (226.05,191.22) and (225.29,192.2) .. (224.36,192.2) .. controls (223.43,192.2) and (222.68,191.22) .. (222.68,190.02) -- cycle ;
%Flowchart: Connector [id:dp9718765436861366] 
\draw  [color={rgb, 255:red, 245; green, 166; blue, 35 }  ,draw opacity=1 ][fill={rgb, 255:red, 245; green, 166; blue, 35 }  ,fill opacity=1 ] (258.28,190.42) .. controls (258.28,189.21) and (259.03,188.23) .. (259.96,188.23) .. controls (260.89,188.23) and (261.65,189.21) .. (261.65,190.42) .. controls (261.65,191.62) and (260.89,192.6) .. (259.96,192.6) .. controls (259.03,192.6) and (258.28,191.62) .. (258.28,190.42) -- cycle ;
%Shape: Moon [id:dp6867909251267638] 
\draw  [color={rgb, 255:red, 245; green, 166; blue, 35 }  ,draw opacity=1 ][fill={rgb, 255:red, 245; green, 166; blue, 35 }  ,fill opacity=1 ] (229.25,202.2) .. controls (231.68,202.2) and (233.65,196.92) .. (233.65,190.4) .. controls (233.65,183.89) and (231.68,178.6) .. (229.25,178.6) .. controls (230.56,180.96) and (231.45,185.36) .. (231.45,190.4) .. controls (231.45,195.45) and (230.56,199.85) .. (229.25,202.2) -- cycle ;
%Shape: Moon [id:dp06639604053498771] 
\draw  [color={rgb, 255:red, 245; green, 166; blue, 35 }  ,draw opacity=1 ][fill={rgb, 255:red, 245; green, 166; blue, 35 }  ,fill opacity=1 ] (246.05,202.2) .. controls (248.48,202.2) and (250.45,196.92) .. (250.45,190.4) .. controls (250.45,183.89) and (248.48,178.6) .. (246.05,178.6) .. controls (247.36,180.96) and (248.25,185.36) .. (248.25,190.4) .. controls (248.25,195.45) and (247.36,199.85) .. (246.05,202.2) -- cycle ;
%Shape: Moon [id:dp5970015477652931] 
\draw  [color={rgb, 255:red, 245; green, 166; blue, 35 }  ,draw opacity=1 ][fill={rgb, 255:red, 245; green, 166; blue, 35 }  ,fill opacity=1 ] (268.05,202.2) .. controls (270.48,202.2) and (272.45,196.92) .. (272.45,190.4) .. controls (272.45,183.89) and (270.48,178.6) .. (268.05,178.6) .. controls (269.36,180.96) and (270.25,185.36) .. (270.25,190.4) .. controls (270.25,195.45) and (269.36,199.85) .. (268.05,202.2) -- cycle ;
%Flowchart: Connector [id:dp2663043143822782] 
\draw  [color={rgb, 255:red, 189; green, 16; blue, 224 }  ,draw opacity=1 ][fill={rgb, 255:red, 189; green, 16; blue, 224 }  ,fill opacity=1 ] (290.68,190.42) .. controls (290.68,189.21) and (291.43,188.23) .. (292.36,188.23) .. controls (293.29,188.23) and (294.05,189.21) .. (294.05,190.42) .. controls (294.05,191.62) and (293.29,192.6) .. (292.36,192.6) .. controls (291.43,192.6) and (290.68,191.62) .. (290.68,190.42) -- cycle ;
%Flowchart: Connector [id:dp20611037018755252] 
\draw  [color={rgb, 255:red, 189; green, 16; blue, 224 }  ,draw opacity=1 ][fill={rgb, 255:red, 189; green, 16; blue, 224 }  ,fill opacity=1 ] (307.88,190.02) .. controls (307.88,188.81) and (308.63,187.83) .. (309.56,187.83) .. controls (310.49,187.83) and (311.25,188.81) .. (311.25,190.02) .. controls (311.25,191.22) and (310.49,192.2) .. (309.56,192.2) .. controls (308.63,192.2) and (307.88,191.22) .. (307.88,190.02) -- cycle ;
%Flowchart: Connector [id:dp29819361907765474] 
\draw  [color={rgb, 255:red, 189; green, 16; blue, 224 }  ,draw opacity=1 ][fill={rgb, 255:red, 189; green, 16; blue, 224 }  ,fill opacity=1 ] (318.68,190.02) .. controls (318.68,188.81) and (319.43,187.83) .. (320.36,187.83) .. controls (321.29,187.83) and (322.05,188.81) .. (322.05,190.02) .. controls (322.05,191.22) and (321.29,192.2) .. (320.36,192.2) .. controls (319.43,192.2) and (318.68,191.22) .. (318.68,190.02) -- cycle ;
%Flowchart: Connector [id:dp12732600644802128] 
\draw  [color={rgb, 255:red, 189; green, 16; blue, 224 }  ,draw opacity=1 ][fill={rgb, 255:red, 189; green, 16; blue, 224 }  ,fill opacity=1 ] (339.48,190.02) .. controls (339.48,188.81) and (340.23,187.83) .. (341.16,187.83) .. controls (342.09,187.83) and (342.85,188.81) .. (342.85,190.02) .. controls (342.85,191.22) and (342.09,192.2) .. (341.16,192.2) .. controls (340.23,192.2) and (339.48,191.22) .. (339.48,190.02) -- cycle ;
%Shape: Moon [id:dp09440678098141064] 
\draw  [color={rgb, 255:red, 189; green, 16; blue, 224 }  ,draw opacity=1 ][fill={rgb, 255:red, 189; green, 16; blue, 224 }  ,fill opacity=1 ] (242.85,202.6) .. controls (240.42,202.6) and (238.45,197.32) .. (238.45,190.8) .. controls (238.45,184.29) and (240.42,179) .. (242.85,179) .. controls (241.53,181.36) and (240.65,185.76) .. (240.65,190.8) .. controls (240.65,195.85) and (241.53,200.25) .. (242.85,202.6) -- cycle ;
%Shape: Moon [id:dp19498091966192144] 
\draw  [color={rgb, 255:red, 189; green, 16; blue, 224 }  ,draw opacity=1 ][fill={rgb, 255:red, 189; green, 16; blue, 224 }  ,fill opacity=1 ] (203.65,202.6) .. controls (201.22,202.6) and (199.25,197.32) .. (199.25,190.8) .. controls (199.25,184.29) and (201.22,179) .. (203.65,179) .. controls (202.33,181.36) and (201.45,185.76) .. (201.45,190.8) .. controls (201.45,195.85) and (202.33,200.25) .. (203.65,202.6) -- cycle ;
%Shape: Moon [id:dp8055487233434944] 
\draw  [color={rgb, 255:red, 0; green, 0; blue, 0 }  ,draw opacity=1 ][fill={rgb, 255:red, 0; green, 0; blue, 0 }  ,fill opacity=1 ] (190.45,202.2) .. controls (188.02,202.2) and (186.05,196.92) .. (186.05,190.4) .. controls (186.05,183.89) and (188.02,178.6) .. (190.45,178.6) .. controls (189.13,180.96) and (188.25,185.36) .. (188.25,190.4) .. controls (188.25,195.45) and (189.13,199.85) .. (190.45,202.2) -- cycle ;
%Shape: Moon [id:dp9552614138219403] 
\draw  [color={rgb, 255:red, 189; green, 16; blue, 224 }  ,draw opacity=1 ][fill={rgb, 255:red, 189; green, 16; blue, 224 }  ,fill opacity=1 ] (176.45,202.2) .. controls (174.02,202.2) and (172.05,196.92) .. (172.05,190.4) .. controls (172.05,183.89) and (174.02,178.6) .. (176.45,178.6) .. controls (175.13,180.96) and (174.25,185.36) .. (174.25,190.4) .. controls (174.25,195.45) and (175.13,199.85) .. (176.45,202.2) -- cycle ;
%Shape: Star [id:dp28910510805600576] 
\draw  [color={rgb, 255:red, 208; green, 2; blue, 27 }  ,draw opacity=1 ][fill={rgb, 255:red, 208; green, 2; blue, 27 }  ,fill opacity=1 ] (232.63,187.18) -- (233.36,188.85) -- (234.99,189.11) -- (233.81,190.4) -- (234.09,192.23) -- (232.63,191.37) -- (231.17,192.23) -- (231.45,190.4) -- (230.27,189.11) -- (231.9,188.85) -- cycle ;

% Text Node
\draw (273.08,200.63) node [anchor=north west][inner sep=0.75pt]  [font=\scriptsize] [align=left] {$\displaystyle \operatorname{med}$};
% Text Node
\draw (209.08,165.03) node [anchor=north west][inner sep=0.75pt]  [font=\tiny,color={rgb, 255:red, 208; green, 2; blue, 27 }  ,opacity=1 ] [align=left] {$\displaystyle \operatorname{\operatorname{medp}}$};

\end{tikzpicture}

% \end{document}

%% file: 2d-l1-lower-bound-instance1.tex
% \documentclass[tikz,border=10pt]{standalone}
% \usepackage{tikz}
% % Other packages or TikZ libraries
% \usetikzlibrary{decorations.pathreplacing}
% \usetikzlibrary{patterns}

% \begin{document}

% Pattern Info
 
\tikzset{
pattern size/.store in=\mcSize, 
pattern size = 5pt,
pattern thickness/.store in=\mcThickness, 
pattern thickness = 0.3pt,
pattern radius/.store in=\mcRadius, 
pattern radius = 1pt}
\makeatletter
\pgfutil@ifundefined{pgf@pattern@name@_o5c8wow0j}{
\pgfdeclarepatternformonly[\mcThickness,\mcSize]{_o5c8wow0j}
{\pgfqpoint{0pt}{0pt}}
{\pgfpoint{\mcSize}{\mcSize}}
{\pgfpoint{\mcSize}{\mcSize}}
{
\pgfsetcolor{\tikz@pattern@color}
\pgfsetlinewidth{\mcThickness}
\pgfpathmoveto{\pgfqpoint{0pt}{\mcSize}}
\pgfpathlineto{\pgfpoint{\mcSize+\mcThickness}{-\mcThickness}}
\pgfpathmoveto{\pgfqpoint{0pt}{0pt}}
\pgfpathlineto{\pgfpoint{\mcSize+\mcThickness}{\mcSize+\mcThickness}}
\pgfusepath{stroke}
}}
\makeatother
\tikzset{every picture/.style={line width=0.75pt}} %set default line width to 0.75pt        

\begin{tikzpicture}[x=0.75pt,y=0.75pt,yscale=-1,xscale=1]
%uncomment if require: \path (0,310); %set diagram left start at 0, and has height of 310

%Straight Lines [id:da09687242022595488] 
\draw    (150.17,180.56) -- (380.34,180.93) ;
%Shape: Circle [id:dp7758524110001784] 
\draw  [fill={rgb, 255:red, 0; green, 0; blue, 0 }  ,fill opacity=1 ] (248.77,180.75) .. controls (248.77,179.93) and (249.44,179.26) .. (250.26,179.26) .. controls (251.08,179.26) and (251.74,179.93) .. (251.74,180.75) .. controls (251.74,181.57) and (251.08,182.23) .. (250.26,182.23) .. controls (249.44,182.23) and (248.77,181.57) .. (248.77,180.75) -- cycle ;
%Shape: Circle [id:dp1314334449048481] 
\draw  [fill={rgb, 255:red, 0; green, 0; blue, 0 }  ,fill opacity=1 ] (299.17,180.35) .. controls (299.17,179.53) and (299.84,178.86) .. (300.66,178.86) .. controls (301.48,178.86) and (302.14,179.53) .. (302.14,180.35) .. controls (302.14,181.17) and (301.48,181.83) .. (300.66,181.83) .. controls (299.84,181.83) and (299.17,181.17) .. (299.17,180.35) -- cycle ;
%Shape: Circle [id:dp46395953788280464] 
\draw  [fill={rgb, 255:red, 0; green, 0; blue, 0 }  ,fill opacity=1 ] (223.97,180.75) .. controls (223.97,179.93) and (224.64,179.26) .. (225.46,179.26) .. controls (226.28,179.26) and (226.94,179.93) .. (226.94,180.75) .. controls (226.94,181.57) and (226.28,182.23) .. (225.46,182.23) .. controls (224.64,182.23) and (223.97,181.57) .. (223.97,180.75) -- cycle ;
%Shape: Diamond [id:dp7963241948901209] 
\draw  [color={rgb, 255:red, 245; green, 166; blue, 35 }  ,draw opacity=1 ][dash pattern={on 4.5pt off 4.5pt}] (250.26,80.67) -- (349.66,180.75) -- (250.26,280.82) -- (150.86,180.75) -- cycle ;
%Shape: Diamond [id:dp9928802092831213] 
\draw  [color={rgb, 255:red, 189; green, 16; blue, 224 }  ,draw opacity=1 ][dash pattern={on 4.5pt off 4.5pt}] (300.66,105.02) -- (375.7,180.35) -- (300.66,255.68) -- (225.62,180.35) -- cycle ;
%Shape: Diamond [id:dp30484187003572005] 
\draw  [color={rgb, 255:red, 126; green, 211; blue, 33 }  ,draw opacity=1 ][dash pattern={on 4.5pt off 4.5pt}] (225.46,130.26) -- (275.72,180.75) -- (225.46,231.23) -- (175.2,180.75) -- cycle ;
%Shape: Circle [id:dp8510967456039776] 
\draw  [color={rgb, 255:red, 208; green, 2; blue, 27 }  ,draw opacity=1 ][fill={rgb, 255:red, 208; green, 2; blue, 27 }  ,fill opacity=1 ] (173.82,179.87) .. controls (173.82,179.05) and (174.49,178.38) .. (175.31,178.38) .. controls (176.13,178.38) and (176.79,179.05) .. (176.79,179.87) .. controls (176.79,180.69) and (176.13,181.35) .. (175.31,181.35) .. controls (174.49,181.35) and (173.82,180.69) .. (173.82,179.87) -- cycle ;
%Shape: Circle [id:dp41302839144468406] 
\draw  [color={rgb, 255:red, 208; green, 2; blue, 27 }  ,draw opacity=1 ][fill={rgb, 255:red, 208; green, 2; blue, 27 }  ,fill opacity=1 ] (346.69,180.75) .. controls (346.69,179.93) and (347.35,179.26) .. (348.17,179.26) .. controls (348.99,179.26) and (349.66,179.93) .. (349.66,180.75) .. controls (349.66,181.57) and (348.99,182.23) .. (348.17,182.23) .. controls (347.35,182.23) and (346.69,181.57) .. (346.69,180.75) -- cycle ;
%Shape: Polygon [id:ds8709301033901957] 
\draw  [color={rgb, 255:red, 155; green, 155; blue, 155 }  ,draw opacity=1 ][pattern=_o5c8wow0j,pattern size=6pt,pattern thickness=0.75pt,pattern radius=0pt, pattern color={rgb, 255:red, 155; green, 155; blue, 155}][dash pattern={on 0.84pt off 2.51pt}] (200.34,156.53) -- (195.94,180.93) -- (199.54,204.93) -- (156.74,180.53) -- cycle ;
%Shape: Circle [id:dp6444145185257905] 
\draw  [fill={rgb, 255:red, 0; green, 0; blue, 0 }  ,fill opacity=1 ] (198.86,156.53) .. controls (198.86,155.71) and (199.52,155.05) .. (200.34,155.05) .. controls (201.16,155.05) and (201.83,155.71) .. (201.83,156.53) .. controls (201.83,157.36) and (201.16,158.02) .. (200.34,158.02) .. controls (199.52,158.02) and (198.86,157.36) .. (198.86,156.53) -- cycle ;
%Shape: Circle [id:dp051231207440478044] 
\draw  [fill={rgb, 255:red, 0; green, 0; blue, 0 }  ,fill opacity=1 ] (153.77,180.53) .. controls (153.77,179.71) and (154.44,179.05) .. (155.26,179.05) .. controls (156.08,179.05) and (156.74,179.71) .. (156.74,180.53) .. controls (156.74,181.36) and (156.08,182.02) .. (155.26,182.02) .. controls (154.44,182.02) and (153.77,181.36) .. (153.77,180.53) -- cycle ;
%Shape: Circle [id:dp42583607010529767] 
\draw  [fill={rgb, 255:red, 0; green, 0; blue, 0 }  ,fill opacity=1 ] (194.46,179.45) .. controls (194.46,178.63) and (195.12,177.96) .. (195.94,177.96) .. controls (196.76,177.96) and (197.43,178.63) .. (197.43,179.45) .. controls (197.43,180.27) and (196.76,180.93) .. (195.94,180.93) .. controls (195.12,180.93) and (194.46,180.27) .. (194.46,179.45) -- cycle ;
%Shape: Circle [id:dp6238201060517444] 
\draw  [fill={rgb, 255:red, 0; green, 0; blue, 0 }  ,fill opacity=1 ] (198.06,206.42) .. controls (198.06,205.6) and (198.72,204.93) .. (199.54,204.93) .. controls (200.36,204.93) and (201.03,205.6) .. (201.03,206.42) .. controls (201.03,207.24) and (200.36,207.91) .. (199.54,207.91) .. controls (198.72,207.91) and (198.06,207.24) .. (198.06,206.42) -- cycle ;

% Text Node
\draw (220.57,182.96) node [anchor=north west][inner sep=0.75pt]   [align=left] {{\scriptsize 2}};
% Text Node
\draw (246.57,182.96) node [anchor=north west][inner sep=0.75pt]   [align=left] {{\scriptsize 1}};
% Text Node
\draw (299.77,181.76) node [anchor=north west][inner sep=0.75pt]   [align=left] {{\scriptsize 3}};
% Text Node
\draw (176.97,197.56) node [anchor=north west][inner sep=0.75pt]   [align=left] {$\displaystyle \mathcal{R}_{1}$};
% Text Node
\draw (147.37,177.97) node [anchor=north west][inner sep=0.75pt]   [align=left] {{\tiny X}};
% Text Node
\draw (205.37,145.97) node [anchor=north west][inner sep=0.75pt]   [align=left] {{\tiny Y}};
% Text Node
\draw (197.94,182.45) node [anchor=north west][inner sep=0.75pt]   [align=left] {{\tiny Z}};

\end{tikzpicture}

% \end{document}

%% file: 2d-l1-lower-bound-instance2.tex
% \documentclass[tikz,border=10pt]{standalone}
% \usepackage{tikz}
% % Other packages or TikZ libraries
% \usetikzlibrary{decorations.pathreplacing}
% \usetikzlibrary{patterns}

% \begin{document}

% Pattern Info
 
\tikzset{
pattern size/.store in=\mcSize, 
pattern size = 5pt,
pattern thickness/.store in=\mcThickness, 
pattern thickness = 0.3pt,
pattern radius/.store in=\mcRadius, 
pattern radius = 1pt}
\makeatletter
\pgfutil@ifundefined{pgf@pattern@name@_hi1kn8wmh}{
\pgfdeclarepatternformonly[\mcThickness,\mcSize]{_hi1kn8wmh}
{\pgfqpoint{0pt}{0pt}}
{\pgfpoint{\mcSize}{\mcSize}}
{\pgfpoint{\mcSize}{\mcSize}}
{
\pgfsetcolor{\tikz@pattern@color}
\pgfsetlinewidth{\mcThickness}
\pgfpathmoveto{\pgfqpoint{0pt}{\mcSize}}
\pgfpathlineto{\pgfpoint{\mcSize+\mcThickness}{-\mcThickness}}
\pgfpathmoveto{\pgfqpoint{0pt}{0pt}}
\pgfpathlineto{\pgfpoint{\mcSize+\mcThickness}{\mcSize+\mcThickness}}
\pgfusepath{stroke}
}}
\makeatother
\tikzset{every picture/.style={line width=0.75pt}} %set default line width to 0.75pt        

\begin{tikzpicture}[x=0.75pt,y=0.75pt,yscale=-1,xscale=1]
%uncomment if require: \path (0,310); %set diagram left start at 0, and has height of 310

%Straight Lines [id:da09687242022595488] 
\draw    (150.17,180.56) -- (380.34,180.93) ;
%Shape: Circle [id:dp7758524110001784] 
\draw  [fill={rgb, 255:red, 0; green, 0; blue, 0 }  ,fill opacity=1 ] (248.77,180.75) .. controls (248.77,179.93) and (249.44,179.26) .. (250.26,179.26) .. controls (251.08,179.26) and (251.74,179.93) .. (251.74,180.75) .. controls (251.74,181.57) and (251.08,182.23) .. (250.26,182.23) .. controls (249.44,182.23) and (248.77,181.57) .. (248.77,180.75) -- cycle ;
%Shape: Circle [id:dp1314334449048481] 
\draw  [fill={rgb, 255:red, 0; green, 0; blue, 0 }  ,fill opacity=1 ] (299.17,180.35) .. controls (299.17,179.53) and (299.84,178.86) .. (300.66,178.86) .. controls (301.48,178.86) and (302.14,179.53) .. (302.14,180.35) .. controls (302.14,181.17) and (301.48,181.83) .. (300.66,181.83) .. controls (299.84,181.83) and (299.17,181.17) .. (299.17,180.35) -- cycle ;
%Shape: Circle [id:dp46395953788280464] 
\draw  [fill={rgb, 255:red, 0; green, 0; blue, 0 }  ,fill opacity=1 ] (223.97,180.75) .. controls (223.97,179.93) and (224.64,179.26) .. (225.46,179.26) .. controls (226.28,179.26) and (226.94,179.93) .. (226.94,180.75) .. controls (226.94,181.57) and (226.28,182.23) .. (225.46,182.23) .. controls (224.64,182.23) and (223.97,181.57) .. (223.97,180.75) -- cycle ;
%Shape: Diamond [id:dp7963241948901209] 
\draw  [color={rgb, 255:red, 245; green, 166; blue, 35 }  ,draw opacity=1 ][dash pattern={on 4.5pt off 4.5pt}] (250.26,80.67) -- (349.66,180.75) -- (250.26,280.82) -- (150.86,180.75) -- cycle ;
%Shape: Diamond [id:dp9928802092831213] 
\draw  [color={rgb, 255:red, 189; green, 16; blue, 224 }  ,draw opacity=1 ][dash pattern={on 4.5pt off 4.5pt}] (300.66,129.98) -- (349.62,180.35) -- (300.66,230.71) -- (251.7,180.35) -- cycle ;
%Shape: Diamond [id:dp30484187003572005] 
\draw  [color={rgb, 255:red, 126; green, 211; blue, 33 }  ,draw opacity=1 ][dash pattern={on 4.5pt off 4.5pt}] (225.46,130.26) -- (275.72,180.75) -- (225.46,231.23) -- (175.2,180.75) -- cycle ;
%Shape: Circle [id:dp8510967456039776] 
\draw  [color={rgb, 255:red, 208; green, 2; blue, 27 }  ,draw opacity=1 ][fill={rgb, 255:red, 208; green, 2; blue, 27 }  ,fill opacity=1 ] (173.82,179.87) .. controls (173.82,179.05) and (174.49,178.38) .. (175.31,178.38) .. controls (176.13,178.38) and (176.79,179.05) .. (176.79,179.87) .. controls (176.79,180.69) and (176.13,181.35) .. (175.31,181.35) .. controls (174.49,181.35) and (173.82,180.69) .. (173.82,179.87) -- cycle ;
%Shape: Circle [id:dp41302839144468406] 
\draw  [color={rgb, 255:red, 208; green, 2; blue, 27 }  ,draw opacity=1 ][fill={rgb, 255:red, 208; green, 2; blue, 27 }  ,fill opacity=1 ] (346.69,180.75) .. controls (346.69,179.93) and (347.35,179.26) .. (348.17,179.26) .. controls (348.99,179.26) and (349.66,179.93) .. (349.66,180.75) .. controls (349.66,181.57) and (348.99,182.23) .. (348.17,182.23) .. controls (347.35,182.23) and (346.69,181.57) .. (346.69,180.75) -- cycle ;
%Shape: Polygon [id:ds8709301033901957] 
\draw  [color={rgb, 255:red, 155; green, 155; blue, 155 }  ,draw opacity=1 ][pattern=_hi1kn8wmh,pattern size=6pt,pattern thickness=0.75pt,pattern radius=0pt, pattern color={rgb, 255:red, 155; green, 155; blue, 155}][dash pattern={on 0.84pt off 2.51pt}] (337.14,167.54) -- (356.74,180.74) -- (336.74,192.34) -- (329.54,180.74) -- cycle ;
%Shape: Circle [id:dp6238201060517444] 
\draw  [fill={rgb, 255:red, 0; green, 0; blue, 0 }  ,fill opacity=1 ] (355.26,180.74) .. controls (355.26,179.92) and (355.92,179.26) .. (356.74,179.26) .. controls (357.56,179.26) and (358.23,179.92) .. (358.23,180.74) .. controls (358.23,181.56) and (357.56,182.23) .. (356.74,182.23) .. controls (355.92,182.23) and (355.26,181.56) .. (355.26,180.74) -- cycle ;
%Shape: Circle [id:dp913534062916079] 
\draw  [fill={rgb, 255:red, 0; green, 0; blue, 0 }  ,fill opacity=1 ] (335.66,167.54) .. controls (335.66,166.72) and (336.32,166.06) .. (337.14,166.06) .. controls (337.96,166.06) and (338.63,166.72) .. (338.63,167.54) .. controls (338.63,168.36) and (337.96,169.03) .. (337.14,169.03) .. controls (336.32,169.03) and (335.66,168.36) .. (335.66,167.54) -- cycle ;
%Shape: Circle [id:dp7624793505465607] 
\draw  [fill={rgb, 255:red, 0; green, 0; blue, 0 }  ,fill opacity=1 ] (326.57,180.74) .. controls (326.57,179.92) and (327.24,179.26) .. (328.06,179.26) .. controls (328.88,179.26) and (329.54,179.92) .. (329.54,180.74) .. controls (329.54,181.56) and (328.88,182.23) .. (328.06,182.23) .. controls (327.24,182.23) and (326.57,181.56) .. (326.57,180.74) -- cycle ;
%Shape: Circle [id:dp4417614230504936] 
\draw  [fill={rgb, 255:red, 0; green, 0; blue, 0 }  ,fill opacity=1 ] (335.26,193.83) .. controls (335.26,193.01) and (335.92,192.34) .. (336.74,192.34) .. controls (337.56,192.34) and (338.23,193.01) .. (338.23,193.83) .. controls (338.23,194.65) and (337.56,195.31) .. (336.74,195.31) .. controls (335.92,195.31) and (335.26,194.65) .. (335.26,193.83) -- cycle ;

% Text Node
\draw (220.57,182.96) node [anchor=north west][inner sep=0.75pt]   [align=left] {{\scriptsize 2}};
% Text Node
\draw (246.57,182.96) node [anchor=north west][inner sep=0.75pt]   [align=left] {{\scriptsize 1}};
% Text Node
\draw (299.77,181.76) node [anchor=north west][inner sep=0.75pt]   [align=left] {{\scriptsize 3}};
% Text Node
\draw (346.17,186.76) node [anchor=north west][inner sep=0.75pt]   [align=left] {$\displaystyle \mathcal{R}_{2}$};
% Text Node
\draw (315.37,163.97) node [anchor=north west][inner sep=0.75pt]   [align=left] {{\scriptsize X}};
% Text Node
\draw (340.97,146.37) node [anchor=north west][inner sep=0.75pt]   [align=left] {{\scriptsize Y}};
% Text Node
\draw (364.17,162.77) node [anchor=north west][inner sep=0.75pt]   [align=left] {{\scriptsize Z}};

\end{tikzpicture}

% \end{document}

%% file: vertical-extended-half-diamond.tex
% \documentclass[tikz,border=10pt]{standalone}
% \usepackage{tikz}
% % Add any other packages or TikZ libraries you need for this figure
% \usetikzlibrary{decorations.pathreplacing}

% \begin{document}
\tikzset{every picture/.style={line width=0.75pt}} %set default line width to 0.75pt        

\tikzset{every picture/.style={line width=0.75pt}} %set default line width to 0.75pt        

\begin{tikzpicture}[x=0.75pt,y=0.75pt,yscale=-1,xscale=1]
%uncomment if require: \path (0,310); %set diagram left start at 0, and has height of 310

%Shape: Diamond [id:dp14760109556328704] 
\draw   (333.44,76) -- (368.44,111) -- (333.44,146) -- (298.44,111) -- cycle ;
%Shape: Diamond [id:dp5106623697049492] 
\draw   (300.5,161.4) -- (347,204.65) -- (300.5,247.9) -- (254,204.65) -- cycle ;
%Shape: Diamond [id:dp3434249557518859] 
\draw   (280.5,39.38) -- (341,101.13) -- (280.5,162.88) -- (220,101.13) -- cycle ;
%Shape: Circle [id:dp27464225147512533] 
\draw   (278.88,102.75) .. controls (278.87,101.85) and (279.6,101.12) .. (280.5,101.13) .. controls (281.4,101.13) and (282.12,101.85) .. (282.12,102.75) .. controls (282.13,103.65) and (281.4,104.38) .. (280.5,104.37) .. controls (279.6,104.37) and (278.88,103.65) .. (278.88,102.75) -- cycle ;
%Shape: Circle [id:dp782908265998904] 
\draw   (333.44,111) .. controls (333.44,110.18) and (334.11,109.5) .. (334.94,109.51) .. controls (335.77,109.51) and (336.44,110.18) .. (336.44,111.01) .. controls (336.44,111.84) and (335.77,112.51) .. (334.94,112.51) .. controls (334.11,112.5) and (333.44,111.83) .. (333.44,111) -- cycle ;
%Shape: Circle [id:dp24944265047183] 
\draw   (298.97,204.65) .. controls (298.97,203.8) and (299.65,203.12) .. (300.5,203.12) .. controls (301.34,203.12) and (302.03,203.81) .. (302.03,204.65) .. controls (302.03,205.5) and (301.35,206.18) .. (300.5,206.18) .. controls (299.66,206.18) and (298.97,205.49) .. (298.97,204.65) -- cycle ;
%Shape: Star [id:dp8527628887662962] 
\draw  [color={rgb, 255:red, 208; green, 2; blue, 27 }  ,draw opacity=1 ][fill={rgb, 255:red, 208; green, 2; blue, 27 }  ,fill opacity=1 ] (298.44,106.32) -- (299.69,108.18) -- (302.48,108.48) -- (300.46,109.92) -- (300.94,111.97) -- (298.44,111) -- (295.94,111.97) -- (296.42,109.92) -- (294.4,108.48) -- (297.19,108.18) -- cycle ;
%Straight Lines [id:da014961909085422276] 
\draw [color={rgb, 255:red, 245; green, 166; blue, 35 }  ,draw opacity=1 ][line width=1.5]    (280.5,39.38) -- (279.8,11.4) ;
%Straight Lines [id:da585188382885428] 
\draw [color={rgb, 255:red, 245; green, 166; blue, 35 }  ,draw opacity=1 ][line width=1.5]    (280.5,162.88) -- (281,259.8) ;
%Straight Lines [id:da7358568965360428] 
\draw [color={rgb, 255:red, 245; green, 166; blue, 35 }  ,draw opacity=1 ][line width=1.5]    (280.5,39.38) -- (341,101.13) ;
%Straight Lines [id:da968209216518077] 
\draw [color={rgb, 255:red, 245; green, 166; blue, 35 }  ,draw opacity=1 ][line width=1.5]    (280.5,162.88) -- (341,101.13) ;
%Straight Lines [id:da22715366283529992] 
\draw    (333.44,76) -- (298.44,111) ;
%Straight Lines [id:da6456469025057097] 
\draw [color={rgb, 255:red, 189; green, 16; blue, 224 }  ,draw opacity=1 ][line width=1.5]    (333.44,76) -- (333,10.6) ;
%Straight Lines [id:da2823606853039452] 
\draw [color={rgb, 255:red, 189; green, 16; blue, 224 }  ,draw opacity=1 ][line width=1.5]    (333.44,146) -- (333,259.8) ;
%Straight Lines [id:da1960499119407736] 
\draw [color={rgb, 255:red, 189; green, 16; blue, 224 }  ,draw opacity=1 ][line width=1.5]    (333.44,76) -- (298.44,111) ;
%Straight Lines [id:da09803998249677681] 
\draw [color={rgb, 255:red, 189; green, 16; blue, 224 }  ,draw opacity=1 ][line width=1.5]    (300.94,111.97) -- (333.44,146) ;
%Straight Lines [id:da11563006314340896] 
\draw [color={rgb, 255:red, 126; green, 211; blue, 33 }  ,draw opacity=1 ][line width=1.5]    (300.6,11.4) -- (300.5,161.4) ;
%Straight Lines [id:da007738370810562278] 
\draw [color={rgb, 255:red, 126; green, 211; blue, 33 }  ,draw opacity=1 ][line width=1.5]    (300.5,161.4) -- (347,204.65) ;
%Straight Lines [id:da35619010405048734] 
\draw [color={rgb, 255:red, 126; green, 211; blue, 33 }  ,draw opacity=1 ][line width=1.5]    (347,204.65) -- (300.5,247.9) ;
%Straight Lines [id:da14908569112007675] 
\draw [color={rgb, 255:red, 126; green, 211; blue, 33 }  ,draw opacity=1 ][line width=1.5]    (300.5,247.9) -- (300.6,260.2) ;

% Text Node
\draw (264,104) node [anchor=north west][inner sep=0.75pt]  [font=\scriptsize] [align=left] {X};
% Text Node
\draw (339.75,108.25) node [anchor=north west][inner sep=0.75pt]   [align=left] {{\scriptsize Y}};
% Text Node
\draw (286.65,209.4) node [anchor=north west][inner sep=0.75pt]   [align=left] {{\scriptsize Z}};

\end{tikzpicture}
% \end{document}

%% file: vertical-split-line.tex
% \documentclass[tikz,border=10pt]{standalone}
% \usepackage{tikz}
% % Other packages or TikZ libraries
% \usetikzlibrary{decorations.pathreplacing}

% \begin{document}
\tikzset{every picture/.style={line width=0.75pt}} %set default line width to 0.75pt        

\tikzset{every picture/.style={line width=0.75pt}} %set default line width to 0.75pt        

\begin{tikzpicture}[x=0.75pt,y=0.75pt,yscale=-1,xscale=1]
%uncomment if require: \path (0,310); %set diagram left start at 0, and has height of 310

%Shape: Diamond [id:dp14760109556328704] 
\draw   (333.44,76) -- (368.44,111) -- (333.44,146) -- (298.44,111) -- cycle ;
%Shape: Diamond [id:dp5106623697049492] 
\draw   (300.5,161.4) -- (347,204.65) -- (300.5,247.9) -- (254,204.65) -- cycle ;
%Shape: Diamond [id:dp3434249557518859] 
\draw   (280.5,39.38) -- (341,101.13) -- (280.5,162.88) -- (220,101.13) -- cycle ;
%Shape: Circle [id:dp27464225147512533] 
\draw   (278.88,102.75) .. controls (278.87,101.85) and (279.6,101.12) .. (280.5,101.13) .. controls (281.4,101.13) and (282.12,101.85) .. (282.12,102.75) .. controls (282.13,103.65) and (281.4,104.38) .. (280.5,104.37) .. controls (279.6,104.37) and (278.88,103.65) .. (278.88,102.75) -- cycle ;
%Shape: Circle [id:dp782908265998904] 
\draw   (333.44,111) .. controls (333.44,110.18) and (334.11,109.5) .. (334.94,109.51) .. controls (335.77,109.51) and (336.44,110.18) .. (336.44,111.01) .. controls (336.44,111.84) and (335.77,112.51) .. (334.94,112.51) .. controls (334.11,112.5) and (333.44,111.83) .. (333.44,111) -- cycle ;
%Shape: Circle [id:dp24944265047183] 
\draw   (298.97,204.65) .. controls (298.97,203.8) and (299.65,203.12) .. (300.5,203.12) .. controls (301.34,203.12) and (302.03,203.81) .. (302.03,204.65) .. controls (302.03,205.5) and (301.35,206.18) .. (300.5,206.18) .. controls (299.66,206.18) and (298.97,205.49) .. (298.97,204.65) -- cycle ;
%Shape: Star [id:dp8527628887662962] 
\draw  [color={rgb, 255:red, 208; green, 2; blue, 27 }  ,draw opacity=1 ][fill={rgb, 255:red, 208; green, 2; blue, 27 }  ,fill opacity=1 ] (298.44,106.32) -- (299.69,108.18) -- (302.48,108.48) -- (300.46,109.92) -- (300.94,111.97) -- (298.44,111) -- (295.94,111.97) -- (296.42,109.92) -- (294.4,108.48) -- (297.19,108.18) -- cycle ;
%Straight Lines [id:da22715366283529992] 
\draw    (333.44,76) -- (298.44,111) ;
%Straight Lines [id:da14908569112007675] 
\draw [color={rgb, 255:red, 248; green, 231; blue, 28 }  ,draw opacity=1 ][line width=2.25]    (300.5,247.9) -- (300.6,260.2) ;
%Straight Lines [id:da6974468881451846] 
\draw [color={rgb, 255:red, 248; green, 231; blue, 28 }  ,draw opacity=1 ][line width=2.25]    (333,218.2) -- (300.5,247.9) ;
%Straight Lines [id:da2194199061535459] 
\draw [color={rgb, 255:red, 248; green, 231; blue, 28 }  ,draw opacity=1 ][line width=2.25]    (333,218.2) -- (333,192.2) ;
%Straight Lines [id:da45815453684681307] 
\draw [color={rgb, 255:red, 248; green, 231; blue, 28 }  ,draw opacity=1 ][line width=2.25]    (300.5,161.4) -- (333,192.2) ;
%Straight Lines [id:da3528506844787528] 
\draw [color={rgb, 255:red, 248; green, 231; blue, 28 }  ,draw opacity=1 ][line width=2.25]    (300.5,161.4) -- (300.6,143.4) ;
%Straight Lines [id:da9358431847813297] 
\draw [color={rgb, 255:red, 248; green, 231; blue, 28 }  ,draw opacity=1 ][line width=2.25]    (300.6,143.4) -- (315,128.2) ;
%Straight Lines [id:da2868418370762589] 
\draw [color={rgb, 255:red, 248; green, 231; blue, 28 }  ,draw opacity=1 ][line width=2.25]    (298.44,111) -- (315,128.2) ;
%Straight Lines [id:da9537391923659008] 
\draw [color={rgb, 255:red, 248; green, 231; blue, 28 }  ,draw opacity=1 ][line width=2.25]    (300.46,109.92) -- (324.6,85) ;
%Straight Lines [id:da13778356716750295] 
\draw [color={rgb, 255:red, 248; green, 231; blue, 28 }  ,draw opacity=1 ][line width=2.25]    (300.6,60.2) -- (324.6,85) ;
%Straight Lines [id:da7985963165067842] 
\draw [color={rgb, 255:red, 248; green, 231; blue, 28 }  ,draw opacity=1 ][line width=2.25]    (300.6,60.2) -- (300.2,11.4) ;

% Text Node
\draw (264,104) node [anchor=north west][inner sep=0.75pt]  [font=\scriptsize] [align=left] {X};
% Text Node
\draw (339.75,108.25) node [anchor=north west][inner sep=0.75pt]   [align=left] {{\scriptsize Y}};
% Text Node
\draw (286.65,209.4) node [anchor=north west][inner sep=0.75pt]   [align=left] {{\scriptsize Z}};

\end{tikzpicture}
% \end{document}

%% file: horizontal-extended-half-diamond.tex
% \documentclass[tikz,border=10pt]{standalone}
% \usepackage{tikz}
% % Other packages or TikZ libraries
% \usetikzlibrary{decorations.pathreplacing}

% \begin{document}
\begin{tikzpicture}[x=0.75pt,y=0.75pt,yscale=-1,xscale=1]
%uncomment if require: \path (0,310); %set diagram left start at 0, and has height of 310

%Shape: Diamond [id:dp14760109556328704] 
\draw   (333.44,76) -- (368.44,111) -- (333.44,146) -- (298.44,111) -- cycle ;
%Shape: Diamond [id:dp5106623697049492] 
\draw   (300.5,161.4) -- (347,204.65) -- (300.5,247.9) -- (254,204.65) -- cycle ;
%Shape: Diamond [id:dp3434249557518859] 
\draw   (280.5,39.38) -- (341,101.13) -- (280.5,162.88) -- (220,101.13) -- cycle ;
%Shape: Circle [id:dp27464225147512533] 
\draw   (278.88,102.75) .. controls (278.87,101.85) and (279.6,101.12) .. (280.5,101.13) .. controls (281.4,101.13) and (282.12,101.85) .. (282.12,102.75) .. controls (282.13,103.65) and (281.4,104.38) .. (280.5,104.37) .. controls (279.6,104.37) and (278.88,103.65) .. (278.88,102.75) -- cycle ;
%Shape: Circle [id:dp782908265998904] 
\draw   (333.44,111) .. controls (333.44,110.18) and (334.11,109.5) .. (334.94,109.51) .. controls (335.77,109.51) and (336.44,110.18) .. (336.44,111.01) .. controls (336.44,111.84) and (335.77,112.51) .. (334.94,112.51) .. controls (334.11,112.5) and (333.44,111.83) .. (333.44,111) -- cycle ;
%Shape: Circle [id:dp24944265047183] 
\draw   (298.97,204.65) .. controls (298.97,203.8) and (299.65,203.12) .. (300.5,203.12) .. controls (301.34,203.12) and (302.03,203.81) .. (302.03,204.65) .. controls (302.03,205.5) and (301.35,206.18) .. (300.5,206.18) .. controls (299.66,206.18) and (298.97,205.49) .. (298.97,204.65) -- cycle ;
%Shape: Star [id:dp8527628887662962] 
\draw  [color={rgb, 255:red, 208; green, 2; blue, 27 }  ,draw opacity=1 ][fill={rgb, 255:red, 208; green, 2; blue, 27 }  ,fill opacity=1 ] (298.44,106.32) -- (299.69,108.18) -- (302.48,108.48) -- (300.46,109.92) -- (300.94,111.97) -- (298.44,111) -- (295.94,111.97) -- (296.42,109.92) -- (294.4,108.48) -- (297.19,108.18) -- cycle ;
%Straight Lines [id:da22715366283529992] 
\draw    (333.44,76) -- (298.44,111) ;
%Straight Lines [id:da0014222773404343192] 
\draw [color={rgb, 255:red, 245; green, 166; blue, 35 }  ,draw opacity=1 ][line width=1.5]    (200.6,101) -- (220,101.13) ;
%Straight Lines [id:da6479522440234468] 
\draw [color={rgb, 255:red, 245; green, 166; blue, 35 }  ,draw opacity=1 ][line width=1.5]    (220,101.13) -- (280.5,162.88) ;
%Straight Lines [id:da6185959632708107] 
\draw [color={rgb, 255:red, 245; green, 166; blue, 35 }  ,draw opacity=1 ][line width=1.5]    (280.5,162.88) -- (341,101.13) ;
%Straight Lines [id:da0813743490865787] 
\draw [color={rgb, 255:red, 245; green, 166; blue, 35 }  ,draw opacity=1 ][line width=1.5]    (341,101.13) -- (380.2,101) ;
%Straight Lines [id:da14785629131183287] 
\draw [color={rgb, 255:red, 126; green, 211; blue, 33 }  ,draw opacity=1 ][line width=1.5]    (200.6,204.2) -- (254,204.65) ;
%Straight Lines [id:da44338561062607074] 
\draw [color={rgb, 255:red, 126; green, 211; blue, 33 }  ,draw opacity=1 ][line width=1.5]    (380.2,204.6) -- (347,204.65) ;
%Straight Lines [id:da3191982084170747] 
\draw [color={rgb, 255:red, 126; green, 211; blue, 33 }  ,draw opacity=1 ][line width=1.5]    (254,204.65) -- (300.5,161.4) ;
%Straight Lines [id:da9889446022914942] 
\draw [color={rgb, 255:red, 126; green, 211; blue, 33 }  ,draw opacity=1 ][line width=1.5]    (300.5,161.4) -- (347,204.65) ;
%Straight Lines [id:da8371975107319448] 
\draw [color={rgb, 255:red, 189; green, 16; blue, 224 }  ,draw opacity=1 ][line width=1.5]    (200.6,109.8) -- (298.44,111) ;
%Straight Lines [id:da8880480856795583] 
\draw [color={rgb, 255:red, 189; green, 16; blue, 224 }  ,draw opacity=1 ][line width=1.5]    (298.44,111) -- (333.44,76) ;
%Straight Lines [id:da8122670691424774] 
\draw [color={rgb, 255:red, 189; green, 16; blue, 224 }  ,draw opacity=1 ][line width=1.5]    (368.44,111) -- (379.8,111) ;
%Straight Lines [id:da039649461360194094] 
\draw [color={rgb, 255:red, 189; green, 16; blue, 224 }  ,draw opacity=1 ][line width=1.5]    (333.44,76) -- (368.44,111) ;

% Text Node
\draw (264,104) node [anchor=north west][inner sep=0.75pt]  [font=\scriptsize] [align=left] {X};
% Text Node
\draw (339.75,108.25) node [anchor=north west][inner sep=0.75pt]   [align=left] {{\scriptsize Y}};
% Text Node
\draw (286.65,209.4) node [anchor=north west][inner sep=0.75pt]   [align=left] {{\scriptsize Z}};

\end{tikzpicture}

%% file: horizontal-split-line.tex
% \documentclass[tikz,border=10pt]{standalone}
% \usepackage{tikz}
% % Other packages or TikZ libraries
% \usetikzlibrary{decorations.pathreplacing}

% \begin{document}
\tikzset{every picture/.style={line width=0.75pt}} %set default line width to 0.75pt        

\tikzset{every picture/.style={line width=0.75pt}} %set default line width to 0.75pt        

\begin{tikzpicture}[x=0.75pt,y=0.75pt,yscale=-1,xscale=1]
%uncomment if require: \path (0,310); %set diagram left start at 0, and has height of 310

%Shape: Diamond [id:dp14760109556328704] 
\draw   (333.44,76) -- (368.44,111) -- (333.44,146) -- (298.44,111) -- cycle ;
%Shape: Diamond [id:dp5106623697049492] 
\draw   (300.5,161.4) -- (347,204.65) -- (300.5,247.9) -- (254,204.65) -- cycle ;
%Shape: Diamond [id:dp3434249557518859] 
\draw   (280.5,39.38) -- (341,101.13) -- (280.5,162.88) -- (220,101.13) -- cycle ;
%Shape: Circle [id:dp27464225147512533] 
\draw   (278.88,102.75) .. controls (278.87,101.85) and (279.6,101.12) .. (280.5,101.13) .. controls (281.4,101.13) and (282.12,101.85) .. (282.12,102.75) .. controls (282.13,103.65) and (281.4,104.38) .. (280.5,104.37) .. controls (279.6,104.37) and (278.88,103.65) .. (278.88,102.75) -- cycle ;
%Shape: Circle [id:dp782908265998904] 
\draw   (333.44,111) .. controls (333.44,110.18) and (334.11,109.5) .. (334.94,109.51) .. controls (335.77,109.51) and (336.44,110.18) .. (336.44,111.01) .. controls (336.44,111.84) and (335.77,112.51) .. (334.94,112.51) .. controls (334.11,112.5) and (333.44,111.83) .. (333.44,111) -- cycle ;
%Shape: Circle [id:dp24944265047183] 
\draw   (298.97,204.65) .. controls (298.97,203.8) and (299.65,203.12) .. (300.5,203.12) .. controls (301.34,203.12) and (302.03,203.81) .. (302.03,204.65) .. controls (302.03,205.5) and (301.35,206.18) .. (300.5,206.18) .. controls (299.66,206.18) and (298.97,205.49) .. (298.97,204.65) -- cycle ;
%Shape: Star [id:dp8527628887662962] 
\draw  [color={rgb, 255:red, 208; green, 2; blue, 27 }  ,draw opacity=1 ][fill={rgb, 255:red, 208; green, 2; blue, 27 }  ,fill opacity=1 ] (298.44,106.32) -- (299.69,108.18) -- (302.48,108.48) -- (300.46,109.92) -- (300.94,111.97) -- (298.44,111) -- (295.94,111.97) -- (296.42,109.92) -- (294.4,108.48) -- (297.19,108.18) -- cycle ;
%Straight Lines [id:da22715366283529992] 
\draw    (333.44,76) -- (298.44,111) ;
%Straight Lines [id:da8671961389703536] 
\draw [color={rgb, 255:red, 74; green, 144; blue, 226 }  ,draw opacity=1 ][line width=2.25]    (200.6,109.8) -- (228.2,109.8) ;
%Straight Lines [id:da29862634893086015] 
\draw [color={rgb, 255:red, 74; green, 144; blue, 226 }  ,draw opacity=1 ][line width=2.25]    (228.2,109.8) -- (280.5,162.88) ;
%Straight Lines [id:da6508364244822923] 
\draw [color={rgb, 255:red, 74; green, 144; blue, 226 }  ,draw opacity=1 ][line width=2.25]    (280.5,162.88) -- (341,101.13) ;
%Straight Lines [id:da30872328623561196] 
\draw [color={rgb, 255:red, 74; green, 144; blue, 226 }  ,draw opacity=1 ][line width=2.25]    (341,101.13) -- (358.2,101) ;
%Straight Lines [id:da03130939893874629] 
\draw [color={rgb, 255:red, 74; green, 144; blue, 226 }  ,draw opacity=1 ][line width=2.25]    (358.2,101) -- (368.44,111) ;
%Straight Lines [id:da5273214849172158] 
\draw [color={rgb, 255:red, 74; green, 144; blue, 226 }  ,draw opacity=1 ][line width=2.25]    (368.44,111) -- (380.2,111.4) ;

% Text Node
\draw (264,104) node [anchor=north west][inner sep=0.75pt]  [font=\scriptsize] [align=left] {X};
% Text Node
\draw (339.75,108.25) node [anchor=north west][inner sep=0.75pt]   [align=left] {{\scriptsize Y}};
% Text Node
\draw (286.65,209.4) node [anchor=north west][inner sep=0.75pt]   [align=left] {{\scriptsize Z}};

\end{tikzpicture}
% \end{document}

%% file: split-line-intersection-proof.tex
% \documentclass[tikz,border=10pt]{standalone}
% \usepackage{tikz}
% % Other packages or TikZ libraries
% \usetikzlibrary{decorations.pathreplacing}
% \usetikzlibrary{patterns}

% \begin{document}

% Pattern Info
 
\tikzset{
pattern size/.store in=\mcSize, 
pattern size = 5pt,
pattern thickness/.store in=\mcThickness, 
pattern thickness = 0.3pt,
pattern radius/.store in=\mcRadius, 
pattern radius = 1pt}\makeatletter
\pgfutil@ifundefined{pgf@pattern@name@_h5wbbsaex}{
\pgfdeclarepatternformonly[\mcThickness,\mcSize]{_h5wbbsaex}
{\pgfqpoint{-\mcThickness}{-\mcThickness}}
{\pgfpoint{\mcSize}{\mcSize}}
{\pgfpoint{\mcSize}{\mcSize}}
{\pgfsetcolor{\tikz@pattern@color}
\pgfsetlinewidth{\mcThickness}
\pgfpathmoveto{\pgfpointorigin}
\pgfpathlineto{\pgfpoint{\mcSize}{0}}
\pgfpathmoveto{\pgfpointorigin}
\pgfpathlineto{\pgfpoint{0}{\mcSize}}
\pgfusepath{stroke}}}
\makeatother

% Pattern Info
 
\tikzset{
pattern size/.store in=\mcSize, 
pattern size = 5pt,
pattern thickness/.store in=\mcThickness, 
pattern thickness = 0.3pt,
pattern radius/.store in=\mcRadius, 
pattern radius = 1pt}
\makeatletter
\pgfutil@ifundefined{pgf@pattern@name@_8s29nexyd}{
\pgfdeclarepatternformonly[\mcThickness,\mcSize]{_8s29nexyd}
{\pgfqpoint{0pt}{0pt}}
{\pgfpoint{\mcSize+\mcThickness}{\mcSize+\mcThickness}}
{\pgfpoint{\mcSize}{\mcSize}}
{
\pgfsetcolor{\tikz@pattern@color}
\pgfsetlinewidth{\mcThickness}
\pgfpathmoveto{\pgfqpoint{0pt}{0pt}}
\pgfpathlineto{\pgfpoint{\mcSize+\mcThickness}{\mcSize+\mcThickness}}
\pgfusepath{stroke}
}}
\makeatother
\tikzset{every picture/.style={line width=0.75pt}} %set default line width to 0.75pt        

\begin{tikzpicture}[x=0.75pt,y=0.75pt,yscale=-1,xscale=1]
%uncomment if require: \path (0,310); %set diagram left start at 0, and has height of 310

%Straight Lines [id:da26483845035655496] 
\draw [color={rgb, 255:red, 74; green, 144; blue, 226 }  ,draw opacity=1 ]   (180.17,170.2) -- (249.46,170.6) ;
%Straight Lines [id:da4570988227817907] 
\draw [color={rgb, 255:red, 74; green, 144; blue, 226 }  ,draw opacity=1 ]   (249.46,170.6) -- (280.26,141.4) ;
%Straight Lines [id:da40028001651567524] 
\draw [color={rgb, 255:red, 74; green, 144; blue, 226 }  ,draw opacity=1 ]   (280.26,141.4) -- (299.06,159.8) ;
%Straight Lines [id:da22459402982518173] 
\draw  [dash pattern={on 0.84pt off 2.51pt}]  (299.06,159.8) -- (359.86,220.6) ;
%Straight Lines [id:da6992952022111079] 
\draw  [dash pattern={on 0.84pt off 2.51pt}]  (299.06,159.8) -- (360.74,93) ;
%Straight Lines [id:da6609900145110033] 
\draw [color={rgb, 255:red, 248; green, 231; blue, 28 }  ,draw opacity=1 ]   (279.86,51.4) -- (280.26,141.4) ;
%Straight Lines [id:da543400379325052] 
\draw [color={rgb, 255:red, 208; green, 2; blue, 27 }  ,draw opacity=1 ]   (280.26,141.4) -- (289.66,150.6) ;
%Straight Lines [id:da47976267952801843] 
\draw  [dash pattern={on 4.5pt off 4.5pt}]  (289.94,170.2) -- (240.94,220.2) ;
%Straight Lines [id:da5439924936334872] 
\draw [color={rgb, 255:red, 248; green, 231; blue, 28 }  ,draw opacity=1 ][fill={rgb, 255:red, 248; green, 231; blue, 28 }  ,fill opacity=1 ]   (289.66,150.6) -- (289.94,170.2) ;
%Straight Lines [id:da40240894053317766] 
\draw  [dash pattern={on 4.5pt off 4.5pt}]  (289.94,170.2) -- (339.94,220.2) ;
%Shape: Triangle [id:dp661954713743692] 
\draw  [draw opacity=0][pattern=_h5wbbsaex,pattern size=6pt,pattern thickness=0.75pt,pattern radius=0pt, pattern color={rgb, 255:red, 155; green, 155; blue, 155}][dash pattern={on 4.5pt off 4.5pt}] (290.44,170.2) -- (339.94,220.2) -- (240.94,220.2) -- cycle ;
%Shape: Triangle [id:dp555051292007964] 
\draw  [draw opacity=0][pattern=_8s29nexyd,pattern size=6pt,pattern thickness=0.75pt,pattern radius=0pt, pattern color={rgb, 255:red, 155; green, 155; blue, 155}] (299.06,159.8) -- (359.86,93) -- (359.86,220.6) -- cycle ;
%Straight Lines [id:da2644357802221986] 
\draw    (228.57,170.6) ;
\draw [shift={(228.57,170.6)}, rotate = 180] [color={rgb, 255:red, 0; green, 0; blue, 0 }  ][line width=0.75]    (10.93,-3.29) .. controls (6.95,-1.4) and (3.31,-0.3) .. (0,0) .. controls (3.31,0.3) and (6.95,1.4) .. (10.93,3.29)   ;
%Straight Lines [id:da4558544541125865] 
\draw    (279.94,119.4) -- (279.83,123) ;
\draw [shift={(279.77,125)}, rotate = 271.75] [color={rgb, 255:red, 0; green, 0; blue, 0 }  ][line width=0.75]    (10.93,-3.29) .. controls (6.95,-1.4) and (3.31,-0.3) .. (0,0) .. controls (3.31,0.3) and (6.95,1.4) .. (10.93,3.29)   ;

\end{tikzpicture}

% \end{document}

%% file: split-line-intersection.tex
% \documentclass[tikz,border=10pt]{standalone}
% \usepackage{tikz}
% % Other packages or TikZ libraries
% \usetikzlibrary{decorations.pathreplacing}

% \begin{document}
\tikzset{every picture/.style={line width=0.75pt}} %set default line width to 0.75pt        

\tikzset{every picture/.style={line width=0.75pt}} %set default line width to 0.75pt        

\begin{tikzpicture}[x=0.75pt,y=0.75pt,yscale=-1,xscale=1]
%uncomment if require: \path (0,310); %set diagram left start at 0, and has height of 310

%Shape: Diamond [id:dp14760109556328704] 
\draw   (333.44,76) -- (368.44,111) -- (333.44,146) -- (298.44,111) -- cycle ;
%Shape: Diamond [id:dp5106623697049492] 
\draw   (300.5,161.4) -- (347,204.65) -- (300.5,247.9) -- (254,204.65) -- cycle ;
%Shape: Diamond [id:dp3434249557518859] 
\draw   (280.5,39.38) -- (341,101.13) -- (280.5,162.88) -- (220,101.13) -- cycle ;
%Shape: Circle [id:dp27464225147512533] 
\draw   (278.88,102.75) .. controls (278.87,101.85) and (279.6,101.12) .. (280.5,101.13) .. controls (281.4,101.13) and (282.12,101.85) .. (282.12,102.75) .. controls (282.13,103.65) and (281.4,104.38) .. (280.5,104.37) .. controls (279.6,104.37) and (278.88,103.65) .. (278.88,102.75) -- cycle ;
%Shape: Circle [id:dp782908265998904] 
\draw   (333.44,111) .. controls (333.44,110.18) and (334.11,109.5) .. (334.94,109.51) .. controls (335.77,109.51) and (336.44,110.18) .. (336.44,111.01) .. controls (336.44,111.84) and (335.77,112.51) .. (334.94,112.51) .. controls (334.11,112.5) and (333.44,111.83) .. (333.44,111) -- cycle ;
%Shape: Circle [id:dp24944265047183] 
\draw   (298.97,204.65) .. controls (298.97,203.8) and (299.65,203.12) .. (300.5,203.12) .. controls (301.34,203.12) and (302.03,203.81) .. (302.03,204.65) .. controls (302.03,205.5) and (301.35,206.18) .. (300.5,206.18) .. controls (299.66,206.18) and (298.97,205.49) .. (298.97,204.65) -- cycle ;
%Shape: Star [id:dp8527628887662962] 
\draw  [color={rgb, 255:red, 208; green, 2; blue, 27 }  ,draw opacity=1 ][fill={rgb, 255:red, 208; green, 2; blue, 27 }  ,fill opacity=1 ] (298.44,106.32) -- (299.69,108.18) -- (302.48,108.48) -- (300.46,109.92) -- (300.94,111.97) -- (298.44,111) -- (295.94,111.97) -- (296.42,109.92) -- (294.4,108.48) -- (297.19,108.18) -- cycle ;
%Straight Lines [id:da22715366283529992] 
\draw    (333.44,76) -- (298.44,111) ;
%Straight Lines [id:da14908569112007675] 
\draw [color={rgb, 255:red, 248; green, 231; blue, 28 }  ,draw opacity=1 ][line width=2.25]    (300.5,247.9) -- (300.6,260.2) ;
%Straight Lines [id:da6974468881451846] 
\draw [color={rgb, 255:red, 248; green, 231; blue, 28 }  ,draw opacity=1 ][line width=2.25]    (333,218.2) -- (300.5,247.9) ;
%Straight Lines [id:da2194199061535459] 
\draw [color={rgb, 255:red, 248; green, 231; blue, 28 }  ,draw opacity=1 ][line width=2.25]    (333,218.2) -- (333,192.2) ;
%Straight Lines [id:da45815453684681307] 
\draw [color={rgb, 255:red, 248; green, 231; blue, 28 }  ,draw opacity=1 ][line width=2.25]    (300.5,161.4) -- (333,192.2) ;
%Straight Lines [id:da3528506844787528] 
\draw [color={rgb, 255:red, 248; green, 231; blue, 28 }  ,draw opacity=1 ][line width=2.25]    (300.5,161.4) -- (300.6,143.4) ;
%Straight Lines [id:da9358431847813297] 
\draw [color={rgb, 255:red, 248; green, 231; blue, 28 }  ,draw opacity=1 ][line width=2.25]    (300.6,143.4) -- (315,128.2) ;
%Straight Lines [id:da2868418370762589] 
\draw [color={rgb, 255:red, 248; green, 231; blue, 28 }  ,draw opacity=1 ][line width=2.25]    (298.44,111) -- (315,128.2) ;
%Straight Lines [id:da9537391923659008] 
\draw [color={rgb, 255:red, 248; green, 231; blue, 28 }  ,draw opacity=1 ][line width=2.25]    (300.46,109.92) -- (324.6,85) ;
%Straight Lines [id:da13778356716750295] 
\draw [color={rgb, 255:red, 248; green, 231; blue, 28 }  ,draw opacity=1 ][line width=2.25]    (300.6,60.2) -- (324.6,85) ;
%Straight Lines [id:da7985963165067842] 
\draw [color={rgb, 255:red, 248; green, 231; blue, 28 }  ,draw opacity=1 ][line width=2.25]    (300.6,60.2) -- (300.2,11.4) ;

% Text Node
\draw (264,104) node [anchor=north west][inner sep=0.75pt]  [font=\scriptsize] [align=left] {X};
% Text Node
\draw (339.75,108.25) node [anchor=north west][inner sep=0.75pt]   [align=left] {{\scriptsize Y}};
% Text Node
\draw (286.65,209.4) node [anchor=north west][inner sep=0.75pt]   [align=left] {{\scriptsize Z}};

%Straight Lines [id:da8671961389703536] 
\draw [color={rgb, 255:red, 74; green, 144; blue, 226 }  ,draw opacity=1 ][line width=2.25]    (200.6,109.8) -- (228.2,109.8) ;
%Straight Lines [id:da29862634893086015] 
\draw [color={rgb, 255:red, 74; green, 144; blue, 226 }  ,draw opacity=1 ][line width=2.25]    (228.2,109.8) -- (280.5,162.88) ;
%Straight Lines [id:da6508364244822923] 
\draw [color={rgb, 255:red, 74; green, 144; blue, 226 }  ,draw opacity=1 ][line width=2.25]    (280.5,162.88) -- (341,101.13) ;
%Straight Lines [id:da30872328623561196] 
\draw [color={rgb, 255:red, 74; green, 144; blue, 226 }  ,draw opacity=1 ][line width=2.25]    (341,101.13) -- (358.2,101) ;
%Straight Lines [id:da03130939893874629] 
\draw [color={rgb, 255:red, 74; green, 144; blue, 226 }  ,draw opacity=1 ][line width=2.25]    (358.2,101) -- (368.44,111) ;
%Straight Lines [id:da5273214849172158] 
\draw [color={rgb, 255:red, 74; green, 144; blue, 226 }  ,draw opacity=1 ][line width=2.25]    (368.44,111) -- (380.2,111.4) ;
%Straight Lines [id:da3606458337079145] 
\draw [color={rgb, 255:red, 208; green, 2; blue, 27 }  ,draw opacity=1 ][line width=2.25]    (300.6,142.2) -- (315,127) ;

\end{tikzpicture}
% \end{document}

%% file: 2d-l1-strategyproof.tex
\tikzset{every picture/.style={line width=0.75pt}} %set default line width to 0.75pt        

\begin{tikzpicture}[x=0.75pt,y=0.75pt,yscale=-1,xscale=1]
%uncomment if require: \path (0,310); %set diagram left start at 0, and has height of 310

%Shape: Diamond [id:dp15536415882639598] 
\draw   (280.2,129.8) -- (311.4,160.2) -- (280.2,190.6) -- (249,160.2) -- cycle ;
%Shape: Circle [id:dp368336808446716] 
\draw  [fill={rgb, 255:red, 0; green, 0; blue, 0 }  ,fill opacity=1 ] (278.4,160.1) .. controls (278.4,159.05) and (279.25,158.2) .. (280.3,158.2) .. controls (281.35,158.2) and (282.2,159.05) .. (282.2,160.1) .. controls (282.2,161.15) and (281.35,162) .. (280.3,162) .. controls (279.25,162) and (278.4,161.15) .. (278.4,160.1) -- cycle ;
%Shape: Circle [id:dp3201909244972535] 
\draw  [fill={rgb, 255:red, 0; green, 0; blue, 0 }  ,fill opacity=1 ] (318.8,130.5) .. controls (318.8,129.45) and (319.65,128.6) .. (320.7,128.6) .. controls (321.75,128.6) and (322.6,129.45) .. (322.6,130.5) .. controls (322.6,131.55) and (321.75,132.4) .. (320.7,132.4) .. controls (319.65,132.4) and (318.8,131.55) .. (318.8,130.5) -- cycle ;
%Shape: Diamond [id:dp5934802766992875] 
\draw  [color={rgb, 255:red, 245; green, 166; blue, 35 }  ,draw opacity=1 ][dash pattern={on 4.5pt off 4.5pt}] (280.2,109.6) -- (331,160.2) -- (280.2,210.8) -- (229.4,160.2) -- cycle ;
%Shape: Diamond [id:dp7427167344581358] 
\draw  [color={rgb, 255:red, 155; green, 155; blue, 155 }  ,draw opacity=1 ][dash pattern={on 0.84pt off 2.51pt}] (280.05,89.8) -- (349.9,160.6) -- (280.05,231.4) -- (210.2,160.6) -- cycle ;
%Straight Lines [id:da32929042849050383] 
\draw    (282.2,160.1) -- (309.4,160.19) ;
\draw [shift={(311.4,160.2)}, rotate = 180.2] [color={rgb, 255:red, 0; green, 0; blue, 0 }  ][line width=0.75]    (10.93,-3.29) .. controls (6.95,-1.4) and (3.31,-0.3) .. (0,0) .. controls (3.31,0.3) and (6.95,1.4) .. (10.93,3.29)   ;
%Straight Lines [id:da8167054730020608] 
\draw    (280.05,160.6) -- (280.19,111.6) ;
\draw [shift={(280.2,109.6)}, rotate = 90.17] [color={rgb, 255:red, 0; green, 0; blue, 0 }  ][line width=0.75]    (10.93,-3.29) .. controls (6.95,-1.4) and (3.31,-0.3) .. (0,0) .. controls (3.31,0.3) and (6.95,1.4) .. (10.93,3.29)   ;
%Straight Lines [id:da2658634147627216] 
\draw    (280.3,160.1) -- (319.09,131.68) ;
\draw [shift={(320.7,130.5)}, rotate = 143.77] [color={rgb, 255:red, 0; green, 0; blue, 0 }  ][line width=0.75]    (10.93,-3.29) .. controls (6.95,-1.4) and (3.31,-0.3) .. (0,0) .. controls (3.31,0.3) and (6.95,1.4) .. (10.93,3.29)   ;

% Text Node
\draw (271.2,161.6) node [anchor=north west][inner sep=0.75pt]   [align=left] {{\tiny (0, 0)}};
% Text Node
\draw (325.6,116.4) node [anchor=north west][inner sep=0.75pt]   [align=left] {{\fontfamily{pcr}\selectfont {\scriptsize t}}};

\end{tikzpicture}